\documentclass[11pt,a4paper]{article} 
\pdfoutput=1
\usepackage[utf8x]{inputenc}     
\usepackage{jheppub}
\usepackage{enumerate}
\usepackage{epsfig} 
\usepackage{float} 
\usepackage[caption = false]{subfig}
\usepackage{wasysym} 
\usepackage{mathrsfs} 
\usepackage{amsfonts} 
\usepackage{amsbsy} 
\usepackage{amscd} 
\usepackage{cancel}
\usepackage{pstricks} 
\usepackage{multirow} 
\usepackage{tikz}
\usepackage{color}
\usepackage{xcolor}
\usepackage{slashed}
\usepackage{array}
\usepackage{tablefootnote}
\usepackage[normalem]{ulem}

\usetikzlibrary{arrows,positioning,shapes.geometric} 
\usepackage[compat=1.1.0]{tikz-feynman}          
\usepackage[font=small,labelfont=bf]{caption}
\usepackage{diagbox}
\usepackage{multirow}
\usepackage{tabularx}
\usepackage{soul}  
\usepackage{listings} 
\usepackage{cleveref}
\usepackage{color}

\definecolor{somethingC}{rgb}{0.07, 0.04, 0.56} 

\usepackage{amsthm}

 \newcommand{\ds}[1]{{\color[HTML]{008080} #1}}

\newcommand{\stkout}[1]{\ifmmode\text{\sout{\ensuremath{#1}}}\else\sout{#1}\fi}
\title{Precise probing and discrimination of third-generation scalar leptoquarks} 
 
\author[a,b]{Anupam Ghosh,}
\author[a]{Partha Konar,}
\author[a]{Debashis Saha,}
\author[a]{Satyajit Seth}

\affiliation[a]{Theoretical Physics Division, Physical Research Laboratory, Shree Pannalal Patel Marg, Ahmedabad, 380009, Gujarat, India}
\affiliation[b]{Discipline of Physics, Indian Institute of Technology, Palaj, Gandhinagar, 382424, Gujarat, India}

\emailAdd{anupam@prl.res.in}
\emailAdd{konar@prl.res.in}
\emailAdd{debasaha@prl.res.in}
\emailAdd{seth@prl.res.in}

\abstract{We explore the pair production of third-generation scalar leptoquark at the Large Hadron Collider to next-to-leading order accuracy in QCD, matched to parton shower for a precise probing of the stemming  model. We propose to tag two boosted top-like fatjets produced from the decay of heavy leptoquarks in association with notably large missing transverse momentum and consider them as the potential signal. Such a signal demonstrates the capability of a robust discovery prospect in the multivariate analysis with different high-level observables, including jet substructure variables. Various scalar leptoquark models predict different chirality of the top quark appearing from the decay of the leptoquark carrying same electromagnetic charge. We make use of the polarization variables sensitive to the top quark polarization in order to identify the underlying theory.}
\preprint{\today}

\keywords{Leptoquark, QCD corrections, Boosted top, Jet substructure, Top polarization}

\begin{document}
\maketitle

\newpage

\section{Introduction}
\label{intro}
Leptoquark (LQ) is a hypothetical particle that couples to quark and lepton together. It carries both baryon number and lepton number and provides a means to unify quarks and leptons. It can appear in many interesting scenarios beyond the Standard Model, for example, Pati-Salam model~\cite{Pati:1974yy, Pati:1973uk}, Grand unified theory ~\cite{Georgi:1974sy, Langacker:1980js}, Composite model~\cite{Wudka:1985ef} {\em etc.}, and therefore it remains as a very active area in experimental searches.
In some of these models, the baryon number gets violated and that allows protons to decay. But the strong constraints from the non-observation of proton decay so far have pushed the masses of the leptoquark to a very high scale, typically around $10^{16}$ GeV.
However, imposing baryon number or lepton number conservation one gets a set of leptoquarks in the Buchm{\"u}ller-R{\"u}ckl-Wyler (BRW) framework~\cite{Buchmuller:1986zs}, which allows leptoquark masses to be in a range accessible to the collider searches. Also, such leptoquarks are favourable to explain anomalies observed in the B-meson decays in BaBar~\cite{BaBar:2013mob}, Belle~\cite{BELLE:2019xld,Belle:2019gij,Belle:2019oag} and LHCb  experiments~\cite{LHCb:2015gmp,LHCb:2023zxo}. Note that recent results from the LHCb with $\rm 9\ fb^{-1}$ of data have made the anomaly disappear in the measurement of neutral current observables, {\em i.e.} $R_K$ and $R_{K^*}$ \footnote{$R_K=\frac{\Gamma(B \rightarrow K \mu^+ \mu^-)}{\Gamma(B \rightarrow K e^+ e^-)}$ and $R_{K^*}=\frac{\Gamma(B \rightarrow K^* \mu^+ \mu^-)}{\Gamma(B \rightarrow K^* e^+ e^-)}$}~\cite{LHCb:2022zom, LHCb:2022qnv}. 
Although these new results imply no lepton flavour universality violation in the flavour-changing neutral current (FCNC) from the decay of B-meson, they do not exclude the possibility of the existence of a TeV scale leptoquark.
 Such results only indicate that, if a TeV-scale leptoquark exists, either it may not couple to the bottom and strange quarks at the same time together with a lepton, or the couplings are such that they get cancelled in the ratio of the decay widths. Further implications of the LHCb results on the parameter space of different leptoquark models in various production channels have been discussed in \cite{Desai:2023jxh}.



In low-energy experiments, leptoquarks may be probed indirectly because they appear as off-shell states. Here optimal ratios are formed to reduce the uncertainty due to hadronic activities. However, the number of theory parameters that enter through such ratios also increases. Instead, in the present and future high-energy collider experiments, leptoquarks can be searched directly and indirectly looking into a specific production channel. In order to explain the observation of anomaly, the leptoquark is needed to be coupled to fermions of different generations, in addition to the same generation quark and lepton. 
Although in most of the previous direct searches leptoquark couplings were considered generation-wise, recent experimental studies are extended to include different cross generation fermions \cite{ATLAS:2020dsk, ATLAS:2022wcu}. 
Bounds on the first and second generation scalar leptoquarks are obtained at the LHC considering production of either two charged light leptons of the same flavour or one charged light lepton with sizeable missing transverse energy together with a pair of jets. The limits on the parameters of the different components of the leptoquark model are obtained assuming different branching fractions.
At 95\% CL, ATLAS collaboration has constrained the mass of first two generation scalar leptoquarks up to 1400 GeV assuming 100\% branching into certain decay modes with 36.1 $\rm fb^{-1}$ data~\cite{ATLAS:2019ebv}. The CMS collaboration has also excluded masses below 1430 GeV and 1530 GeV for the first and second generation respectively, with 35.9 $\rm fb^{-1}$ data using the same branching fraction~\cite{CMS:2018lab, CMS:2018ncu}. For the recent bounds on the vector leptoquark from collider searches see Ref.~\cite{ATLAS:2023uox,CMS:2020wzx,CMS:2018qqq}.

In this work, we focus on the third-generation scalar leptoquarks. Phenomenology of such leptoquarks are studied widely in different channels~\cite{Gripaios:2010hv,Chandak:2019iwj, Bhaskar:2021gsy, Belanger:2021smw} and they are also searched by the ATLAS and CMS collaborations~\cite{ATLAS:2019qpq,ATLAS:2023uox,CMS:2020wzx,CMS:2022zks}.
{In a recent analysis \cite{ATLAS:2020dsf}, the ATLAS collaboration did a cut-based analysis and extracted the limit for the up-type third-generation scalar leptoquark, assuming LQ decaying into a top quark and neutrino with a $100\%$ branching ratio. Their analysis put a lower limit of 1240 $\rm GeV$ on the LQ mass at $95\%$ CL for an integrated luminosity of 139 $\rm fb^{-1}$ at the 13 $\rm TeV$ LHC. This paper presents an alternative search strategy considering two top-like fatjets plus significant missing energy in the final state with a sophisticated multivariate analysis of the NLO+PS signal events including jet substructure variables.} Given the already constrained parameter space, a relatively heavy leptoquark would naturally produce top quark at the boosted region once produced from its decay. Thus, it is prudent to identify such top quarks as a top-like fatjet from its hadronic decay. Note that the corresponding leptonic decay mode not only suffers from branching ratio suppression, but also identifying such leptons inside a jetty signature is a challenging task and therefore it affects the efficiency significantly.
We observe that our result is consistent with the existing search and find that the third-generation LQ can be discovered with a significance of $\geq 5 \sigma$ for masses below 1380 GeV with 3000 $\rm fb^{-1}$ data at the High-Luminosity LHC (HL-LHC).

Further, we put limit on the LQ mass up to which HL-LHC can exclude such LQ models with $95\%$ confidence level. For the third generation leptoquark, $t\bar{t}$ plus missing energy channel was also used in Ref.~\cite{Vignaroli:2018lpq}, where the authors found that $Z+$jets is the main background, while $t\bar{t}+$jets is the negligible one. However, the mono-boson background can be controlled substantially by enforcing at least one b-tagging inside the leading or subleading top-like fatjet. Such a demand in our analysis brings the mono-boson background in a similar footing as $t\bar{t}+$jets, thereby improving the result significantly.

{Once discovered, the next goal would be distinguishing scalar leptoquarks of the same electromagnetic charge. Therefore we also analyze distinguishing different scalar leptoquark models based on the same final state signature at the LHC. One proposal has been made to determine different leptoquark types of the same spin and different electromagnetic charges by measuring jet charge~\cite{Bandyopadhyay:2020wfv}.
We show that in the context of third-generation up-type leptoquark, measuring the polarization of the top quark resulting from the leptoquark decay can be an efficient way to distinguish scalar leptoquark models of the same electromagnetic charge without requiring the measurement of jet charge. In this work, for the first time, we use polarization variables to distinguish two scalar leptoquark models, considering all the backgrounds.}

As the top quark decays before it hadronizes, its spin information can be obtained from its decay products\footnote{Other quarks form bound states before their decay and hence lose their spin information.}\cite{Barger:1988jj}. Top quark polarization has been studied for more than last thirty years~\cite{Hagiwara:1985yu, Kane:1991bg,Dalitz:1991wa,Sumino:2005pg,Godbole:2006tq,Bernreuther:2008ju,Shelton:2008nq,Godbole:2009dp,Godbole:2010kr,Bhattacherjee:2012ir, Bernreuther:2014dla,PrasathV:2014omf,Bhattacharya:2020aid}.
 Determination of the polarization of boosted top-quark is studied in~\cite{Krohn:2009wm}. The possibility of distinguishing two models in the $t\bar{t}\tau\bar{\tau}$ channel was explored before for scalar leptoquark in Ref.~\cite{Papaefstathiou:2011kd} without signal-to-background study. The prospect of distinguishing a scalar leptoquark from the background based on polarization variables at the LHC was shown to be small~\cite{Roy:2018nwc}. The potential of discriminating two specific BSM scenarios in mono-top search at the LHC using top polarization has been explored in~\cite{Allahverdi:2015mha}. 

We set our probe strategy based on two chosen leptoquark models namely $S_3$ and $R_2$, that produce same pair production cross-section; but the top quark is produced as left and right chiral for these models, respectively.
We find the difference in the kinematic distributions of these two models due to different chirality rendering minimal effect in separating the signal from the background. It leads to the almost identical mass limit for exclusion and discovery potential of these two models. However, one can use the polarization variables like the ratio of the b-jet energy to the reconstructed top jet energy to distinguish two models at 14 TeV LHC and a futuristic 27 TeV collider (HE-LHC).

We consider the signal events at the next-to-leading order (NLO) in QCD matched to parton shower (PS) for reduced scale uncertainties and realistic results. By matching the fixed-order (FO) NLO correction with the parton shower (PS)~\cite{Kramer:2004df,Mandal:2015lca}, we get more accurate results for different kinematic distributions as it resums large leading logarithms in the collinear region. We show the effect of NLO+PS calculations on different kinematic distributions of the leptoquarks. 

The rest of the paper is organized as follows. In Sec.~\ref{sec:model}, we describe the third-generation scalar leptoquark models. 
 In Sec.~\ref{sec:nlops}, we show the effect of NLO calculations. We study the impact of parton shower over the fixed-order (FO) NLO calculation, k-factor variation in differential distributions, and reduction of scale uncertainties at the NLO+PS accuracy. In Sec.~\ref{sec:collider}, we describe our search strategy and provide details on multivariate analysis used to discriminate the signal and the background.
In Sec.~\ref{sec:distinguigh_models}, we discuss how the polarization observables can be instrumental in distinguishing two above mentioned models. Finally, we summarize and conclude in Sec.~\ref{sec:conclsn}.

\section{The models}
\label{sec:model}
Under the Standard Model (SM) gauge group $SU(3)_c \otimes SU(2)_L \otimes U(1)_Y$, there are total six species of scalar leptoquarks, namely $S_3$, $R_2$, $\tilde{R}_2$, $\tilde{S}_1$, ${S}_1$ and $\bar{S}_1$. Since a quark transforms as a triplet of $SU(3)_c$, a leptoquark should also transform as the same multiplet of $SU(3)_c$ in order to form gauge invariant interaction terms. In Tab.~\ref{tab:models}, we show the SM quantum numbers of all scalar leptoquarks. The subscripts on the model name denote their $SU(2)_L$ quantum numbers. If two or more models have same $SU(2)_L$ quantum number but different hypercharges, a tilde or bar is used to identify them. The component fields of the electroweak multiplets are written in the third column of the table with superscripts denoting their electric charges. In this article we are interested in studying the third generation scalar leptoquarks only. Various decay channels of the component fields for third generation leptoquarks are written inside parentheses.

It is interesting to notice that in only two models fields transform as  {$\bf 3$} and for the rest of the models they transform as {$\bf \bar{3}$} under $SU(3)_c$. Let us, for example, consider the $S_3$ model in which fields transform as {$\bf \bar{3}$}. The reason behind this transformation is that the fields in $S_3$ should couple to a quark doublet Q and lepton doublet L, as it transforms as {$\bf 3$} under $SU(2)_L$\footnote{2$\otimes 2= 3 \oplus 1$}. But it can couple to only $\bar{Q^C}L$, not with $\bar{Q}L$, since the latter is zero . As $\bar{Q^C}$ transforms as {$\bf 3$}, $S_3$ would transform as {$\bf \bar{3}$} \footnote{For $R_2$ model, fields transform as {$\bf 2$} under $SU(2)_L$ and therefore one fermion needs to be doublet, while the other one needs to be singlet. The doublet and singlet are left and right handed respectively. Hence, in this case, interaction with charge conjugated quark field vanishes, while the one without charge conjugation survives. This explains why $R_2$ transforms as {$\bf 3$} under $SU(3)_c$.}. Obviously the conjugate of $S_3$ transforms as {$\bf 3$}, however for that lepton doublet precedes the quark doublet in the Lagrangian. Here we label a leptoquark as field (as opposed to the conjugate field) if in the interaction term quark precedes lepton. Transformation properties of the other leptoquarks under $SU(3)_c$ can be understood in a similar way.


\begin{table}[!tb]
\begin{center}
\resizebox{\columnwidth}{!}{%
\begin{tabular}{|c|c|c|c|c||c||}
\hline
Models & {\tiny ($SU(3)_c, SU(2)_L, U(1)_Y$)} & Components \& Decay \\
   \hline
$S_3$   & ({\bf $\bar{3}, 3, \frac{1}{3}$})& $S_3^{\frac{4}{3}}(\tilde{b}, \tau^{+})$, $S_3^{\frac{1}{3}}((\tilde{t}, \tau^{+}), (\tilde{b}, \tilde{\nu}_\tau))$, $S_3^{-\frac{2}{3}}(\tilde{t}, \tilde{\nu}_\tau)$ \\
   \hline
$R_2$   & ($3, 2, \frac{7}{6}$) & $R_2^{\frac{5}{3}}(t , \tau^{+})$, $R_2^{\frac{2}{3}} ((t , \tilde{\nu}_\tau), (b,  \tau^{+}))$ \\
   \hline
$\tilde{R}_2$   & ($3, 2, \frac{1}{6}$) & $\tilde{R}_2^{\frac{2}{3}}((t , \tilde{N}_\tau),(b , \tau^{+}))$, $\tilde{R}_2^{-\frac{1}{3}}((b , \tilde{\nu}_\tau),(b ,  \tilde{N}_\tau))$ \\
   \hline
$\tilde{S}_1$   & ($\bar{3}, 1, \frac{4}{3}$) & $\tilde{S}_1^{\frac{4}{3}}(\tilde{b}, \tau^{+})$ \\
   \hline
${S}_1$   & ($\bar{3}, 1, \frac{1}{3}$) & $S_1^{\frac{1}{3}}((\tilde{t}, \tau^{+}), (\tilde{b}, \tilde{\nu}_\tau), (\tilde{b}, \tilde{N}_\tau))$ \\
   \hline
$\bar{S}_1$   &  ($\bar{3}, 1, -\frac{2}{3}$)  & $\bar{S}_1^{-\frac{2}{3}}(\tilde{t}, \tilde{N}_\tau)$  \\
   \hline
  \end{tabular}
}
\caption{All the possible scalar leptoquark models which give gauge invariant terms in the Lagrangian under the SM gauge group transformations. To learn about the naming convention used for the models, see the text.
}
\label{tab:models}
 \end{center}
\end{table}


One might be interested to probe third generation up-type scalar leptoquark component fields which have $\frac{2}{3}e$ electric charge. There are four such component fields, namely $S_3^{-\frac{2}{3}}$, $R_2^{\frac{2}{3}}$, $\tilde{R}_2^{\frac{2}{3}}$, and $\bar{S}_1^{-\frac{2}{3}}$. At the LHC, the first two fields can give two top fatjets plus missing energy as the signature, whereas the last two, depending on the right handed heavy neutrino decay mechanisms, will give more complicated and model dependent signatures. In this paper, we are interested to study the phenomenology of the $S_3^{-\frac{2}{3}}$ and $R_2^{\frac{2}{3}}$ fields only. 

%

The kinetic term for the generic scalar Leptoquark (S)  can be written as,
\begin{align}
{\cal{L}}_{\rm{kin}} =(D_\mu S)^\dagger (D^\mu S)  - M_S^2 S^\dagger S \,. \label{eq:Lkin}
\end{align}
Here the covariant derivative $D_\mu$ is given as,
\begin{align}
D_\mu = \partial_\mu -i g_s \lambda^a G^a_\mu\,,\label{eq:CovarDer}
\end{align}
where $g_s$ is the strong coupling, $\lambda^a$ and $G^a$ ($a=1,...,8$) denote the Gellman matrices and gluon fields respectively. The above Lagrangian gives rise to the following two vertices -- ({\em i}) gluon-LQ-LQ, ({\em ii}) gluon-gluon-LQ-LQ. The Feynman rules for these vertices are independent of the type of leptoquarks.

The quantum numbers of leptoquark $S_3$ is such that it can allow diquark coupling. However, without baryon or lepton number conservation, for TeV-scale leptoquark, this coupling has to be too tiny as otherwise it would lead to proton decay
\footnote{The models $R_2$ and $\tilde{R}_2$ which do not allow any diquark coupling are called genuine leptoquark. The rest four scalar leptoquark models allow diquark couplings.}.  As this coupling is too constrained, in our analysis they do not play any role\footnote{The diquark coupling can also be forbidden by demanding either baryon number or lepton number conservation.}. The interaction terms for the third generation scalar leptoquarks $S_3$ and $R_2$ of charge $\frac{2}{3}e$, with a quark and a lepton, are given by~\cite{Dorsner:2016wpm},
\begin{align}
{{\cal{L}}_{\rm{Int}}^{S_3^{\tiny \frac{2}{3}}}} &= y_{S_{LL}} * \bar{t_L^C}\ v_{\tau}\  S_3^{-\frac{2}{3}} + h.c., \label{eq:LInt} \\
{{\cal{L}}_{\rm{Int}}^{R_2^{\tiny \frac{2}{3}}}} &= y_{R_{RL}} * \bar{t_R}v_{\tau}\  R_2^{\frac{2}{3}}   +  y_{R_{LR}} * \bar{b_L}\tau_R\  R_2^{\frac{2}{3}} + h.c., \label{eq:LInt2}
\end{align}
where ``RL" in $y_{R_{RL}}$ signifies that the chiralities of the quark and lepton are right-handed and left-handed, respectively. Other subscripts also carry the same convention. As $S_3^{\tiny \frac{2}{3}}$ has only one decay channel ({\em i.e.,} $S_3^{\tiny \frac{2}{3}}\to t_L\nu_\tau$), it has 100\% branching fraction for it. Although $R_2^{\tiny \frac{2}{3}}$ has two decay channels, in our present analysis we shall assume 100\% branching fraction to its $t_R\tilde{\nu}_\tau$ decay mode, which can easily be scaled to other values as required\footnote{For other branching fraction, the production cross-section of leptoquark pair will also depend on $y_{R_{LR}}$ in the five flavor scheme, since a t-channel production diagram will appear when $y_{R_{LR}}$ is non-zero. However, in Ref~\cite{Mandal:2015lca} it has been shown that the dependence of the cross-section on this parameter is quite small.}. 

\section{Pair production at NLO+PS accuracy}
\label{sec:nlops}
\begin{figure}[!tb]
\begin{center}
\includegraphics[angle=0,scale=0.8]{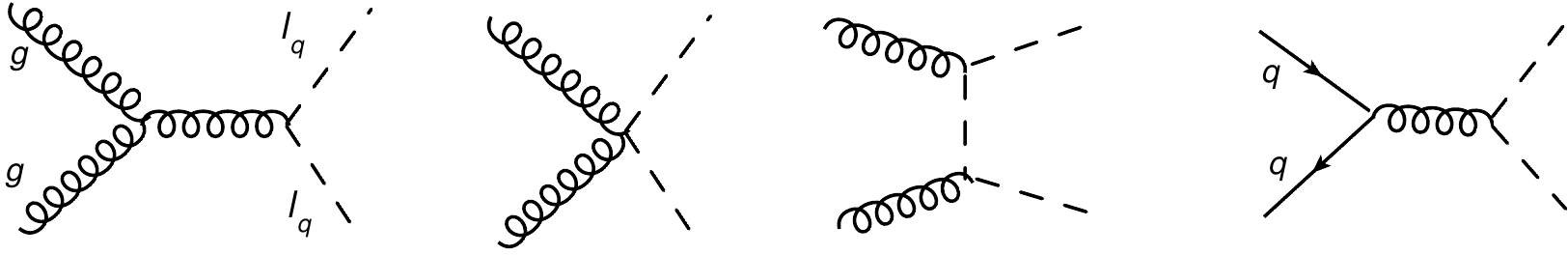}\\
\vspace{0.3cm}
\includegraphics[angle=0,scale=0.8]{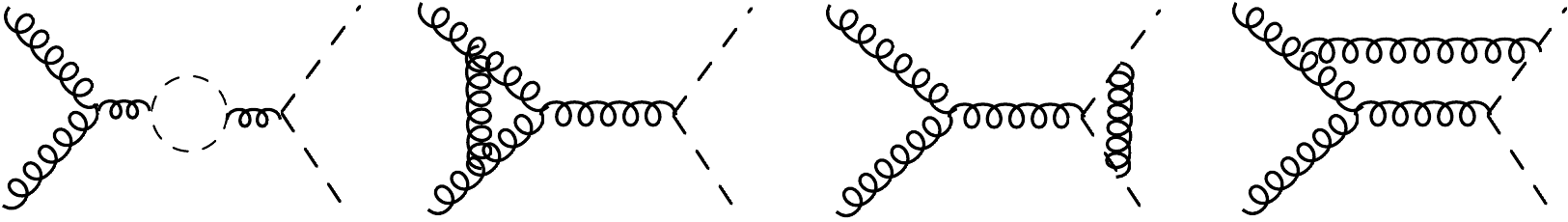}
 \vspace{0.0cm}
\caption{In the upper row, all possible prototype born diagrams are shown. In the lower row, only few prototype virtual diagrams are shown.} 
 \label{fig:feynman_dia}
\end{center}
\end{figure}

We consider signal events at the NLO in QCD matched to parton shower. The production of events at the NLO(FO) QCD accuracy requires calculating amplitudes of LO, virtual and real-emission Feynman diagrams. We show all the LO and a few virtual Feynman diagrams for the pair production of scalar leptoquarks in Fig.~\ref{fig:feynman_dia}. The real-emission diagrams are not shown which are tree level diagrams with an extra gluon or light quark. The diagrams are drawn using {\it JaxoDraw} package~\cite{Binosi:2003yf}. 
The events are produced using {\it MadGraph5\_aMC@NLO}~\cite{Alwall:2014hca}.  For the signal, we first write the model in {\it FeynRules}~\cite{Alloul:2013bka}  and use {\it NLOCT}~\cite{Degrande:2014vpa} package\footnote{The NLOCT package calculates the UV and R2 terms of the OPP method\cite{Ossola:2006us}.} to produce the UFO model~\cite{Degrande:2011ua}. This UFO model is then used in {\it MadGraph5\_aMC@NLO} to generate events at the NLO(FO) accuracy. To account for the infrared divergence in real emission processes, {\it MadGraph5\_aMC@NLO} uses the FKS subtraction method~\cite{Frixione:1995ms,Frixione:1997np}. 
\\

In Tab.~\ref{tab:cross-section}, we show the cross-sections for the production of pair of scalar leptoquarks of mass $\rm M_{LQ}= 1300\ GeV$ at 14 TeV LHC at LO and NLO(FO). The cross-sections for both the models $S_3^{\frac{2}{3}}$ and $R_2^{\frac{2}{3}}$ are same up to Monte Carlo uncertainty, as expected from the discussion in the previous section. Corrections due to the NLO QCD effects are around $10\%$. We have used {\it NNPDF23\_lo\_as\_0119\_qed} and {\it NNPDF23\_nlo\_as\_0119\_qed} parton distribution functions, respectively, for the LO and NLO calculations. The partonic center of mass energy is used as the central choice for the renormalisation and factorization scales. For the scale variation study, we vary the renormalization and factorization scales up and down by a factor of two, resulting in total nine points including the central choice. The upper and lower envelopes of the variations of the cross-section due to these different choices of scales are shown as the percentage change from the central cross-section in the superscript and subscript, respectively. From the table, we see the NLO QCD correction here reduces the scale uncertainty by around a factor of two. 
\\

\begin{table}[H]
\begin{center}
 \begin{tabular}{|c|c|c|}
\hline
\multirow{2}{*}{\diagbox[height=2\line]{model}{order}} & LO & NLO(FO) \\
            & (fb) & (fb) \\
\hline
 $S_3^{\frac{2}{3}}$ & $0.6621^{+37.8\%}_{-25.8\%}$ & $0.7229^{+14.5\%}_{-14.7\%}$  \\
\hline
 $R_2^{\frac{2}{3}}$ &  $0.6631^{+37.8\%}_{-25.8\%}$ & $0.7163^{+14.9\%}_{-14.8\%}$ \\
\hline
\end{tabular}
\caption{Cross-sections for the pair production of scalar leptoquarks of mass $\rm M_{LQ}= 1300\ GeV$ at the 14 TeV LHC. The scale variation are shown in subscript and superscript.}
 \label{tab:cross-section}
\end{center}
\end{table}

\begin{figure}[H]
\centering
\includegraphics[angle=0,scale=0.7]{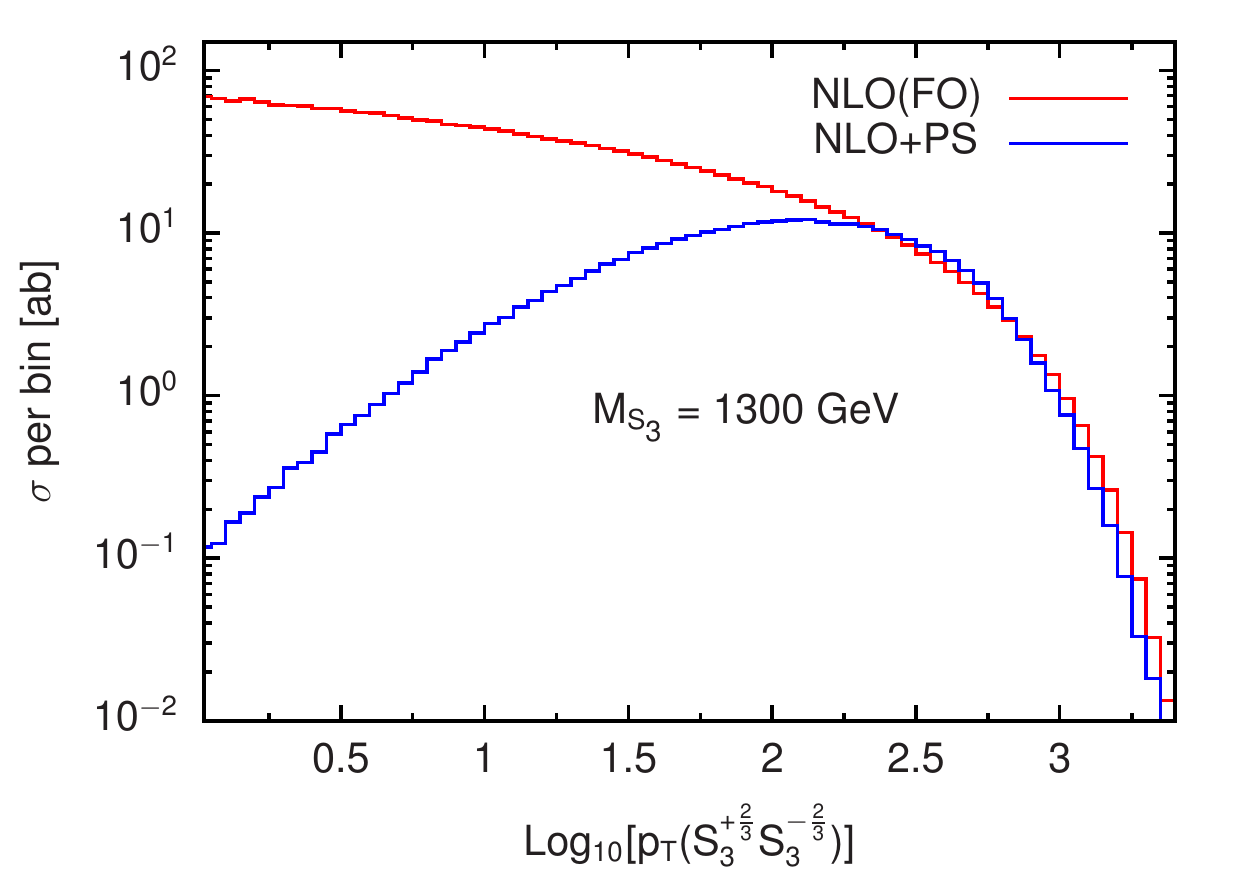}
\caption{Distributions of $\rm{Log_{10}\;[p_T(S_3^{+\frac{2}{3}}S_3^{-\frac{2}{3}})]}$ for NLO(FO) and NLO matched to parton shower.} 
 \label{fig:PSeffectoverNLOFO}
\end{figure}

The NLO(FO) results discussed in the above two paragraphs can give distributions of different kinematic variables using weighted events, but unweighting of the these events cannot be done as the matrix elements are not bounded in this case~\cite{Alwall:2014hca}.
Also in this case, result is not physical for low $p_T$ region. 
However, it can produce unweighted events while matched to the parton shower making use of the MC@NLO formalism~\cite{Frixione:2002ik}. Results at the NLO+PS accuracy give correct description of the low $p_T$ region. 
For showering of events, we use {\it Pythia8}~\cite{Bierlich:2022pfr}. In Fig.~\ref{fig:PSeffectoverNLOFO}, we see that  NLO+PS calculation over the fixed order one  reduces the cross-section at the lower transverse momentum region of the leptoquark pair system $\rm{p_T(S_3^{+\frac{2}{3}}S_3^{-\frac{2}{3}})}$ due to the Sudakov suppression.

\begin{figure}[H]
\centering
\includegraphics[angle=0,width=0.47\linewidth,height=0.43\linewidth]{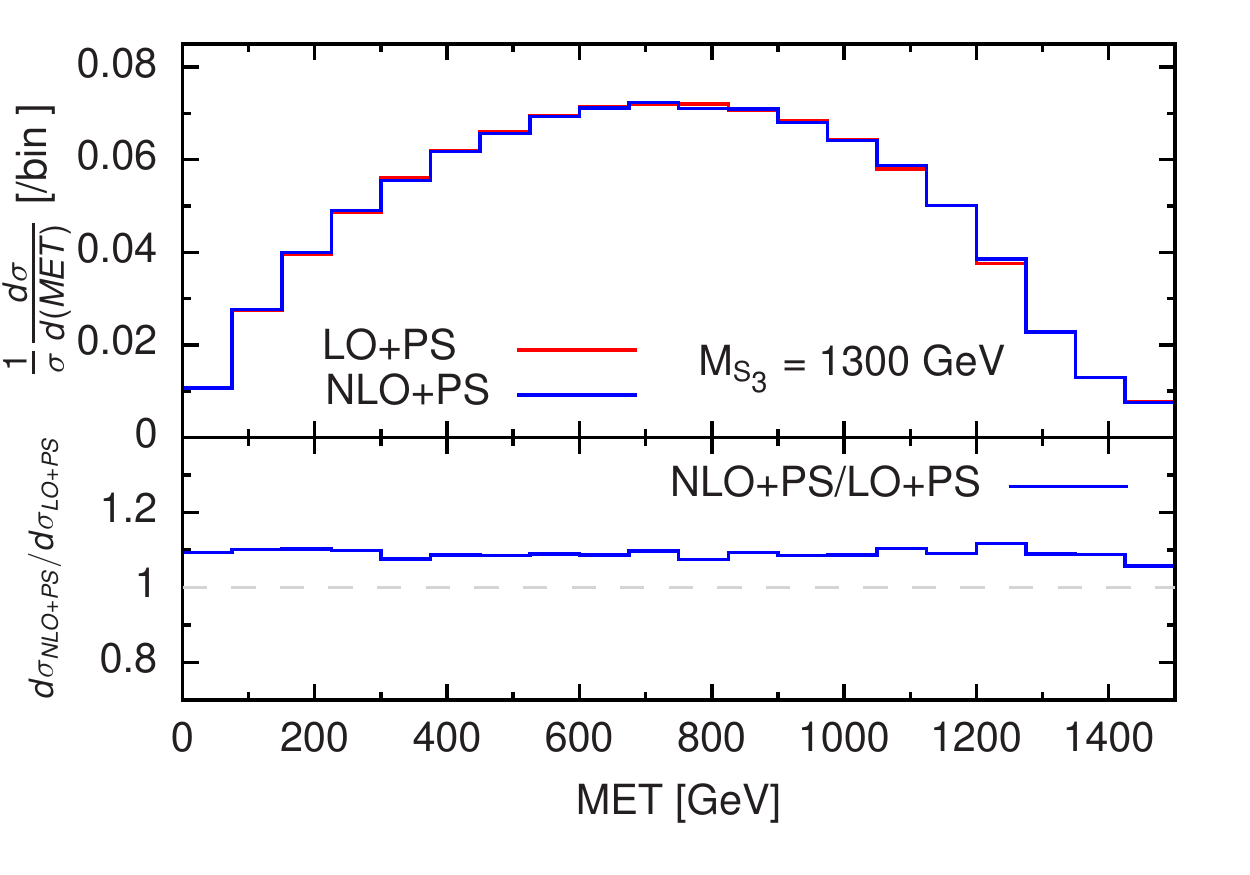}
\includegraphics[angle=0,width=0.47\linewidth,height=0.43\linewidth]{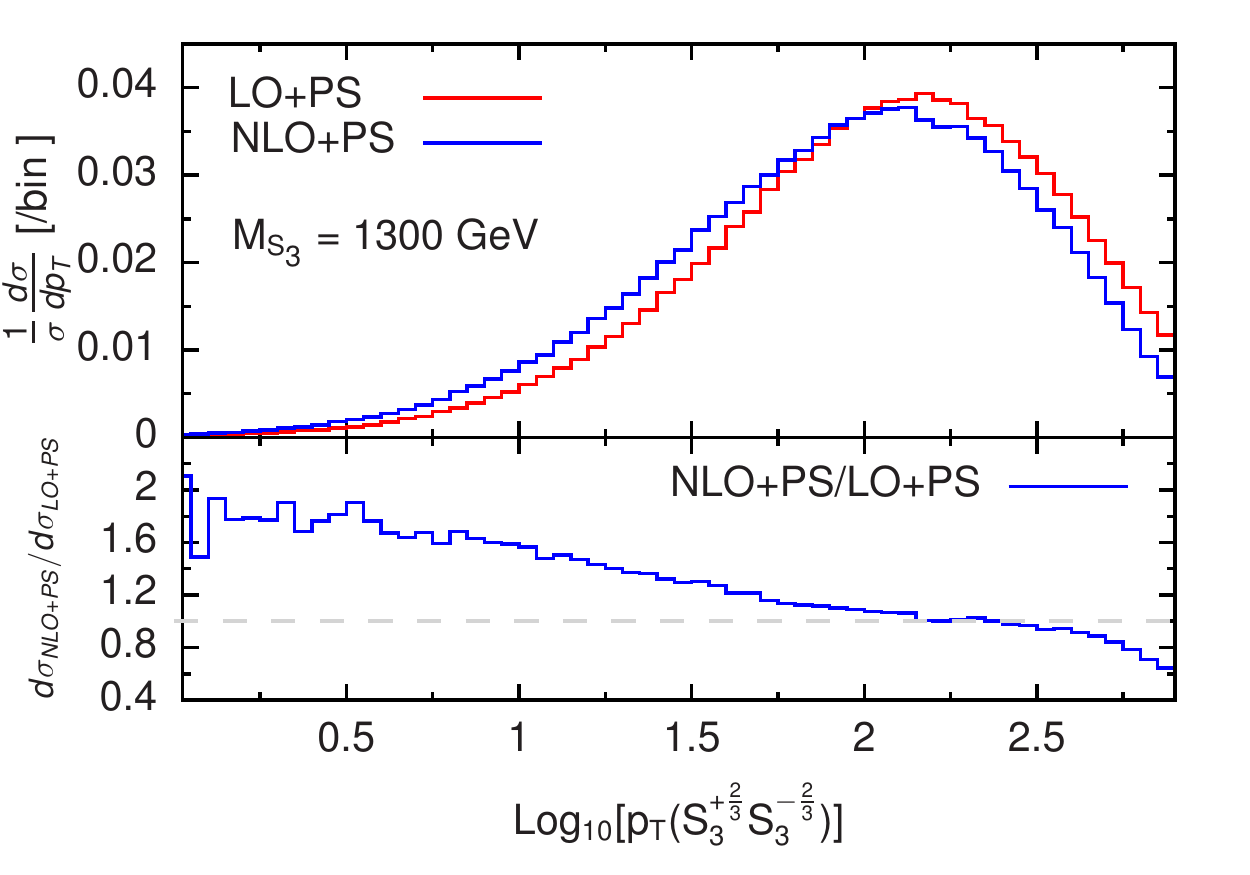}
\caption{The distributions of  MET and $\rm Log_{10}\;[\rm{p_T(S_3^{+\frac{2}{3}}S_3^{-\frac{2}{3}})}$] at LO+PS and NLO+PS.}  
 \label{fig:k-factor}
\end{figure}

In Fig.~\ref{fig:k-factor}, we show LO+PS and NLO+PS normalized distributions of MET and $\rm Log_{10}\;[\rm{p_T(S_3^{+\frac{2}{3}}S_3^{-\frac{2}{3}})}$] on the upper panels of two subfigures. The shapes of MET distributions for LO+PS and NLO+PS are identical and they peak around 700 GeV. For $\rm{p_T(S_3^{+\frac{2}{3}}S_3^{-\frac{2}{3}})}$ distribution in the right figure, the peak for NLO+PS is slightly shifted towards left of LO+PS one and they peak in the range of 100-300 GeV. On the lower panels, we show the $k$-factor for differential distribution, i.e. the ratio of differential NLO+PS cross-section to LO+PS one. In the left figure for MET, we see that for the shown range the k-factor at different bins stays nearly same and takes a value around 1.1. In the right figure, the differential k-factor is not flat for $\rm Log_{10}\;[\rm{p_T(S_3^{+\frac{2}{3}}S_3^{-\frac{2}{3}})}]$ and therefore scaling the leading order events by a constant k-factor would not give precise results.

\begin{figure}[H]
\begin{center}
\includegraphics[angle=0,width=0.6\linewidth,height=0.43\linewidth]{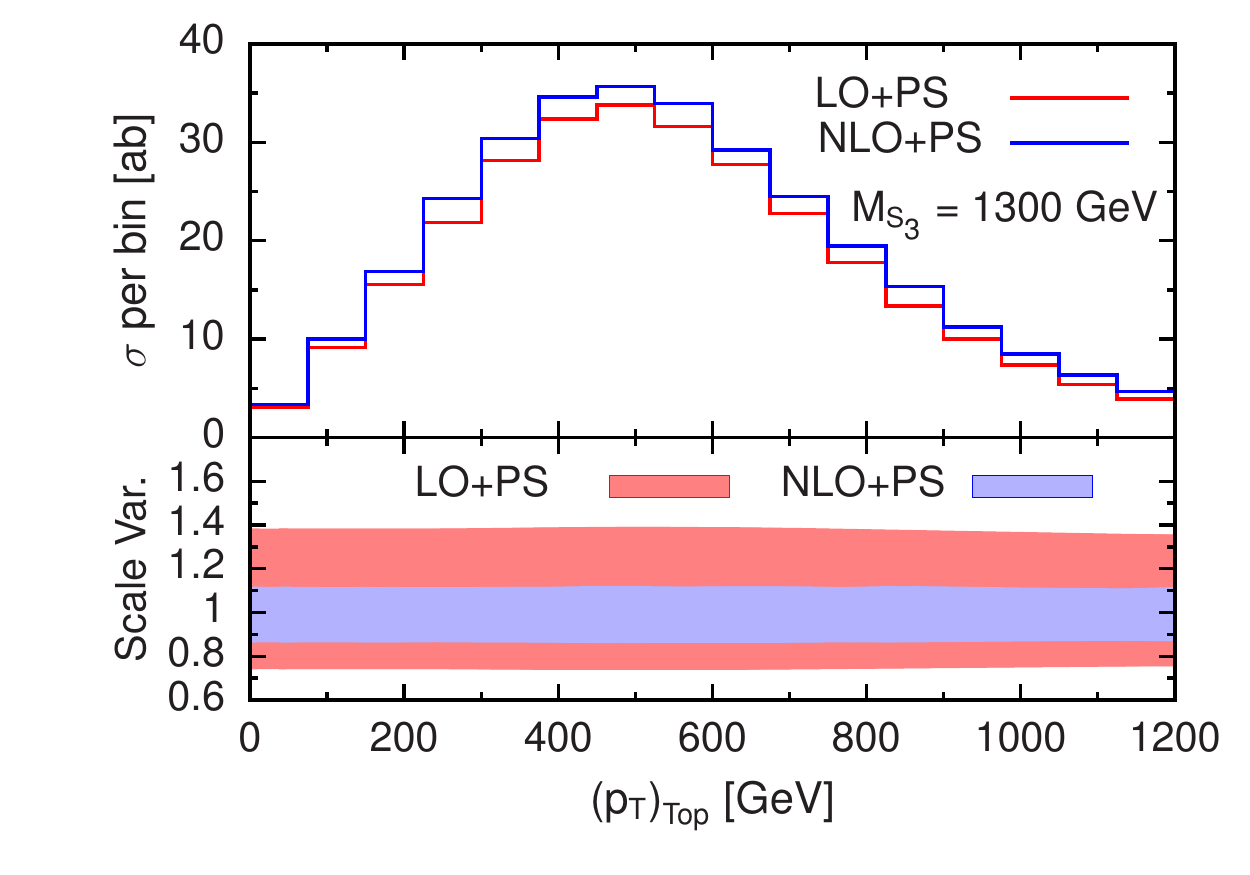}
\caption{In the upper panel, we show distribution of $\rm {(p_T)}_{Top}$ at LO+PS and NLO+PS accuracies. The bands in the lower panel show the scale variation of the distribution with respect to central value. The bands are drawn between the envelopes of the different distributions arising from the different scale choices.}  
 \label{fig:scale_variation}
\end{center}
\end{figure}

On the upper panel of Fig.~\ref{fig:scale_variation}, we show differential distribution of cross-section with respect to the top transverse momentum at the LO+PS and NLO+PS level for the central scale choice. We see that the NLO+PS corrections lead to increased cross-section at every bin. In the lower panel, the effect of scale variation is shown as red and blue bands, where a band is drawn between the upper and lower envelopes of different results for different scale choices. It can be seen that the scale variation of NLO+PS result is significantly smaller compared to the LO+PS one, confirming that the NLO QCD correction leads to more accurate result in addition to the enhancement in the cross-section.

%
%

\section{Collider Analysis}
\label{sec:collider}
We consider pair production of $\frac{2}{3}e$-charged third-generation scalar leptoquarks ($S_3^{2/3}$ and $R_2^{2/3}$) and try to probe them at the 14 TeV LHC with two top-like fatjets plus large missing transverse momentum. Third-generation scalar leptoquark pair production is possible only through gluon fusion and $q \bar{q}$ annihilation, and hence the cross-section is independent of any model-dependent coupling and depends only on the leptoquark mass. We consider NLO QCD corrections matched to parton shower of the LQ pair production channel and few representative diagrams are already shown in Fig.~\ref{fig:feynman_dia}. Equations \ref{eq:LInt} and \ref{eq:LInt2} show the decay modes of $S_3^{2/3}$and $R_2^{2/3}$, respectively. We consider decay of $R_2^{2/3}$ fully into a top quark and a neutrino.
Since the current ATLAS study~\cite{ATLAS:2020dsf} excludes the third-generation LQ of mass lower than 1.24 TeV, the top quark originating from the decay of heavy LQ will have a high boost. The top quark will decay further, and all the decay components will start collimated resulting into a boosted large-radius jet, called top fatjet ($J_t$). We consider the hadronic decay of the top quarks. So, in the final state, we have two boosted top-like fatjets and a significant missing transverse momentum. We use jet substructure variables, missing energy, and other high-level observables to distinguish the signal from the SM background \footnote{Fatjets plus the missing energy signature is also searched to probe different other BSM models in the context of the LHC \cite{Ghosh:2021noq, Ghosh:2022rta}.}. The signal topology is given below,
\begin{equation}
\begin{aligned}
p p \rightarrow S_3^{2/3} S_3^{-2/3} \hspace{1mm} \text{[QCD]} \rightarrow (t \nu_{\tau}) (\bar{t}\bar{\nu}_{\tau}) j \Rightarrow 2 J_t +\slashed{E}_T +X \,, \\
 p p \rightarrow R_2^{2/3} R_2^{-2/3} \hspace{1mm} \text{[QCD]} \rightarrow (t \bar{\nu}_{\tau}) (\bar{t}\nu_{\tau}) j \Rightarrow 2 J_t +\slashed{E}_T +X \,,
\end{aligned}
\label{Eq-signal topology}
\end{equation}
where the top quarks coming from the $S_3^{2/3}$ and $R_2^{2/3}$ decay are respectively left and right chiral.

\subsection{Background simulation}
\label{Background section}

All the background processes that can potentially mimic the signal are included in our analysis. Each background process is generated with two to four additional QCD jets and matched according to the {\sc MLM} scheme~\cite{Mangano:2006rw, Hoeche:2005vzu} with  virtually-ordered Pythia shower. PDF sets, renormalization, and factorization scales that are used in our analysis remain same as  described in Sec.~\ref{sec:nlops}. The showered events are then passed through {\sc Delphes3} \cite{deFavereau:2013fsa} for detector simulation purpose, and we use the default CMS card provided there. Particle-flow towers and tracks are clustered to form anti-kT jets of radius parameter 0.5. Fatjets ($J$ or $J_t$) of radius 1.5 are constructed with the Cambridge-Achen (CA) algorithm~\cite{Dokshitzer:1997in} using {\sc Fastjet 3.2.2} \cite{Cacciari:2011ma}.\\

\noindent \underline{$t\bar{t}+$ jets:}\hspace{0.4cm} One of the main backgrounds for our signal process is the pair production of top quarks when one of the top quarks decays hadronically and the other decay leptonically. The top quark that decays hadronically is reconstructed as top-fatjet. The neutrino from the leptonic decay of the other top quark and the lepton that escapes detection provide missing energy (MET or $\slashed{E}_T$), while another fatjet comes from the QCD radiation or b-jet. Hadronic decay of both top quarks can give two boosted top-fatjets; however, the requirement of significant missing energy reduces this background compared to the previous setup by a factor of 100, since the MET comes from the mis-measurement of the hadronic activities. This background is produced with two additional radiations and matched with the MLM matching scheme.\\

\noindent \underline{$Z+$ jets:}\hspace{0.4cm} Another main background of our signal is the inclusive Z-boson production, where the $Z$-boson decays invisibly. This process is generated with four extra partons, and the {\sc MLM} matching is used. Two fatjets essentially originate from the QCD jets.\\

\noindent \underline{$W+$ jets:}\hspace{0.4cm} It contributes considerably but is smaller than $Z+$ jets background. When the W boson decays leptonically, the missing energy comes from the neutrino and the lepton that escape detection. This background is also generated with four partons following {\sc MLM} matching and here also the fatjets come from the extra radiations. 

Since our analysis requires large missing energy, we generate $Z+$ jets and $W+$ jets backgrounds with a generation-level hard-cut $\slashed{E}_T>100$ GeV for better statistics.\\

\noindent \underline{$tW+$ jets:}\hspace{0.4cm} Single top quark production at the LHC in association with the $W$ boson, contributes considerably as a background, which is generated with two extra parton using {\sc MLM} matching. Top quark decays hadronically to give rise to a boosted top-like fatjet, while another fatjet comes from the QCD radiation. The neutrino with the missing lepton from $W$ decay is the source of the missing energy.\\

\noindent \underline{$VV+$ jets:}\hspace{0.4cm} A small contribution can come from the diboson production, which can be classified into three different categories, $WZ$, $WW$, and $ZZ$, where all of these are matched with two extra partons applying MLM matching scheme. $WZ$ contributes the most among these three, where $Z$ boson decays invisibly to produce missing energy and hadronic decay of the $W$ boson gives one fatjet. Even though $WW$ and $ZZ$ contribute almost negligibly we keep these backgrounds in our analysis. In either case, one of them decays hadronically and the other one decays leptonically ($W$) or invisibly ($Z$). In all these three processes, another fatjet comes due to the QCD radiation. \\

\noindent \underline{$t \bar{t} Z$:}\hspace{0.4cm}  The cross-section of $t \bar{t} Z$ is smaller than any of the above mentioned background processes, but we keep this too in our analysis. This process becomes signal like when $Z$-boson decays invisibly and two tops are reconstructed as top-like fatjets. This process gives almost negligible contribution compared to $Z+$ jets and $t \bar{t}+$ jets backgrounds. We omit $t \bar{t} W$ background since its contribution is found to be even more suppressed.\\

\noindent \underline{QCD background:}\hspace{0.4cm} The di-jet production cross-section is vast at the LHC; even after constructing two fatjets, huge events remain from this background. The requirement of large missing energy gives additional suppression of order 100 since MET here can only occur due to the mis-measurement of hadronic activities. An additional suppression of order 50 comes from the requirement of b-tagged fatjet. So, QCD backgrounds are found to be negligible compared to the other backgrounds and therefore we do not include this in our analysis.\\
\par
The background processes considered in our analysis are normalized with the available higher-order QCD corrected production cross-section, as presented in Tab.~\ref{cross-section:BG}.

\begin{table}[tb!]
\begin{center}
 \begin{tabular}{|l|l|r|}
\hline
 Background & Ref   & $\sigma$ (pb)   \\ 
\hline
$t \bar{t}+$ jets  & \cite{Muselli:2015kba} & 988.57 [$N^3$LO] \\
\hline
$t W+$ jets  & \cite{Kidonakis:2015nna}  & 83.1 [$N^2$LO] \\
\hline
$Z+$ jets & \multirow{2}{*}{\cite{Catani:2009sm, Balossini:2009sa} }  & $6.33\times 10^4$ [$N^2$LO] \\
\cline{1-1} \cline{3-3}
 $W+$ jets &   & $1.95\times 10^5$ [NLO] \\
 \hline
$ZZ+$ jets  & \multirow{3}{*}{\cite{Campbell:2011bn} } & 17.72 [NLO] \\
\cline{1-1} \cline{3-3}
$WW+$ jets &   & 124.31 [NLO] \\
\cline{1-1} \cline{3-3}
$WZ+$ jets &   & 51.82 [NLO] \\
 \hline
 \end{tabular} 
\caption{Higher-order QCD corrected production cross-sections of different background processes at the 14 TeV LHC used in our analysis, where the order of QCD correction is presented in brackets.}
\label{cross-section:BG}
\end{center}
\end{table}

\subsection{Construction of Jet Substructure Variables}
\label{High level Variables}

Jet substructure variables provide good efficiencies when analyzing boosted topologies. The substructure variables that we use in our analysis are listed below.\\
\par
\noindent\textbf{Pruned Jet Mass:}\hspace{0.4cm} Jet mass is a good variable in separating a boosted top-like fatjet from the boosted $W/Z$ boson or the QCD fatjets. Additional soft and wide angle radiations from the underlying QCD interactions can contribute to the fatjet mass. So for realistic predictions, one needs to remove those contributions. Pruning, filtering, and trimming~\cite{Krohn:2009th, Butterworth:2008iy, Ellis:2009su, Ellis:2009me} are different jet grooming techniques and we use pruning in our analysis. The fatjet mass is defined as $M_J=(\sum_{i\epsilon J} p_i)^2$, where the four-momentum of the $i$-th constituent is denoted as $p_i$. After clustering a fatjet using the CA algorithm, we de-cluster its constituents in each recombination step and remove the soft and wide-angle radiations from the fatjet. The merging of $i$-th and $j$-th proto-jets into the fatjet is vetoed, and the softer one is removed, if the following conditions are achieved,
\begin{equation}
Z=\text{min}(P_{Ti},P_{Tj})/(P_{Ti}+P_{Tj})<Z_{\text{cut}}, \hspace{0.3cm} \text{and} \hspace{0.3cm} \Delta R_{ij}>R_{\text{fact}}.
\label{EQ.PrunedMass}
\end{equation} 
The angular separation between two proto-jets is $\Delta R_{ij}$, and we choose $R_{\text{fact}}=0.86 \sim \frac{m_{\text{top}}}{P_{T,\text{top}}}$~\cite{Ellis:2009me}. $Z$ and $P_{Ti}$ are the softness parameter and the transverse momentum of the $i$-th proto-jet respectively. We set $Z_{\text{cut}}=0.1$~\cite{Ellis:2009su} in our analysis.\\
\par
\noindent\textbf{N-subjettiness ratio:}\hspace{0.4cm} N-subjettiness is a jet shape variable that measures how the energy of a fatjet is distributed around different subjet axes and is defined as follows~\cite{Thaler:2010tr, Thaler:2011gf},
\begin{equation}
\tau_N=\dfrac{1}{\mathcal{N}_0} \sum_i  P_{T,i} \hspace{1mm} \text{min} \{ \Delta R_{i,1}, \Delta R_{i,2}, \cdots \Delta R_{i,N} \}.
\label{EQ.N-subjetti}
\end{equation}
The summation runs over all the constituent particles of the jet. $\mathcal{N}_0$ is the normalization factor, defined as $\mathcal{N}_0=\sum\limits_i  P_{T,i} R$, where $P_{T,i}$ is the transverse momentum of the $i$-th constituent of the jet of radius $R$. $\Delta R_{i,K}=\sqrt{(\Delta \eta)^2+(\Delta \phi)^2}$ is the angular separation of the $i$-th constituent of the jet from its $K^{\text{th}}$-subjet axis in the pseudorapidity-azimuthal angle, {\em i.e.,} $\eta-\phi$ plane. Rather than $\tau_N$, the ratio $\frac{\tau_N}{\tau_{N-1}}$ is a more effective discriminating variable between N-prong fatjets and SM background~\cite{Thaler:2010tr}. Our analysis uses $\tau_{32}=\frac{\tau_3}{\tau_2}$ and $\tau_{31}=\frac{\tau_3}{\tau_1}$ to differentiate top-fatjets from the SM background.

\subsection{Event Selection}
\label{subsec:Event Selection}

\underline{Baseline-Selection Criteria:}\hspace{0.4cm} We apply the following pre-selection cuts (C1) to select events for further analysis.
\begin{itemize}
\item The radius parameter of the top fatjet is $R\sim \dfrac{2m_t}{P_T}$, where $P_T$ and $m_t$ are the transverse momenta and top quark's mass, respectively. For each event, we reconstruct at least two fatjets using CA algorithm of radius parameter 1.5 with minimum transverse momentum $P_T(J_0),P_T(J_1)>200$ GeV
\item The missing energy of each event should be greater than 100 GeV
\item Since lepton is not present in the final state of our signal, we veto the events which contain any lepton of transverse momentum $P_T(l)>10$ GeV and pseudorapidity $|\eta(l)| <2.4$
\item A minimal cut on the azimuthal separation between any fatjet and the missing momentum $\Delta \phi(J_i,\slashed{E}_T)>0.2$ is applied to minimise the hadronic mis-measurement contribution
\end{itemize}
%
\begin{table}[htbp!]
\begin{center}
\resizebox{\columnwidth}{!}{
\begin{tabular}{|c|c|c||c|c|c|c|c|c|c|c|c|}
\hline
\multirow{3}{*}{Cuts} & {$S_3$} & {$R_2$}  & {$Z$}  & {$W$} & {$t \bar{t}$} & {$tW$} & {$WZ$}& {$WW$}  & {$ZZ$} & {$t \bar{t} Z$} & {tot}  \\
&  &   & {+jets}  & {+jets} & {+jets} & {+jets} & {+jets}& {+jets}  & {+jets} &  &  {BG}\\
& (fb) & (fb) & (fb) & (fb) & (fb) & (fb) & (fb) & (fb) & (fb) & (fb) & (fb) \\
\hline \hline
\multirow{2}{*}{C1} & 0.2315 & 0.232 & 2517.99  & 1366.91 & 690.65
& 366.91 & 93.53 & 25.90 & 11.51 & 5.24 & 5078.64\\
& [$100\%$] & [$100\%$] & [$100\%$] & [$100\%$] & [$100\%$] 
& [$100\%$] & [$100\%$] & [$100\%$] & [$100\%$] & [$100\%$] & [$100\%$] \\
\hline
\multirow{2}{*}{C2} & 0.2258 & 0.2262 & 1640.29  & 762.59 & 302.16
& 152.52 & 58.35 & 11.51 & 6.973 & 3.96 & 2938.36\\
& [$97.54\%$] & [$97.5\%$] & [$65.14\%$] & [$55.79\%$] & [$43.75\%$] 
& [$41.57\%$] & [$62.39\%$] & [$44.44\%$] & [$60.58\%$] & [$75.57\%$] &[$57.86\%$] \\
\hline
\multirow{2}{*}{C3} & 0.1810 & 0.1801 & 241.73  & 117.99 & 230.94
& 114.39 & 10.79 & 2.45 & 1.92 & 3.28 & 723.48\\
& [$78.19\%$] & [$77.63\%$] & [$9.60\%$] & [$8.63\%$] & [$33.44\%$] 
& [$31.18\%$] & [$11.54\%$] & [$9.46\%$] & [$16.69\%$] & [$62.60\%$] &[$14.25\%$] \\
\hline
\multirow{2}{*}{C4} & 0.1047 & 0.1033 & 25.38  & 17.33 & 64.23 
& 27.45 & 1.24 & 0.33 & 0.2 & 1.474 & 137.634\\
& [$45.23\%$] & [$44.53\%$] & [$1.01\%$] & [$1.27\%$] & [$9.30\%$] 
& [$7.48\%$] & [$1.33\%$] & [$1.27\%$] & [$1.74\%$] & [$28.13\%$] &[$2.71\%$] \\
\hline
 \end{tabular}    
} 
\caption{The expected number of events (in fb, multiplying with the luminosity gives the expected event numbers) and cut efficiency for the signal $S_3$ and $R_2$ (1.3 TeV mass of leptoquark for both models) and all the background processes that contribute to the fatjets $+\slashed{E}_T$ final state after implementing the corresponding cuts at the 14 TeV LHC are shown. The effectiveness of different kinematic cuts can be followed from top to bottom after applying (C1) Preselection cuts, (C2) $\slashed{E}_T>150$ GeV, (C3) requiring at least one b-tag within $J_0$ or $J_1$, and finally (C4) $ M_{J_0}, M_{J_1}>120$ GeV. After applying C4 cut, the remaining events are passed for the multivariate analysis.
}
\label{tab:cut-flow}
\end{center}
\end{table}

\noindent\underline{Final selection cuts:}\hspace{0.4cm} After the primary selection, we apply the following cuts before passing events for multivariate analysis (MVA).
\begin{itemize}
\item[(C2)] Missing energy cut is raised from 100 GeV to 150 GeV, which reduces the background sharply.
\item[(C3)] We tag the leading $b$-jet inside $J_0$ or $J_1$.
\item[(C4)] We demand pruned mass of both the leading $M_{J_0}$ and subleading $M_{J_1}$ fatjets to be greater than  $120$ GeV.
\end{itemize}

Tab.~\ref{tab:cut-flow} displays the cut flow along with the cut efficiencies, anticipated number of events (in fb, multiplying with the luminosity gives the expected event numbers)  for the signal and the background processes for the 14 TeV LHC. One can see that the higher missing energy cut, b-tagging within a fatjet, and the pruned fatjet masses are very effective in significantly reducing backgrounds while maintaining good signal acceptance. The principal backgrounds $Z+$ jets and $W+$ jets are drastically reduced when a b-jet is tagged within the leading or subleading fatjet, and their effects are nearly identical to that of the $t\bar{t}+jets$ background (see the rows up to C3 in table \ref{tab:cut-flow}). 
%
\begin{figure}[htbp!]
\centering
\subfloat[] {\label{fig:FJ_MJ0} \includegraphics[width=0.31\textwidth]{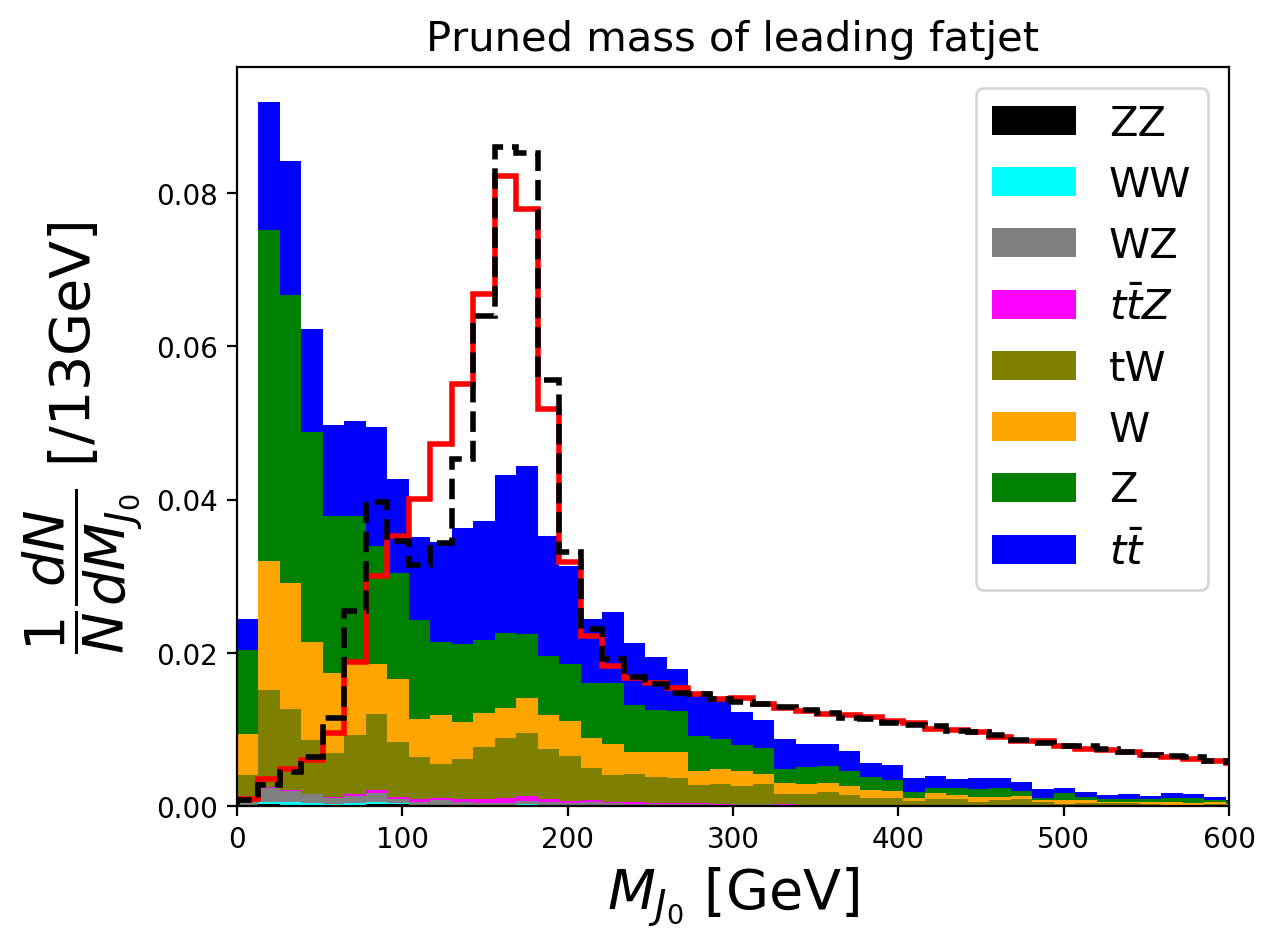}} 
\subfloat[] {\label{fig:FJ_MJ1} \includegraphics[width=0.31\textwidth]{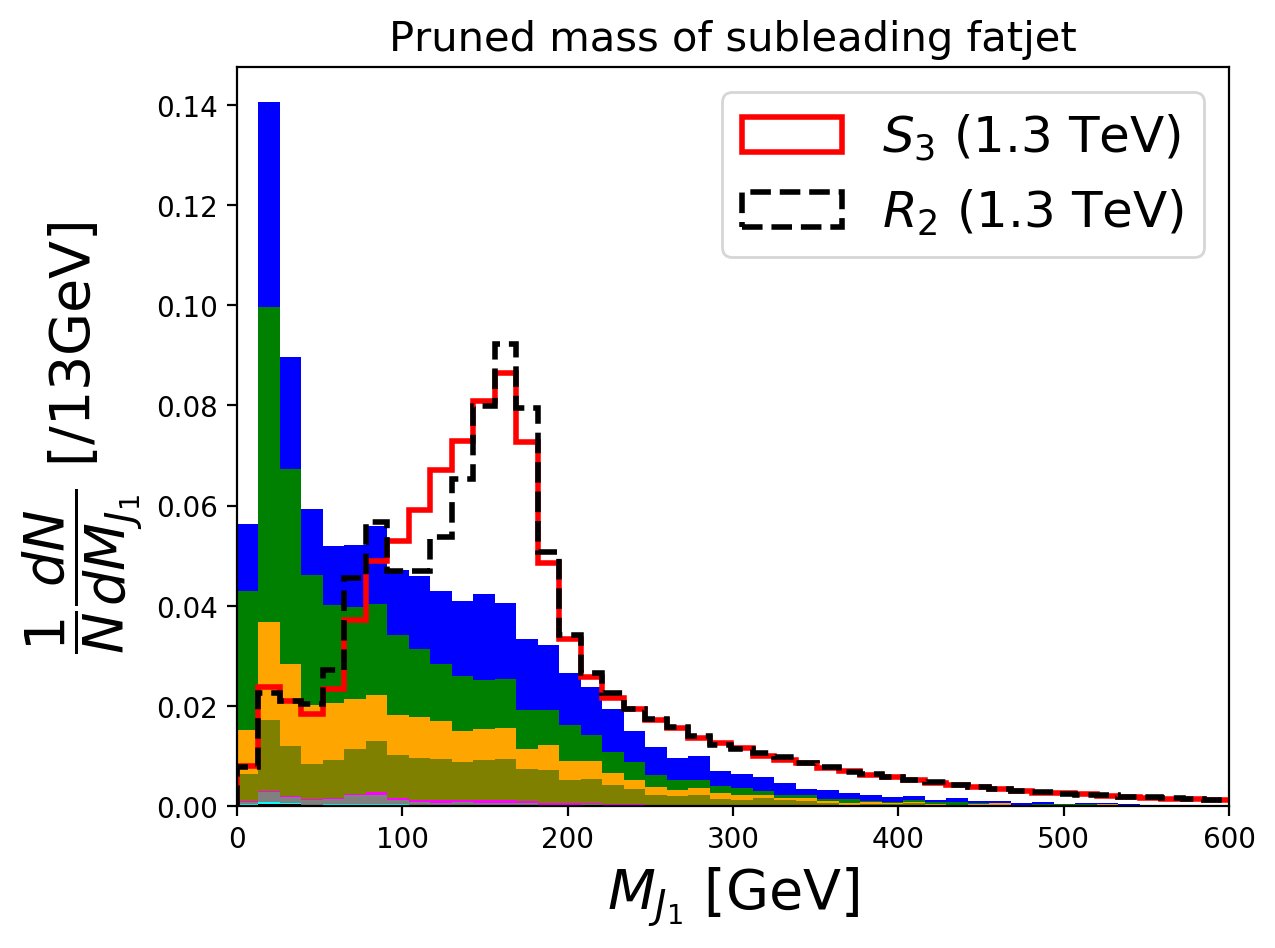}}
\subfloat[] {\label{fig:PTJ0} \includegraphics[width=0.31\textwidth]{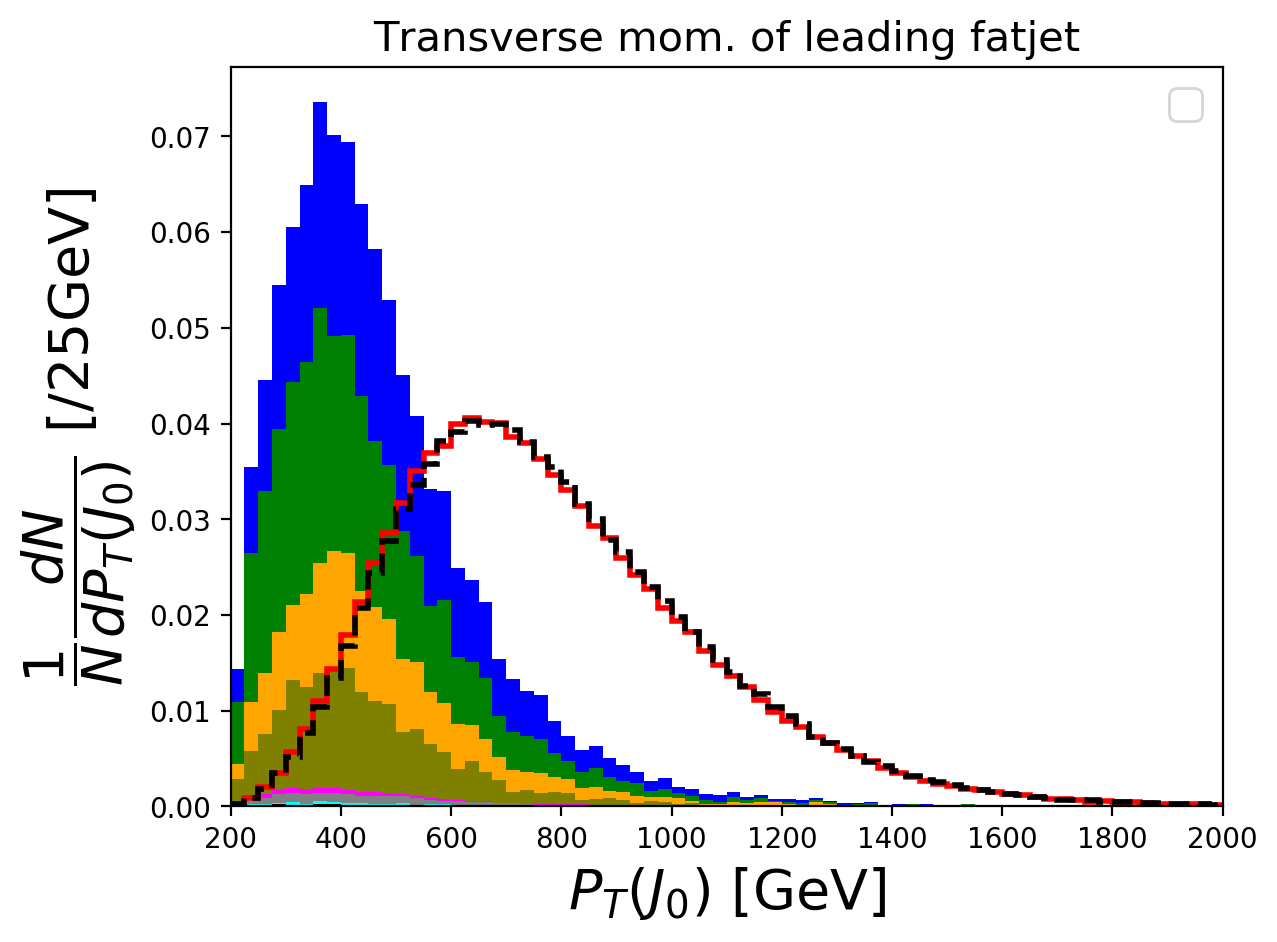}}\\
\subfloat[] {\label{fig:PTJ1} \includegraphics[width=0.31\textwidth]{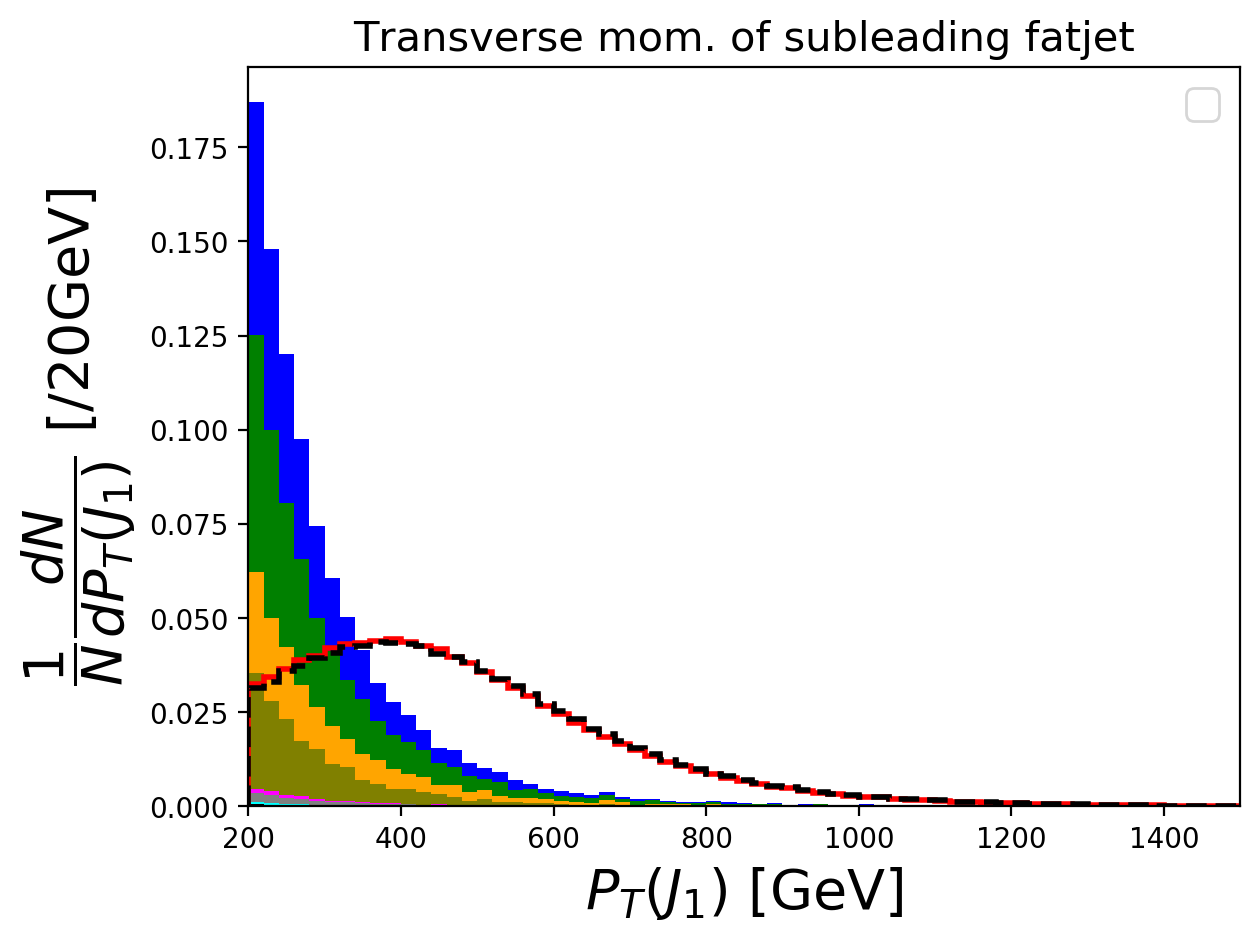}}
\subfloat[] {\label{fig:FJ_Meff} \includegraphics[width=0.31\textwidth]{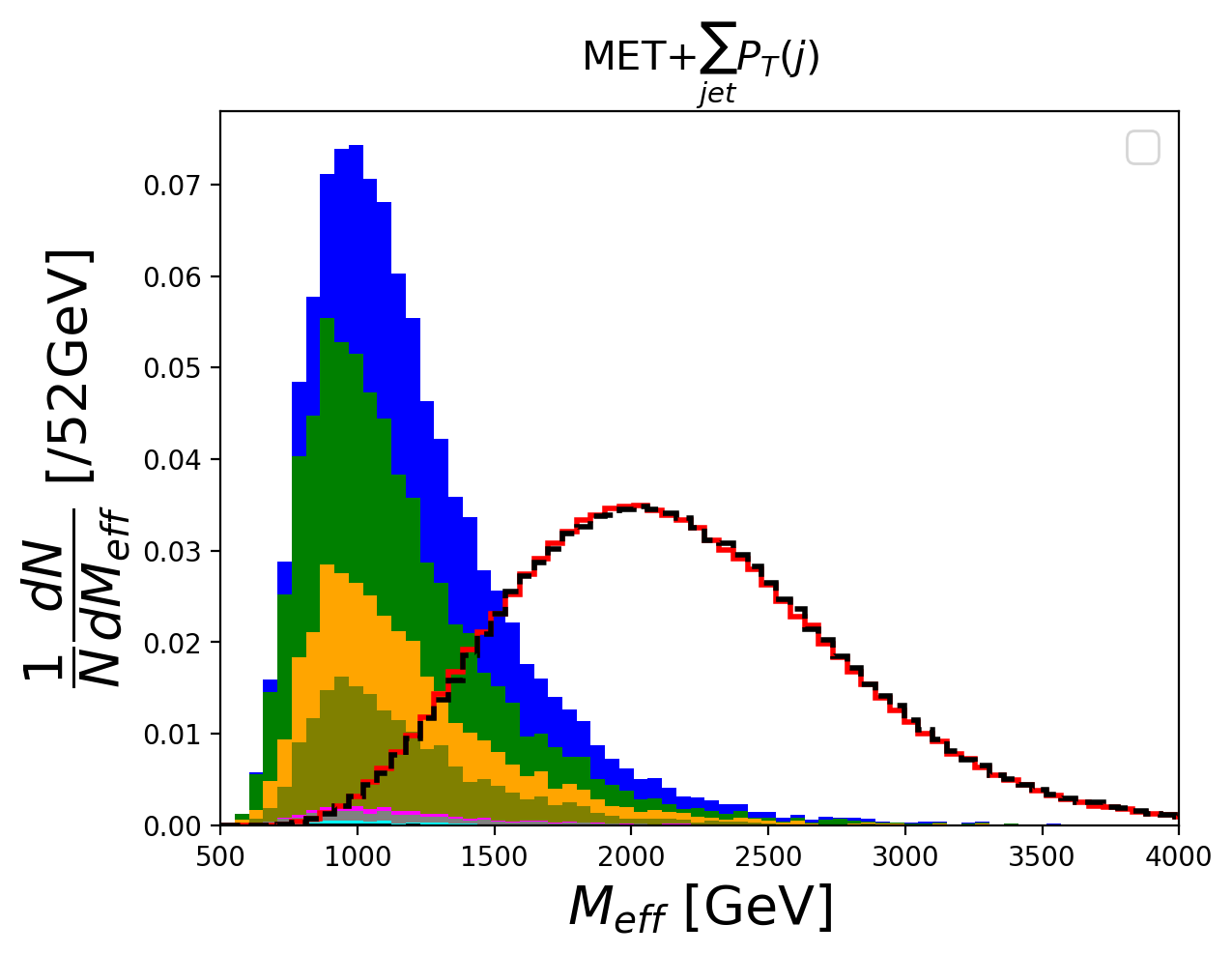}}
\subfloat[] {\label{fig:Shat} \includegraphics[width=0.31\textwidth]{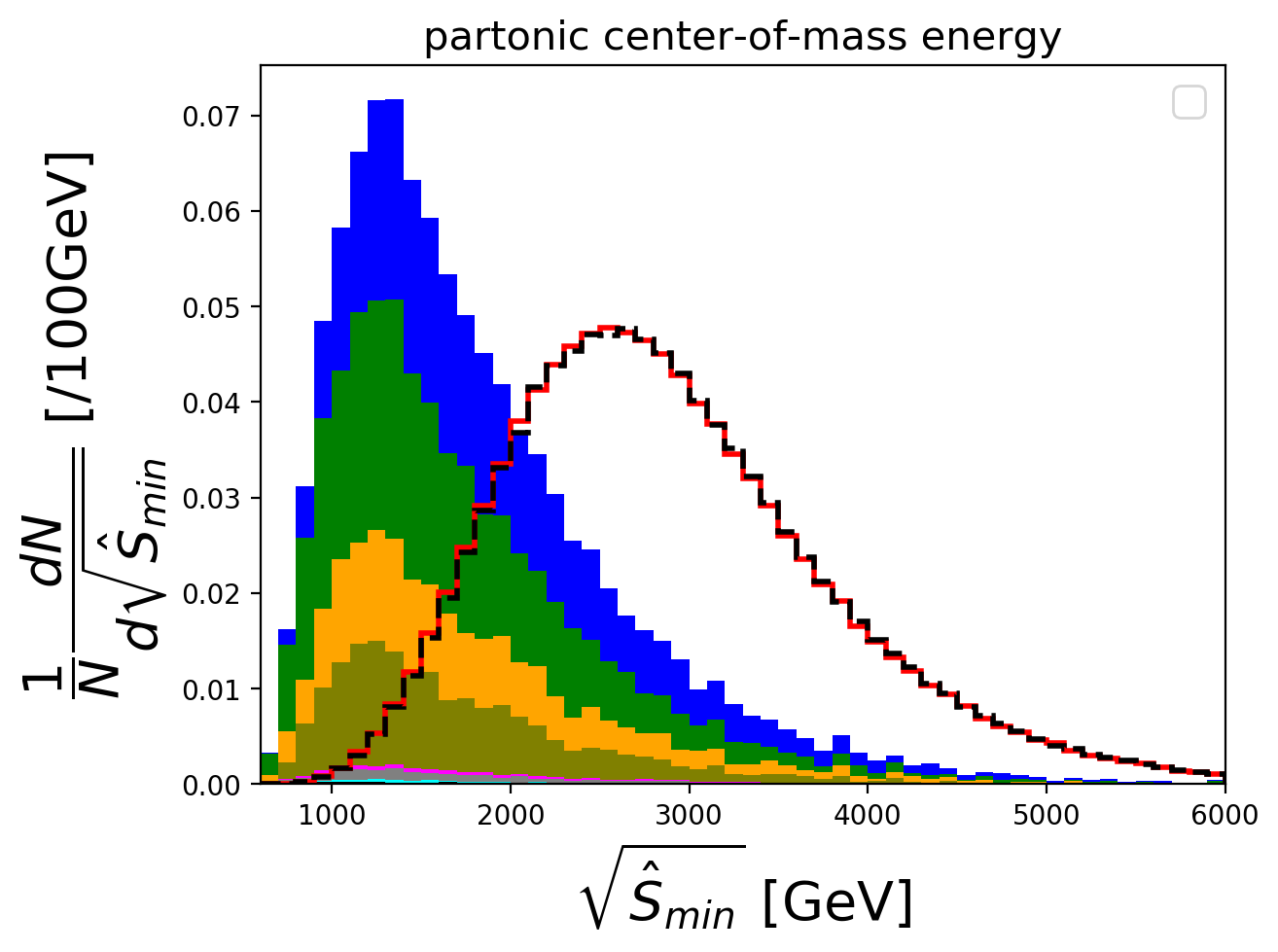}}\\
\subfloat[] {\label{fig:FJ_tau32_J0} \includegraphics[width=0.31\textwidth]{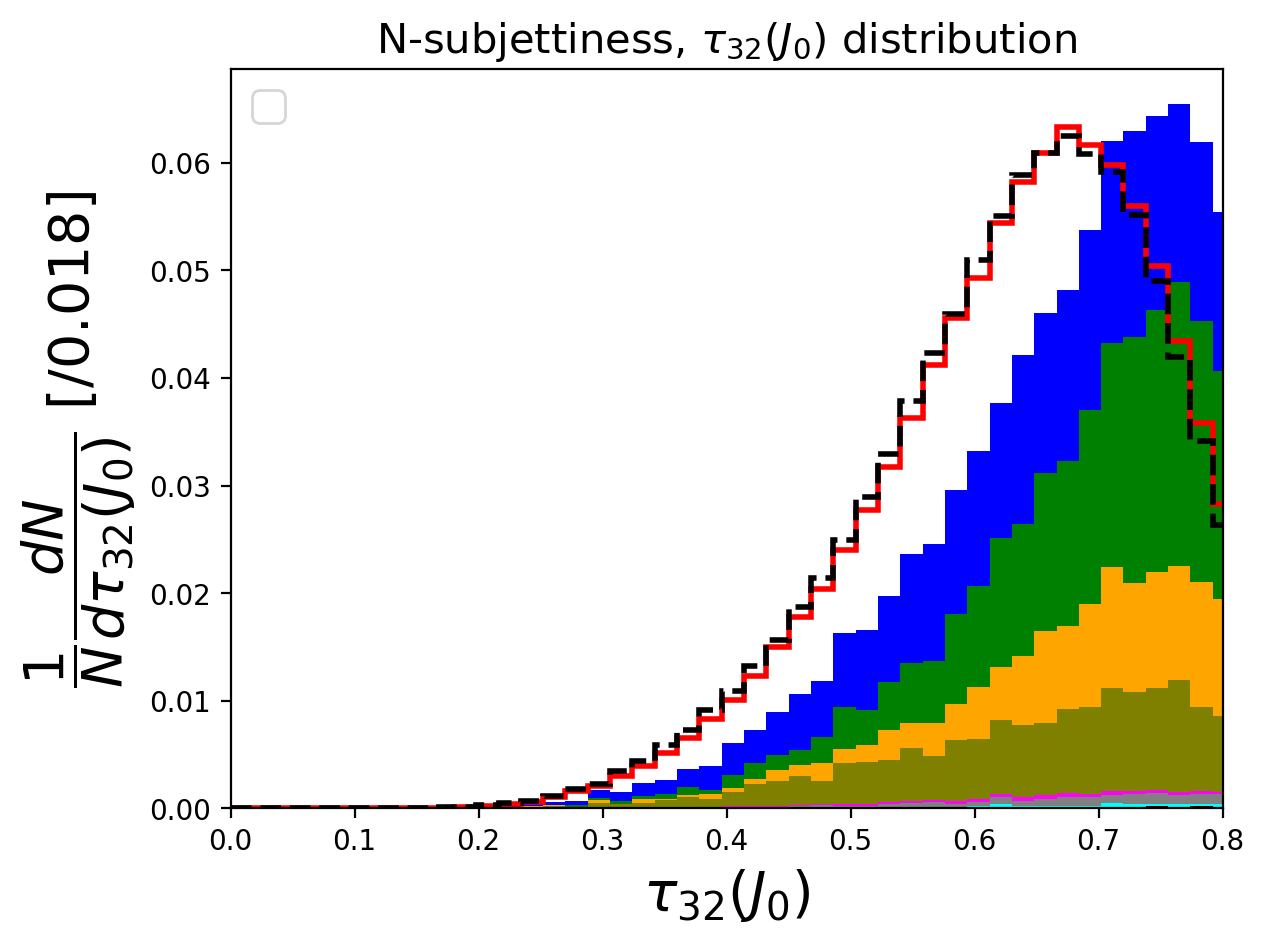}}
\subfloat[] {\label{fig:FJ_tau32_J1} \includegraphics[width=0.31\textwidth]{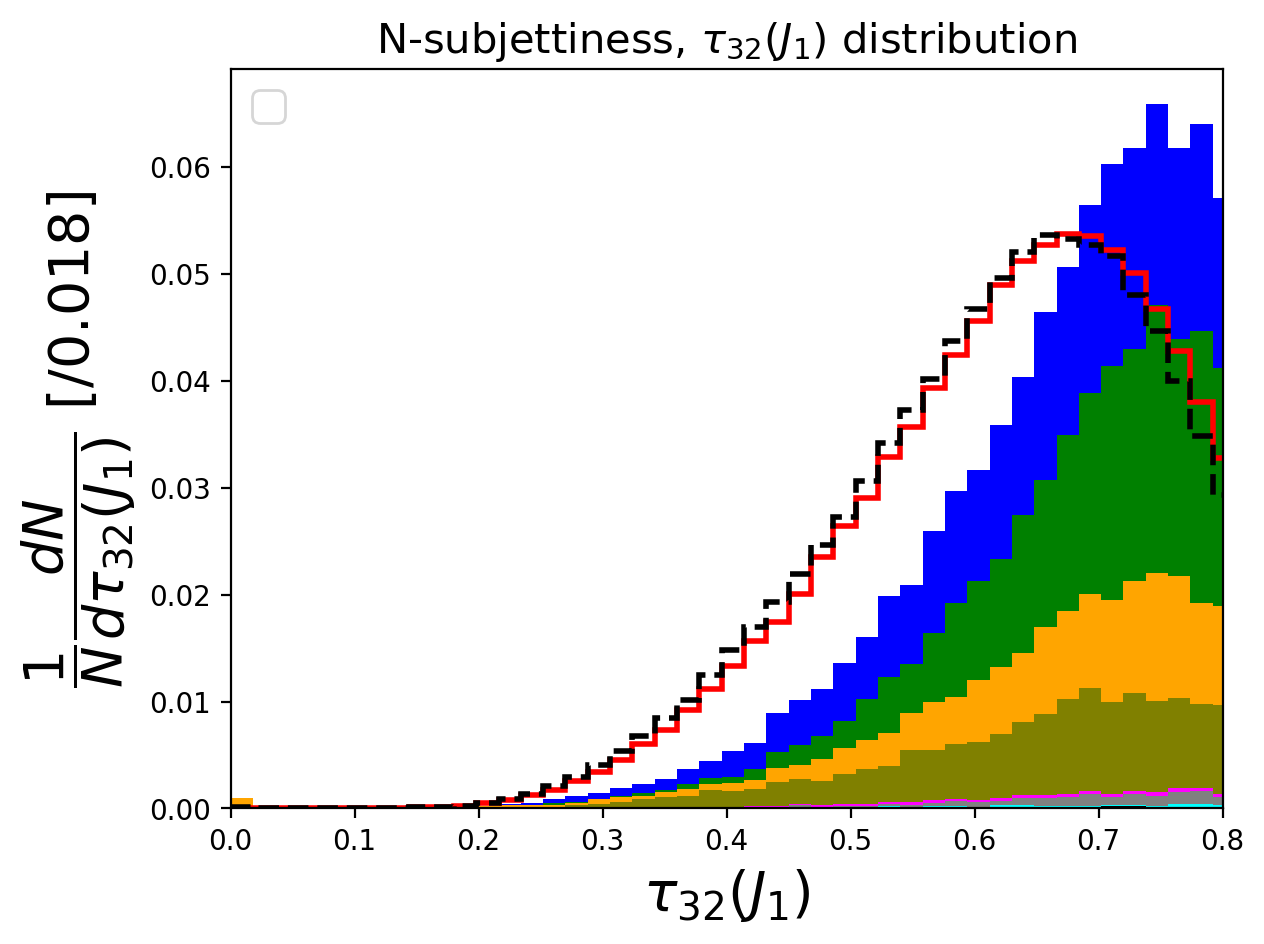}} 
\subfloat[] {\label{fig:FJ_tau31_J0} \includegraphics[width=0.31\textwidth]{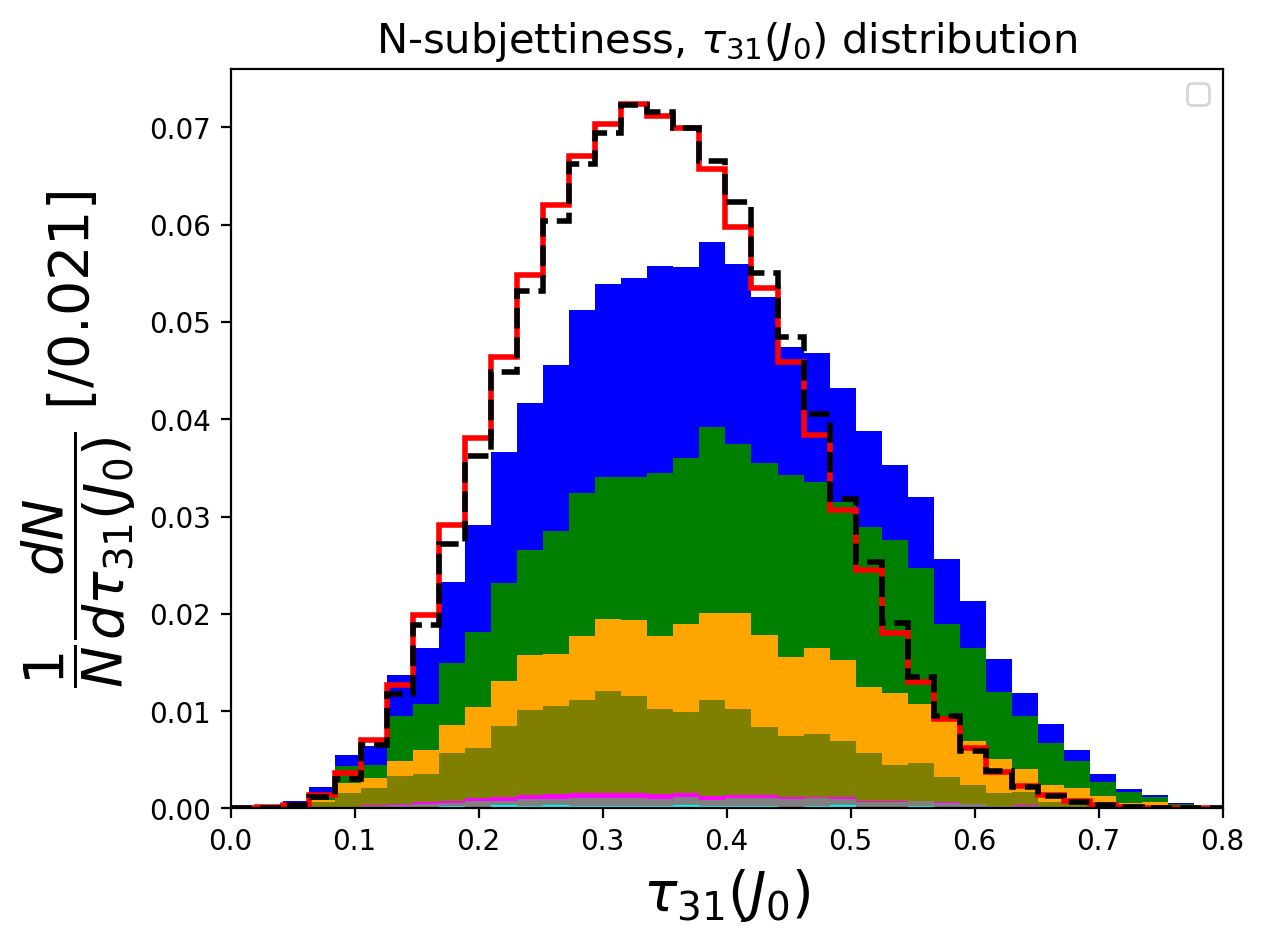}}\\
\subfloat[] {\label{fig:FJ_tau31_J1} \includegraphics[width=0.31\textwidth]{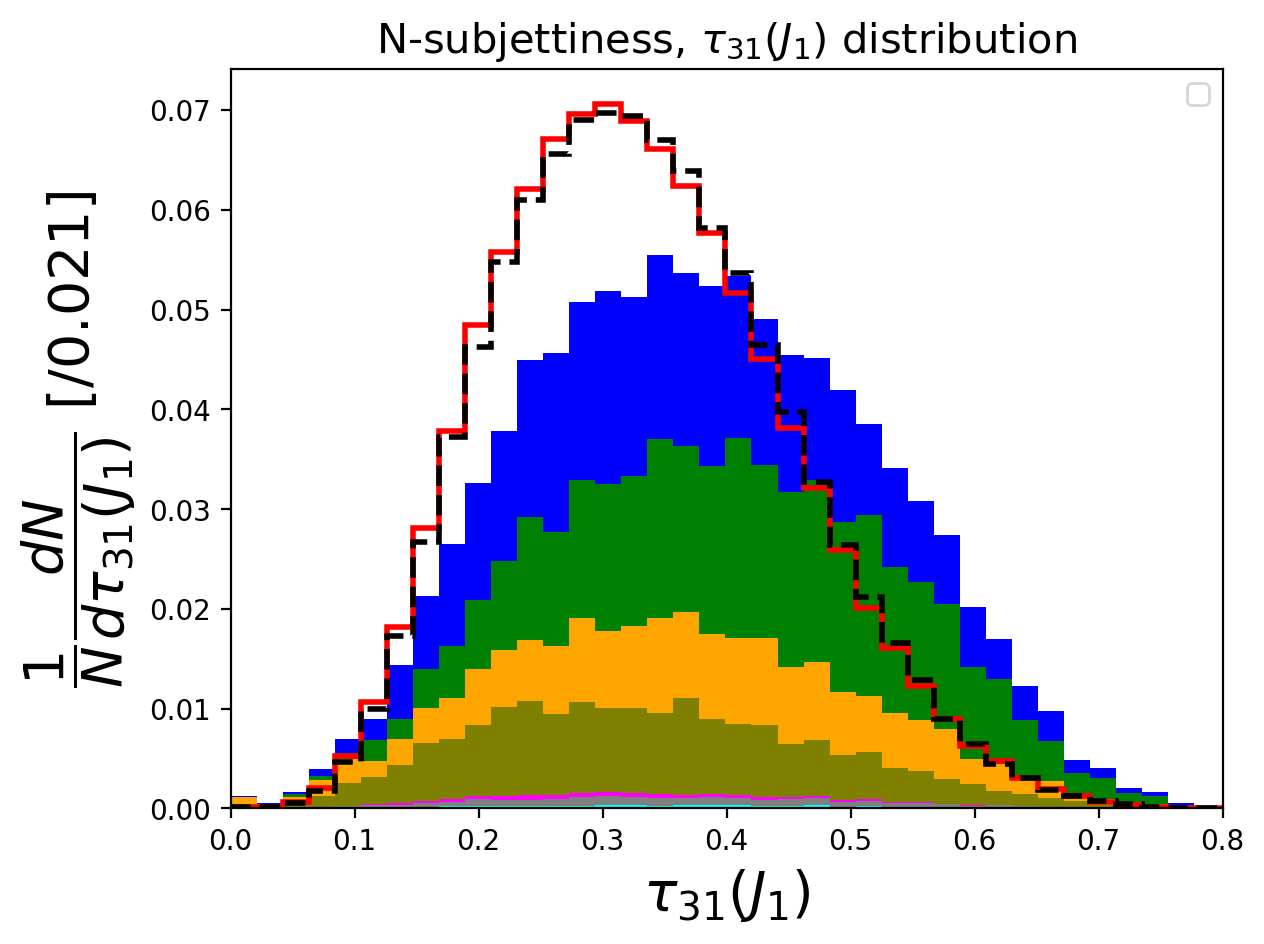}}
\subfloat[] {\label{fig:FJ_MET} \includegraphics[width=0.31\textwidth]{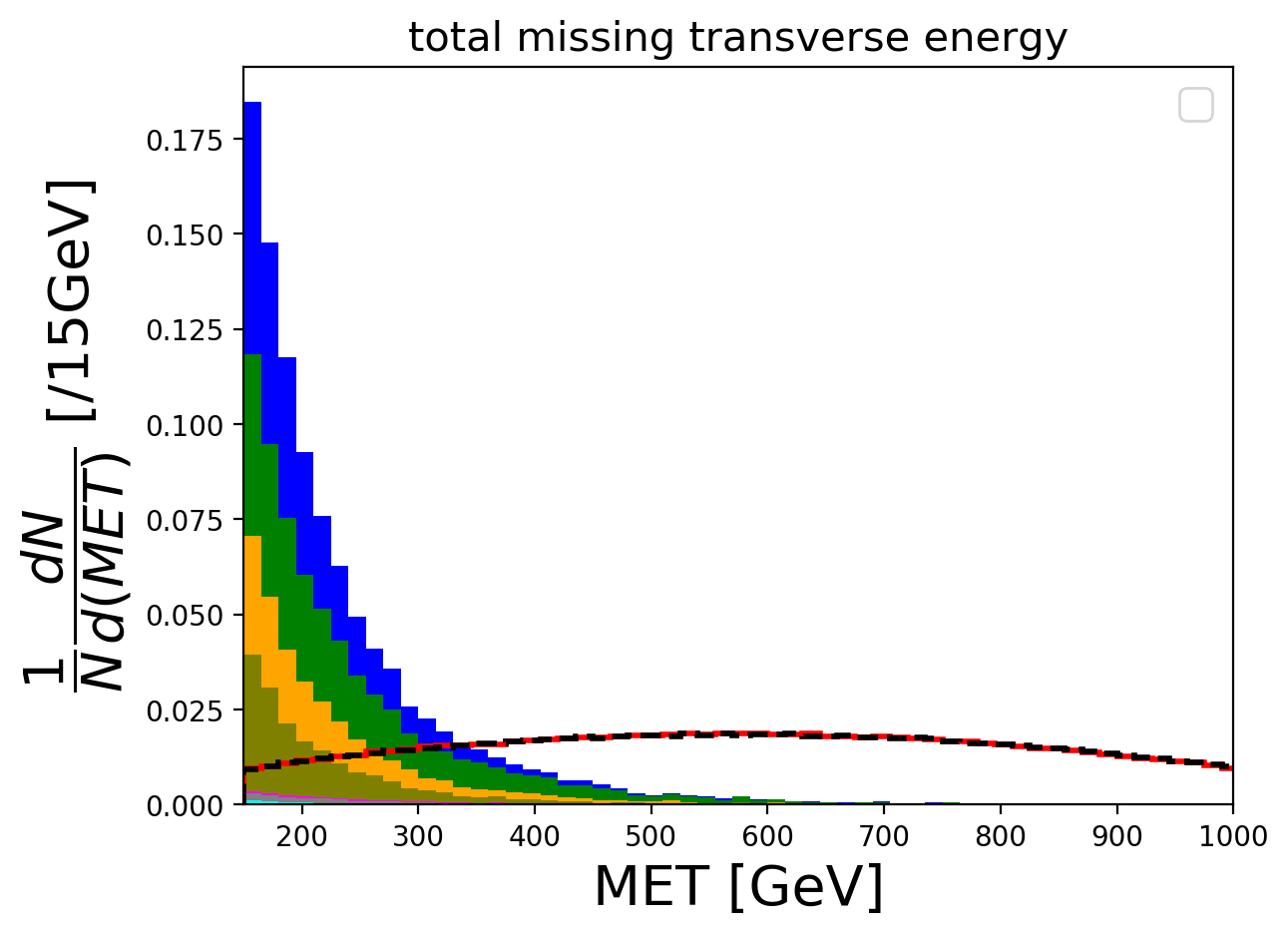}}
\subfloat[] {\label{fig:FJ_Phi_J0ET} \includegraphics[width=0.31\textwidth]{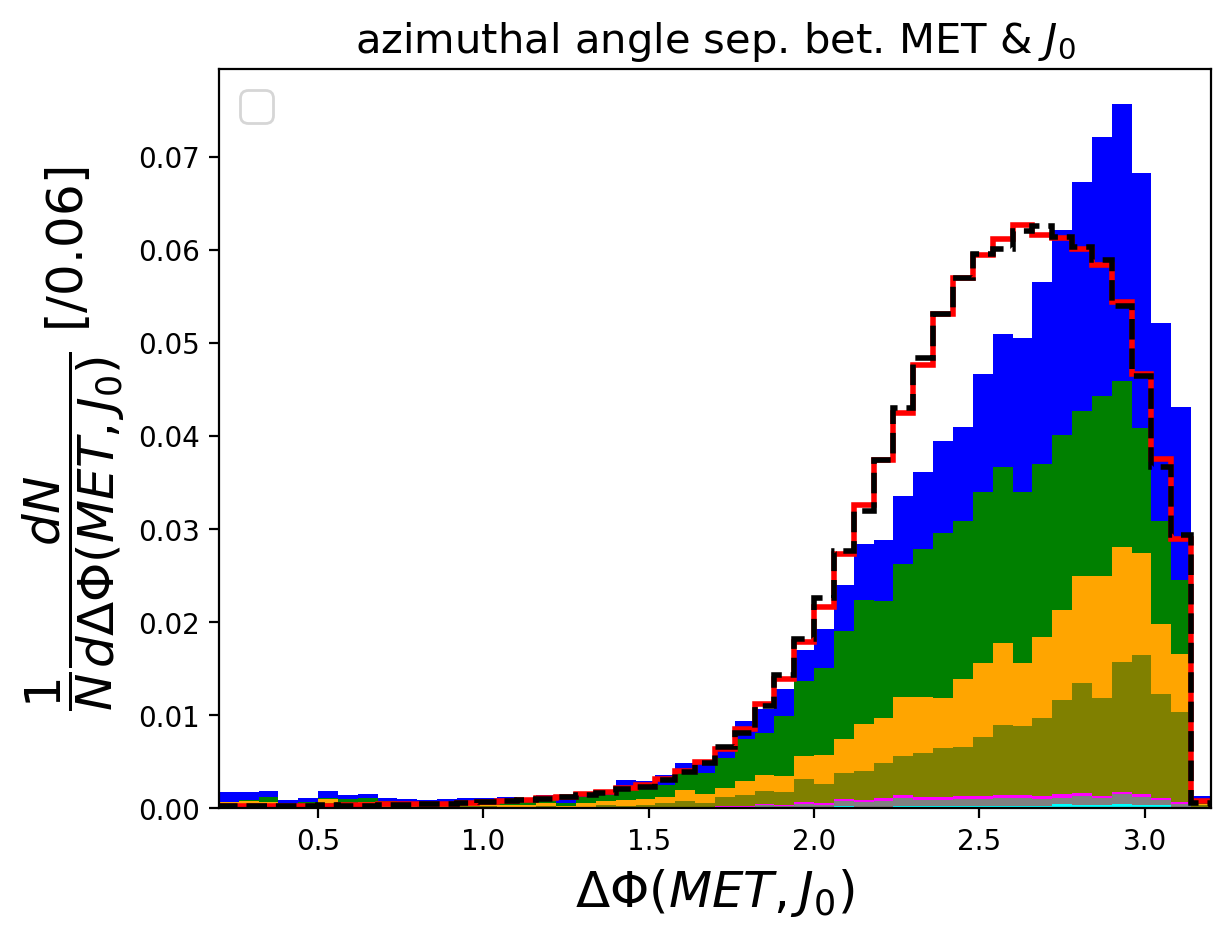}}\\
\subfloat[] {\label{fig:FJ_Phi_J1ET} \includegraphics[width=0.31\textwidth]{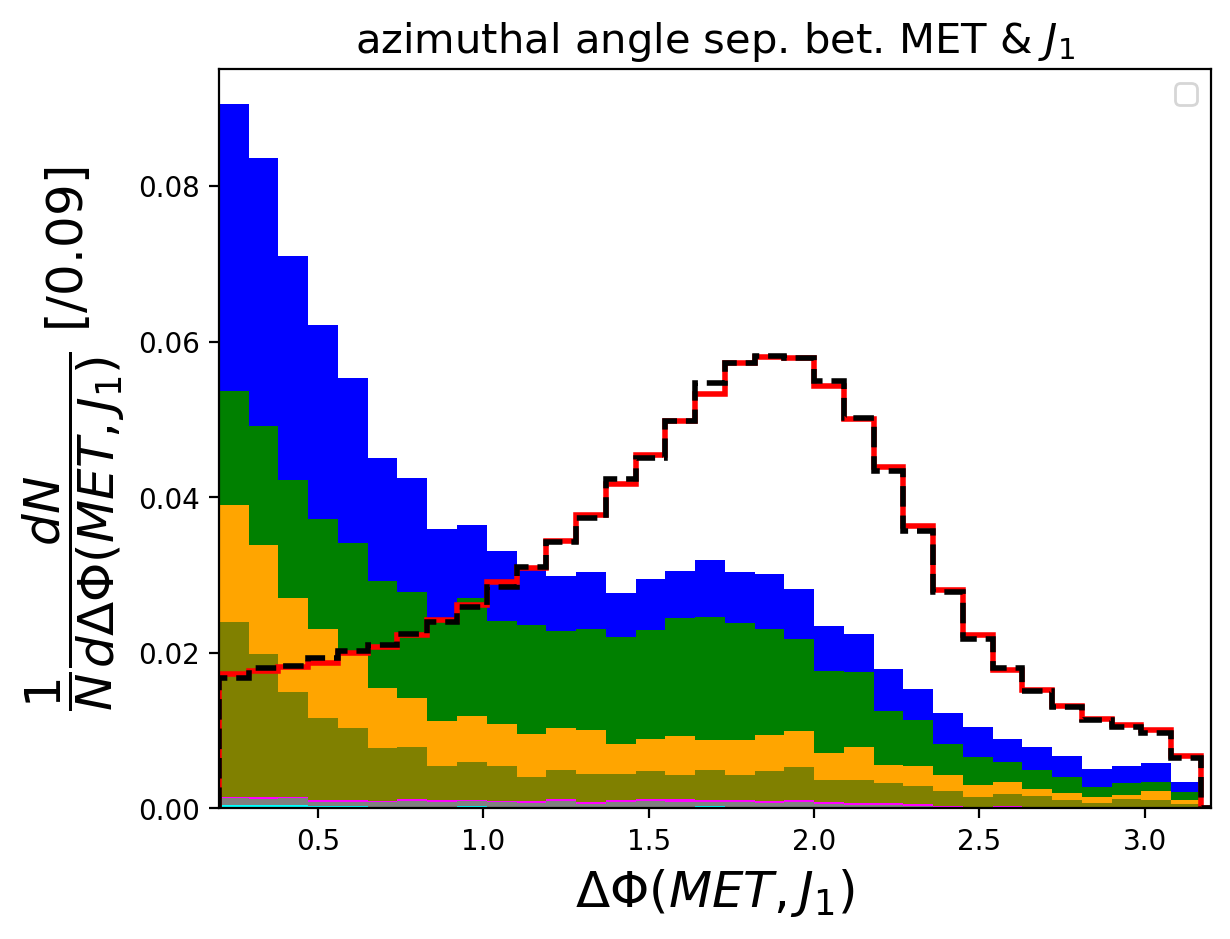}}
\subfloat[] {\label{fig:FJ_RJ0J1} \includegraphics[width=0.31\textwidth]{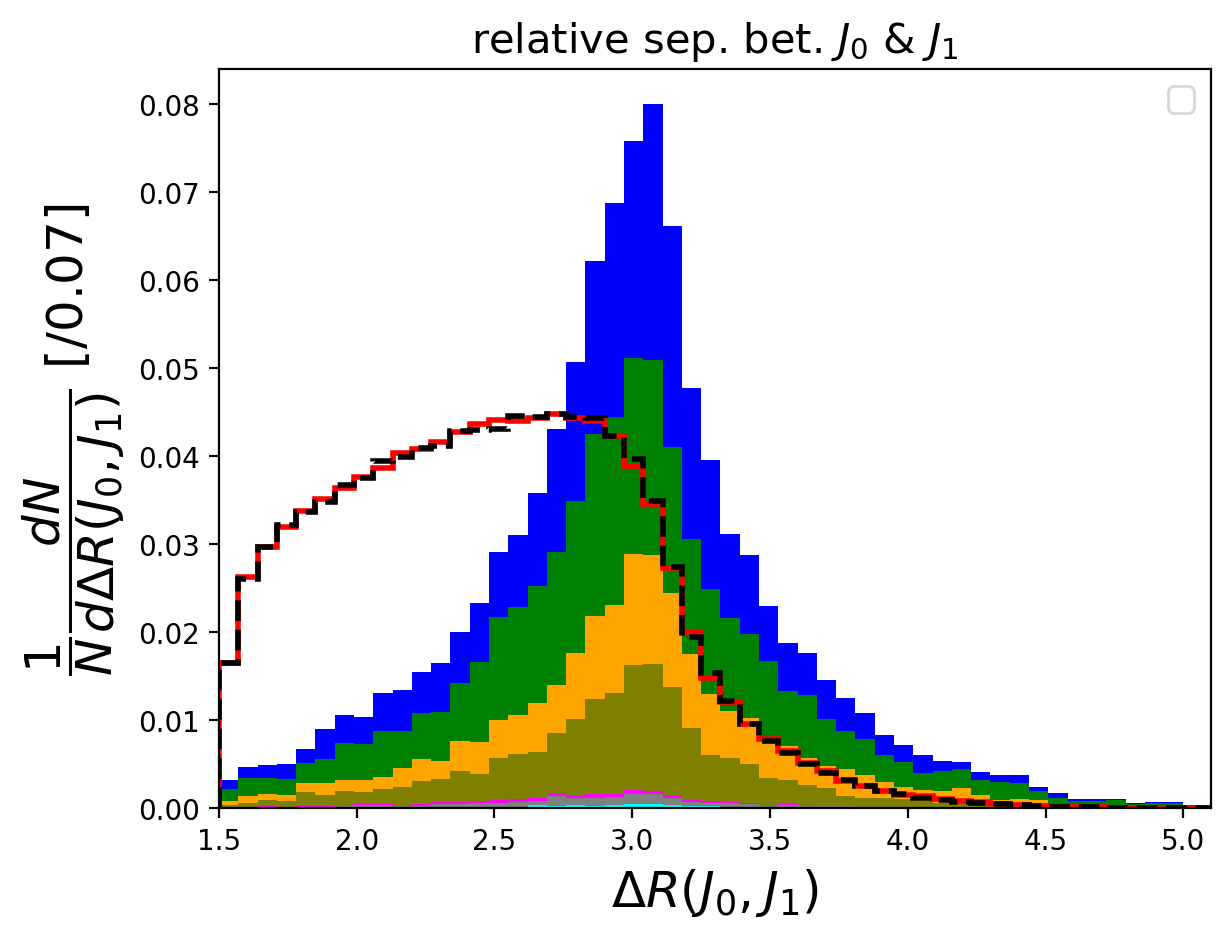}}
\subfloat[] {\label{fig:FJ_MT2} \includegraphics[width=0.31\textwidth]{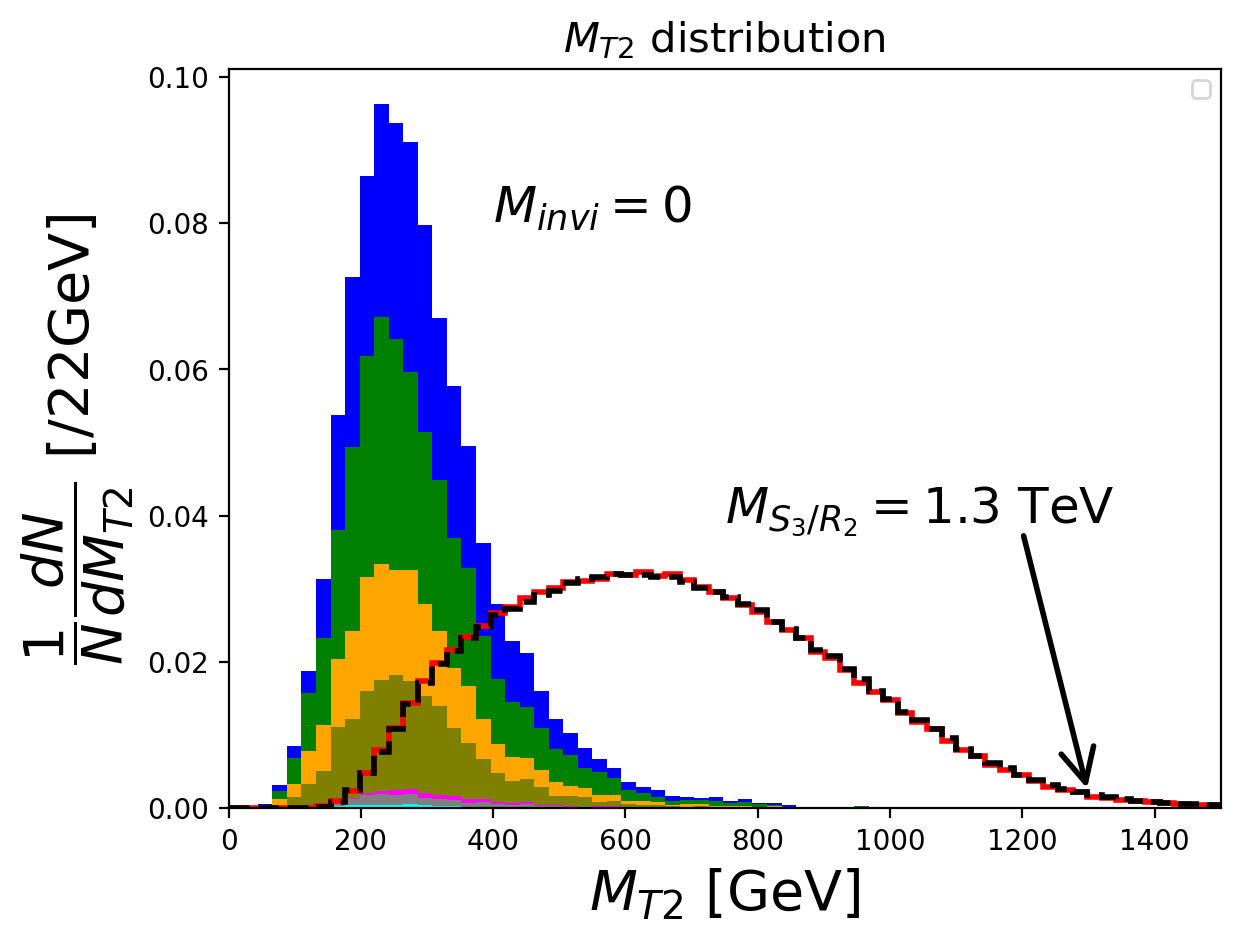}}
\caption{\scriptsize{After imposing $\slashed{E}_T>150$ GeV and b-tagging inside $J_0$ or $J_1$, together with preselection cuts as indicated in the text, the normalized distribution of kinematic variables of the signal $S_3$ (solid red), $R_2$ (dashed black), and bin-wise stacked histogram of all the background processes are shown.}} 
\label{fig:Distri-MVA-input}
\end{figure}

The normalised distributions of various kinematic variables of the signal $S_3$ and $R_2$, as well as bin-wise stacked histograms of all background processes after imposing $\slashed{E}_T>150$ GeV and b-tagging inside $J_0$ or $J_1$, together with preselection cuts, are shown in Fig.~\ref{fig:Distri-MVA-input}, where leptoquark mass is set at 1.3 TeV. The contributions of individual background processes are represented by different colours: blue, green, orange, olive, and magenta, for $t \bar{t}$+jets, $Z$+jets, $W$+jets, $tW$+jets, and $t \bar{t} Z$, respectively. Each background process is weighted by its effective cross-section after applying the cuts listed and normalized to the total cross-section.
\par
The distributions of the leading and subleading fatjets' pruned masses are depicted in Fig.~\ref{fig:FJ_MJ0} and Fig.~\ref{fig:FJ_MJ1} respectively. $Z+$ jets background demonstrates no peak in the $M_{J_0}$ and $M_{J_1}$ distributions, near the $Z$-boson mass or around the top mass as expected, since the fatjets  originate from the QCD radiations. One of the tops in the $t\bar{t}+$ jets background decays hadronically and is reconstructed as a top-fatjet, while the other fatjet comes because of the QCD radiation. As a result, the $M_{J_0}$ distribution exhibits a peak near the top mass, but the $M_{J_1}$ distribution does not exhibit a peak near the top mass. 
\par
It is also interesting to note that fatjet mass distributions are slightly different for the two signals while other kinematic variables remain similar. This is a direct implication of two different polarization. The bottom quark and the $W$-boson travel in the opposite direction in the top quark's rest frame to conserve the linear momentum. 
As in the $S_3$ model the top quark is left chiral, the majority of the $b$-quark in the top quark's rest frame lie in the same direction of the boost (this will be further discussed in the next section). This means that the majority of the $W$ boson emerges at an angle greater than 90 degrees to the boost.
However, in the $R_2$ model (top quark is right chiral), most of the $b$-quarks are found in the direction opposite to the boost in the top quark's rest frame. This suggests that most of the W bosons exist around the boost direction. As a result, in the lab frame, the quarks from the hadronic decay of the $W$ boson are more collimated in the $R_2$ model compared to the $S_3$ model. When $W$ and the $b$-quark get combined to form a single large radius three-prong fatjet, the $S_3$ model produces fewer events than the $R_2$. Because the $W$ boson is heavier than the $b$-quark, $S_3$ needs more boost to bring back all the $W$ bosons along the boost direction compared to $R_2$. As a result, $R_2$ model exhibits larger peaks in both the leading and subleading fatjet mass distributions around the top quark mass than the $S_3$ model. Moreover, for the $R_2$ model, we observe also a distinct peak at the $W$-boson mass in either of the fatjet mass distribution. This is because most $R_2$ events carry $W$ bosons along the boost direction in the top quark's rest frame, and in lab frame decay products of W boson, are more collimated compared to $S_3$ events. However, we see more $S_3$ events than $R_2$ between the $W$ boson and top quark mass because the overall cross-section is the same for both models.
\par
Figures \ref{fig:PTJ0} and \ref{fig:PTJ1} respectively depict the transverse momentum of $J_0$ and $J_1$. From these distributions, we can observe that the signal is substantially harder than the background. Fig.~\ref{fig:FJ_Meff} displays the $M_{\text{eff}}$ distribution, where $M_{\text{eff}}$ is the scalar sum of the total transverse momentum of the visible jets plus MET.
\begin{equation}
M_{\text{eff}}=\slashed{E}_T+\sum |\overrightarrow{P}_{iT}| \,,
\label{EQ-Meff}
\end{equation}
where the summation runs over all the visible jets and $\overrightarrow{P}_{iT}$ is the transverse momentum of the $i$-th jet. Global and inclusive quantities are used to define $\sqrt{\hat{s}_{\text{min}}}$~\cite{Konar:2008ei}, the minimum partonic center-of-mass energy, and its distribution is shown in Fig.~\ref{fig:Shat}. Neutrinos are the missing particles in our system, and the definition of $\sqrt{\hat{s}_{\text{min}}}$ is given by
\begin{equation}
\sqrt{\hat{s}_{\text{min}}} = \sqrt{E^2-P_Z^2}+ \slashed{E}_T
\label{EQ-Shat}
\end{equation}
where $E$ and $P_Z$ are the total energy and longitudinal component of the total visible momentum in the event, respectively. Here visible means all the visible objects in the detector, e.g., jets, electrons, photons, and muons. The signal has a peak towards a larger value of $\sqrt{\hat{s}_{\text{min}}}$ compared to the background since the signal requires more partonic center-of-mass energy to produce two heavy LQs that subsequently decay into the top quark and neutrino.
\par
The N-subjettiness variables, $\tau_{32}$, for both the leading and subleading fatjets are shown in figures \ref{fig:FJ_tau32_J0} and \ref{fig:FJ_tau32_J1}. $\tau_{N}$ tries to quantify the number of subjets inside the fatjet. 
One would anticipate a smaller value of $\tau_{32}$ for a boosted top-fatjet since the value of $\tau_{3}$ for a three-prong fatjet is small and the value of $\tau_{2}$ is large, therefore their ratio produces a smaller value. In contrast, backgrounds are mostly QCD dominated (1-prong) or coming from the weak bosons (2-prong), so the value of  $\tau_{2}$ is small for both QCD jets and fatjets originating from weak bosons, giving larger $\tau_{32}$. The distributions show that the signal has considerably lower $\tau_{32}$ values \footnote{Although the signal peaks at a lower value of $\tau_{32}$ than the background, the peak emerges at roughly 0.6, which is rather substantial.  The three subjets of the top quark are highly collimated, therefore the $\tau_{2}$ value is also small for the top fatjets, which causes the three-prong top fatjets peak to arise for the signal at a significantly large value of $\tau_{32}$. } than the backgrounds, indicating that the signal has a more three-prong structure than the background. 
Different chirality of the top quarks accounts for the slight difference in these distributions for $S_3$ and $R_2$ models. The distributions of $\tau_{31}$ for $J_0$ and $J_1$ are shown in figures \ref{fig:FJ_tau31_J0} and \ref{fig:FJ_tau31_J1}. The distributions show that both the signal and the background peak at a lower value of $\tau_{31}$, indicating that it is not as good as $\tau_{32}$ for distinguishing the signal from the background. 
\par
The distribution of missing transverse momentum is shown in Fig.~\ref{fig:FJ_MET}, where the background can be seen to drop sharply for large MET. In the case of signal, both the neutrinos from the decay of LQs, have equal access to the phase space, resulting in a nearly uniform distribution of the missing transverse momentum. Figures \ref{fig:FJ_Phi_J0ET}, \ref{fig:FJ_Phi_J1ET}, and \ref{fig:FJ_RJ0J1} show, respectively, the distributions of the azimuthal separation of the leading and subleading fatjets from the $\slashed{E}_T$ and the relative separation between the fatjets in the $\eta-\phi$ plane. The distribution of $M_{T2}$~\cite{Lester:1999tx,Barr:2011xt} is shown in Fig.~\ref{fig:FJ_MT2}. $M_{T2}$ is useful in measuring the mass of the parent particle, which is pair-produced at the collider, and subsequently decays into one visible object and one missing particle from the end-point of the distribution, and it is defined as follows
\begin{equation}
M_{T2} =   \min_{\scriptsize{\overrightarrow{p_{1T}}^{\text{invi}}+\overrightarrow{p_{2T}}^{\text{invi}}=\slashed{E}_T}}   [ \max \{  M_{T}^{(1)} , M_{T}^{(2)}  \} ]. 
\label{EQ.MT2}
\end{equation}
$M_{T}^{(i)}$ $(i=1,2)$ are the transverse masses of the LQ and anti-LQ as defined below, 
\begin{equation}
(M_{T}^{(i)})^2=m_i^2+M_{\text{invi}}^2+2(E_{iT} E_{iT}^{\text{invi}} - \overrightarrow{p_{iT}}\cdot \overrightarrow{p_{iT}}^{\text{invi}}), \hspace{1cm} \{ i=1,2 \} \,.
\label{EQ.LQ-TransMass}
\end{equation}
Since LQ decays into a top quark and massless neutrino, we set $M_{\text{invi}}^2 = M_{\nu}^2=0$ and $E_{iT}^{\text{invi}}=|\overrightarrow{p_{iT}}^{\text{invi}}|$, where $\overrightarrow{p_{iT}}^{\text{invi}}$ is the transverse momentum of an individual neutrino. $\overrightarrow{p_{iT}}^{\text{invi}}$ is constrained by the measured missing transverse momentum, 
\begin{equation}
\overrightarrow{p_{1T}}^{\text{invi}}+\overrightarrow{p_{2T}}^{\text{invi}}=\overrightarrow{\slashed{E}_T}.
\end{equation}
$m_i$, and $\overrightarrow{P_{iT}}$ ($i=1,2$) are the reconstructed mass and the transverse momentum of the (sub)leading top-fatjets, respectively. $E_{iT}$ is the transverse energy of the fatjets defined as $E_{iT}=\sqrt{m_i^2+\overrightarrow{P_{iT}}^2}$. One can observe from the distribution Fig.~\ref{fig:FJ_MT2} that its end point correctly predicts the mass of the LQ (1.3 TeV). Since the SM particles have masses that are significantly less than the LQ mass, the background and signal distributions are quite well separated. So this variable not only predicts LQ mass, but also helps in background reduction.   

\subsection{Multivariate Analysis}
\label{MVA analysis}

\begin{figure}[htbp!]
\centering
  \subfloat {\label{correlation_S3_S}\includegraphics[width=0.495\textwidth]{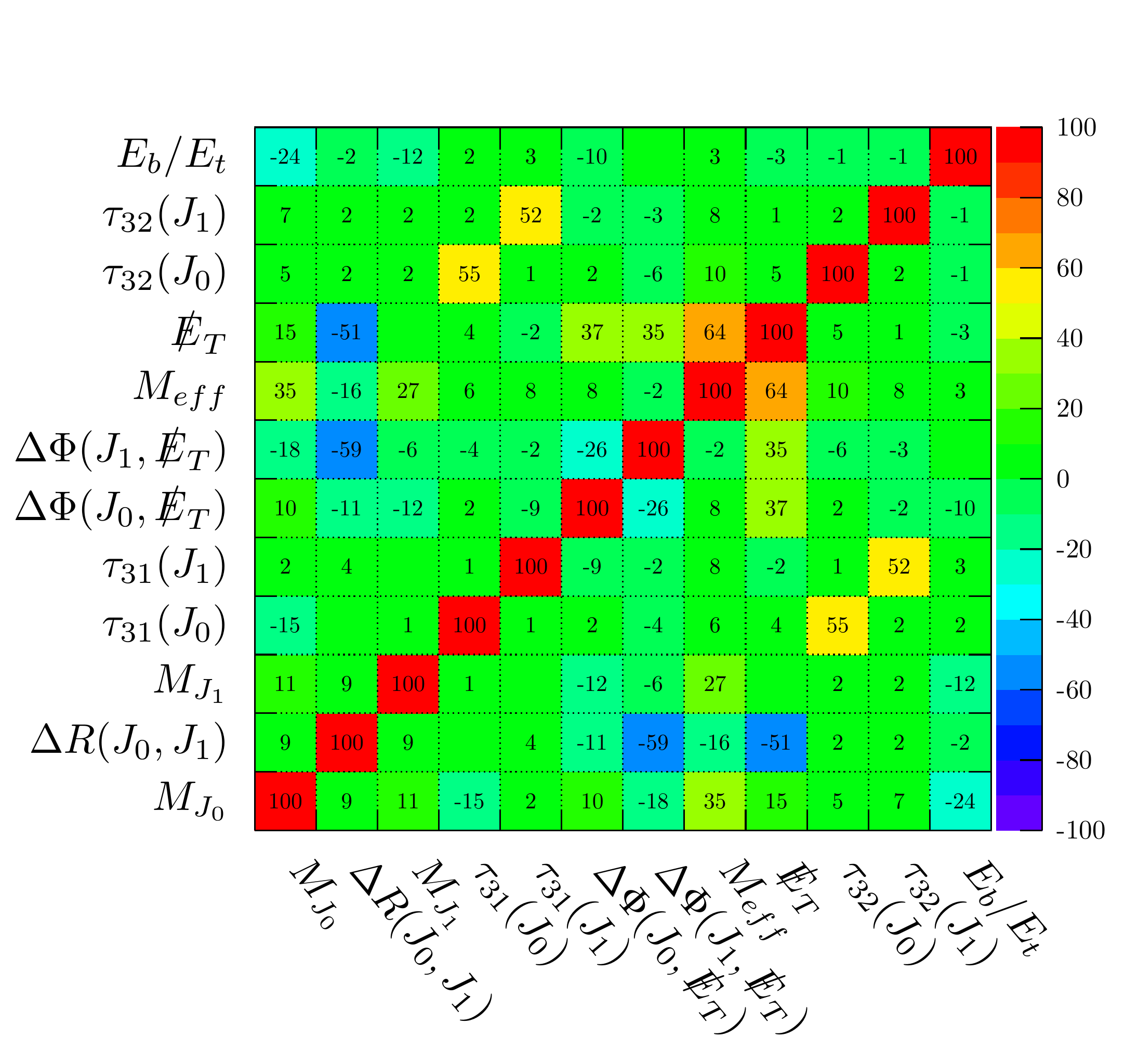}} 
  \subfloat {\label{correlation_S3_B}\includegraphics[width=0.495\textwidth]{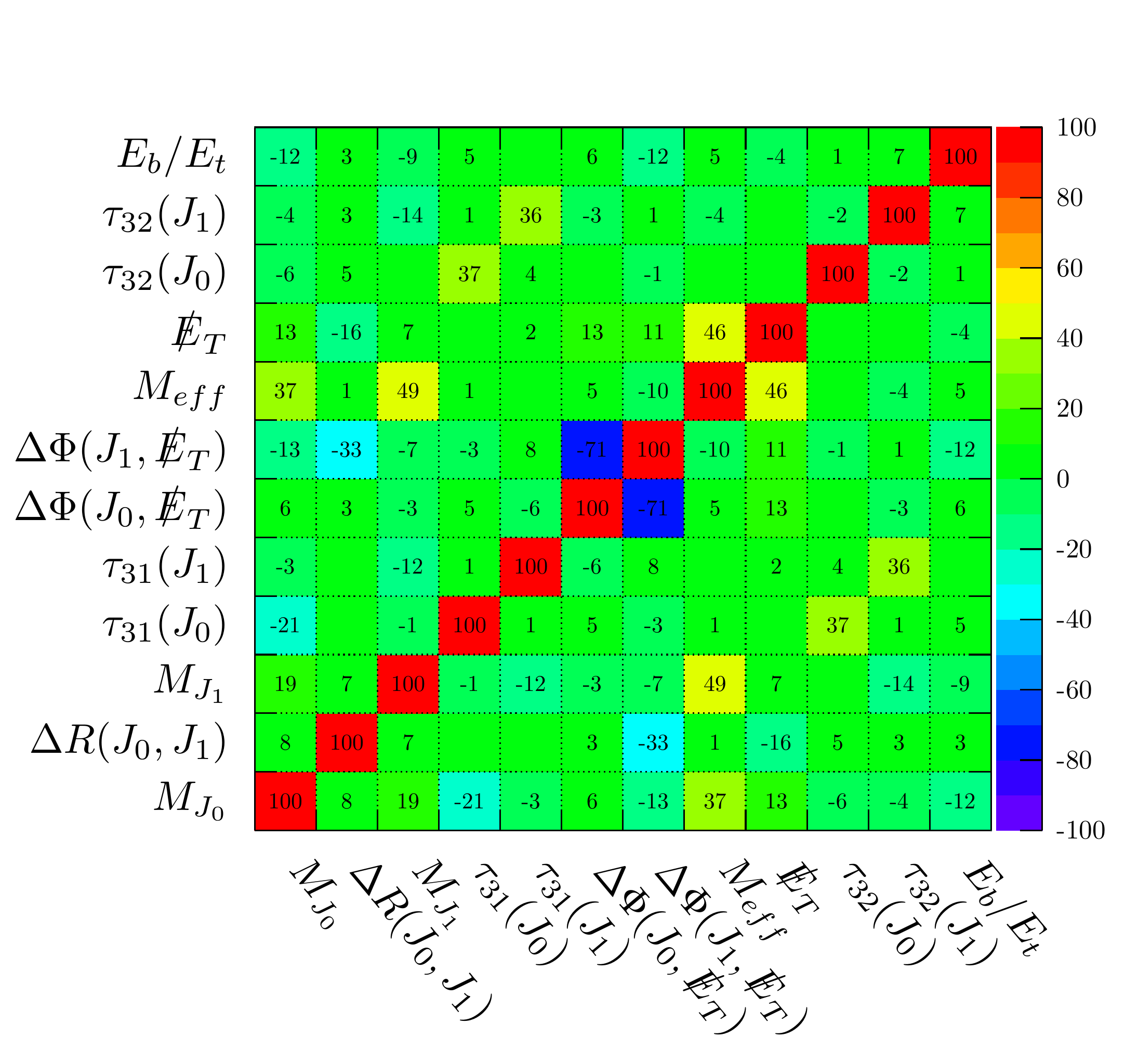}}\\
    \subfloat {\label{correlation_R2_S}\includegraphics[width=0.495\textwidth]{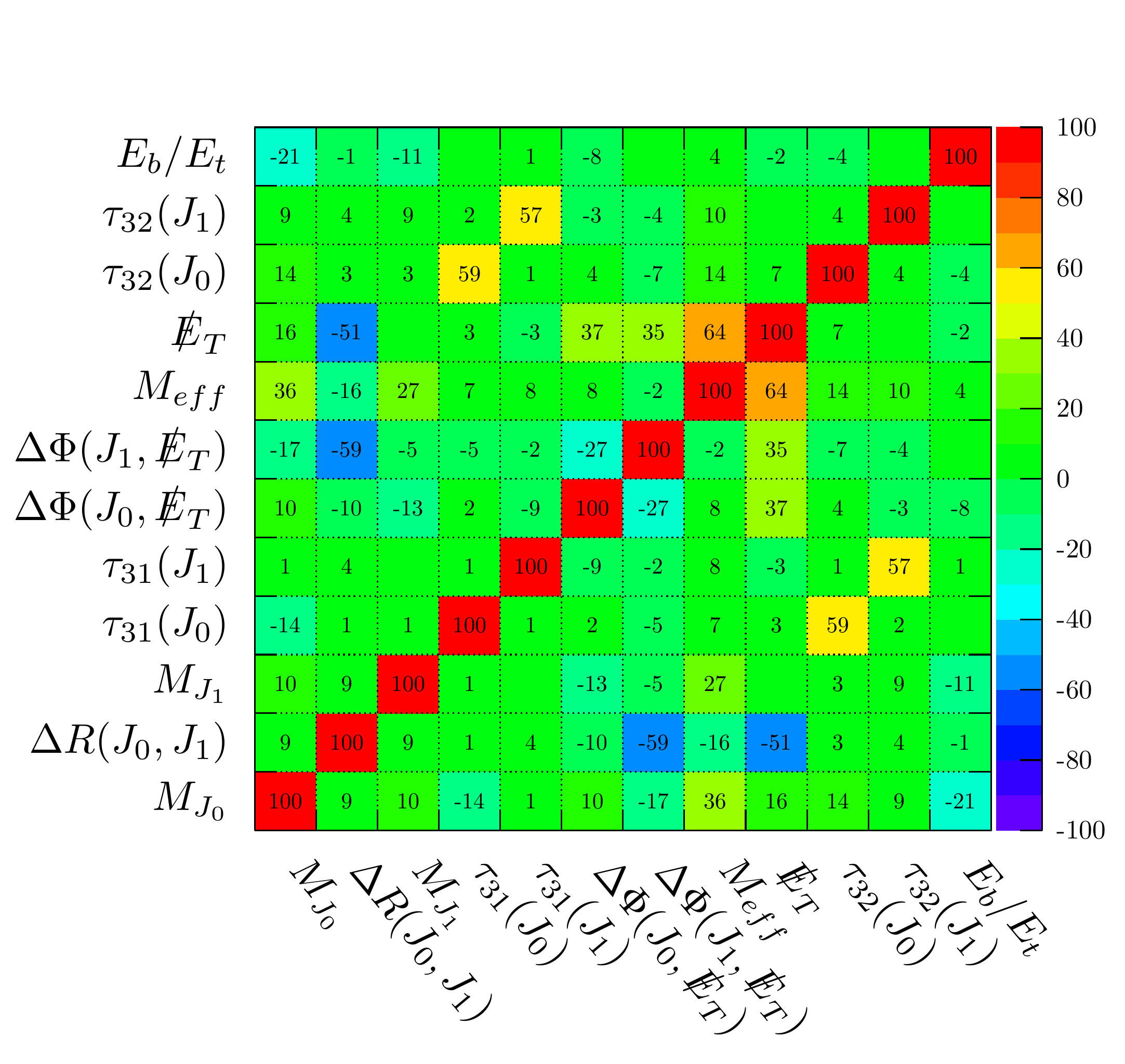}} 
  \subfloat {\label{correlation_R2_B}\includegraphics[width=0.495\textwidth]{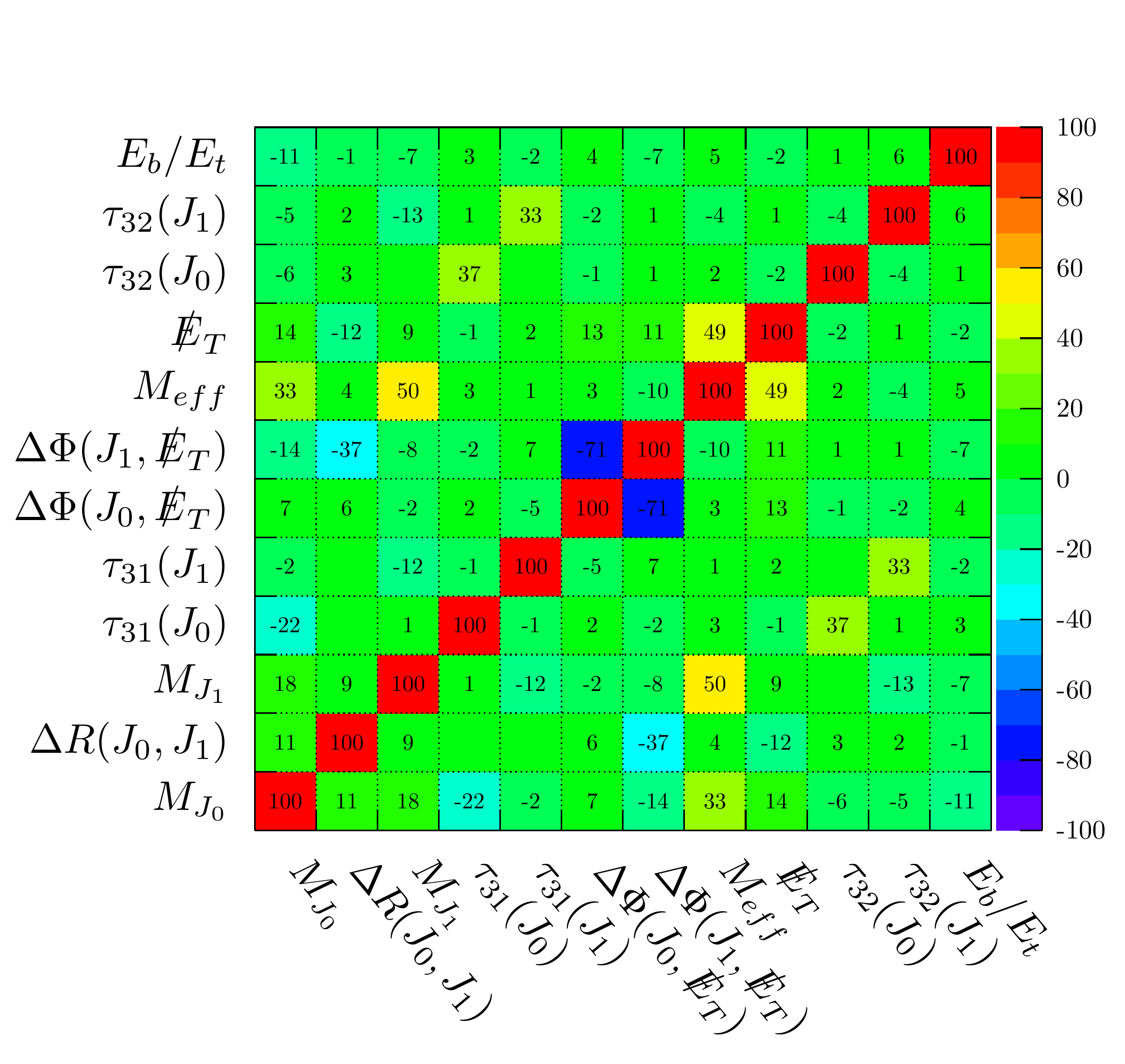}}
  \caption{Linear correlation coefficients ($\%$) between different variables for signal $S_3$ (top left panel) and corresponding background (top right panel); same for signal $R_2$ (bottom left panel) and corresponding background (bottom right panel). Positive and negative coefficients show that two variables are correlated or anti-correlated, respectively. Missing entries indicate an insignificant correlation of less than one.}
  \label{Fig:correlation}
\end{figure}

In the previous subsection, distribution of several observables (without C4 cut), that can be used as input variables for sophisticated multivariate analysis using the gradient boosting technique, are described. For MVA input, we use a loose-cut (up to C4), as mentioned in the preceding subsection. The last row of Tab.~\ref{tab:cut-flow} shows the estimated amount of signals (in fb) from two models, the contribution of different background processes, and the total background at the 14 TeV LHC after applying MVA selection cut (C4). For MVA, we use the adaptive Boosted Decision Tree (BDT) algorithm and construct two statistically independent signal and background event samples. The background is the weighted sum of individual SM background processes. MVA picks a subset of kinematic variables from a larger collection based on the linear correlation among the variables and their relative importance in distinguishing the signal from the background. 
\par
As expected by Eq. \ref{EQ-Meff}, we notice that $P_T(J_0)$ and $P_T(J_1)$ have large correlations with $M_{\text{eff}}$, and $\sqrt{\hat{s}_{\text{min}}}$ also exhibits high correlations with $M_{\text{eff}}$ due to their linear dependence on MET, as shown by Eqs. \ref{EQ-Meff} and \ref{EQ-Shat}. We keep $M_{\text{eff}}$ because of its high relative importance compared to $P_T(J_0)$, $P_T(J_1)$, and $\sqrt{\hat{s}_{\text{min}}}$. A high correlation exists between $M_{T2}$ and MET; however, we retain MET because it has the highest relative importance than any other variables in separating the signal from the background. Although $M_{\text{eff}}$ and MET exhibit a significant correlation in both the signal and background (as predicted by Eq.~\ref{EQ-Meff}), we keep them both in our study since they have exceptionally high separation powers to distinguish the signal from the background. Fig.~\ref{Fig:correlation} exhibits the linear correlation coefficients between different variables for signal $S_3$ (top left panel) and the corresponding background (top right panel). The bottom left and bottom right panels depict the signal $R_2$ and its corresponding background. Positive and negative coefficients indicate whether two variables are correlated or anti-correlated. In the TMVA package \cite{Hocker:2007ht}, the linear correlation coefficient is calculated using the following formula, 
\begin{equation}
\rho(x,y) = \dfrac{\text{cov}(x,y)}{\sigma_x \sigma_y},
\label{EQ:correlation}
\end{equation}
where the covariance between $x$ and $y$ is $\text{cov}(x,y) = \langle xy\rangle-\langle x\rangle\langle y\rangle$ and $\sigma_{x}$, $\sigma_{y}$ are the standard deviation of these variables. 
\par
\begin{figure}[htbp!]
\centering
  \subfloat {\label{output_S3}\includegraphics[width=0.495\textwidth]{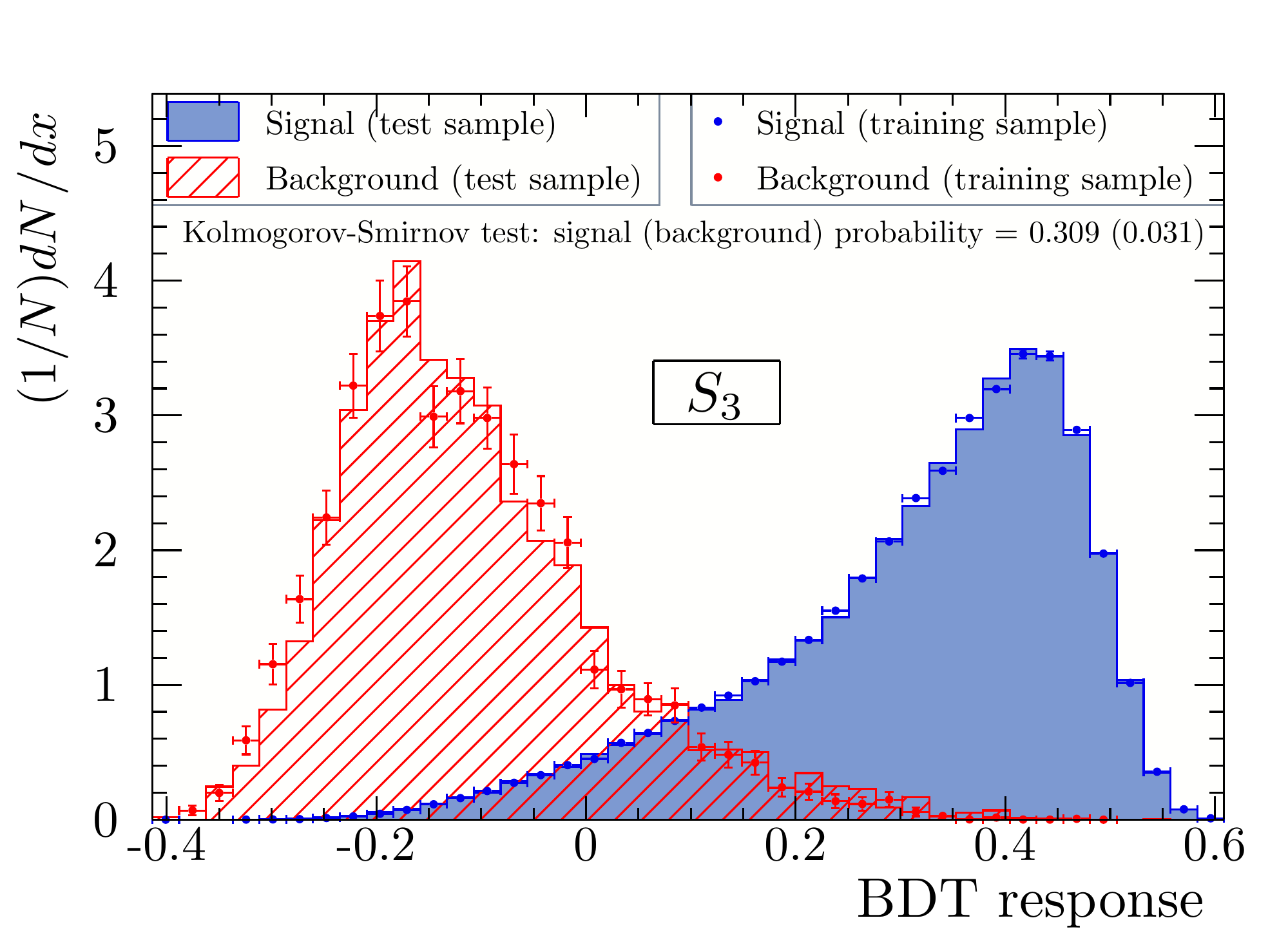}} 
  \subfloat {\label{significance_S3}\includegraphics[width=0.495\textwidth]{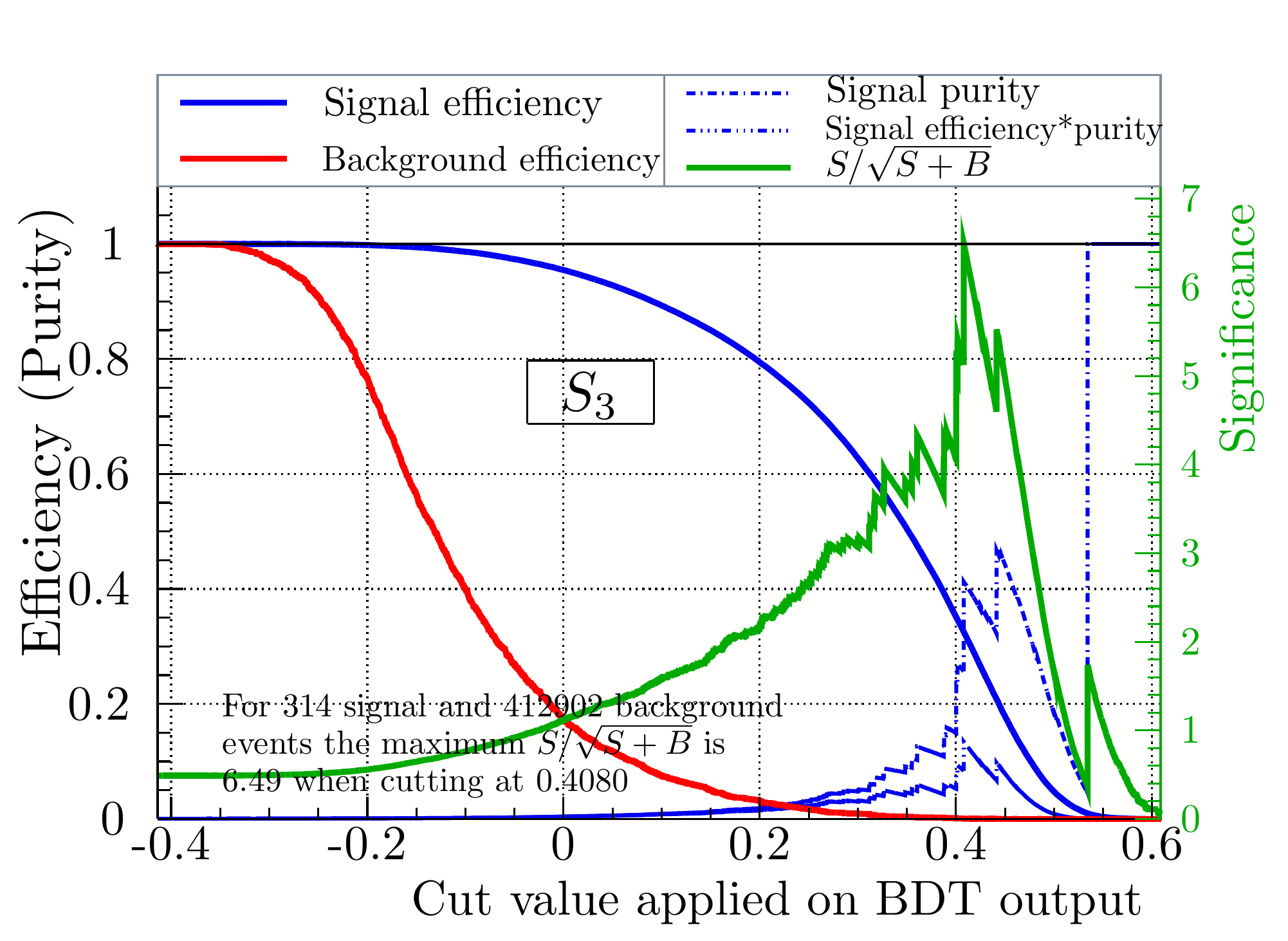}}\\
  \subfloat {\label{output_R2}\includegraphics[width=0.495\textwidth]{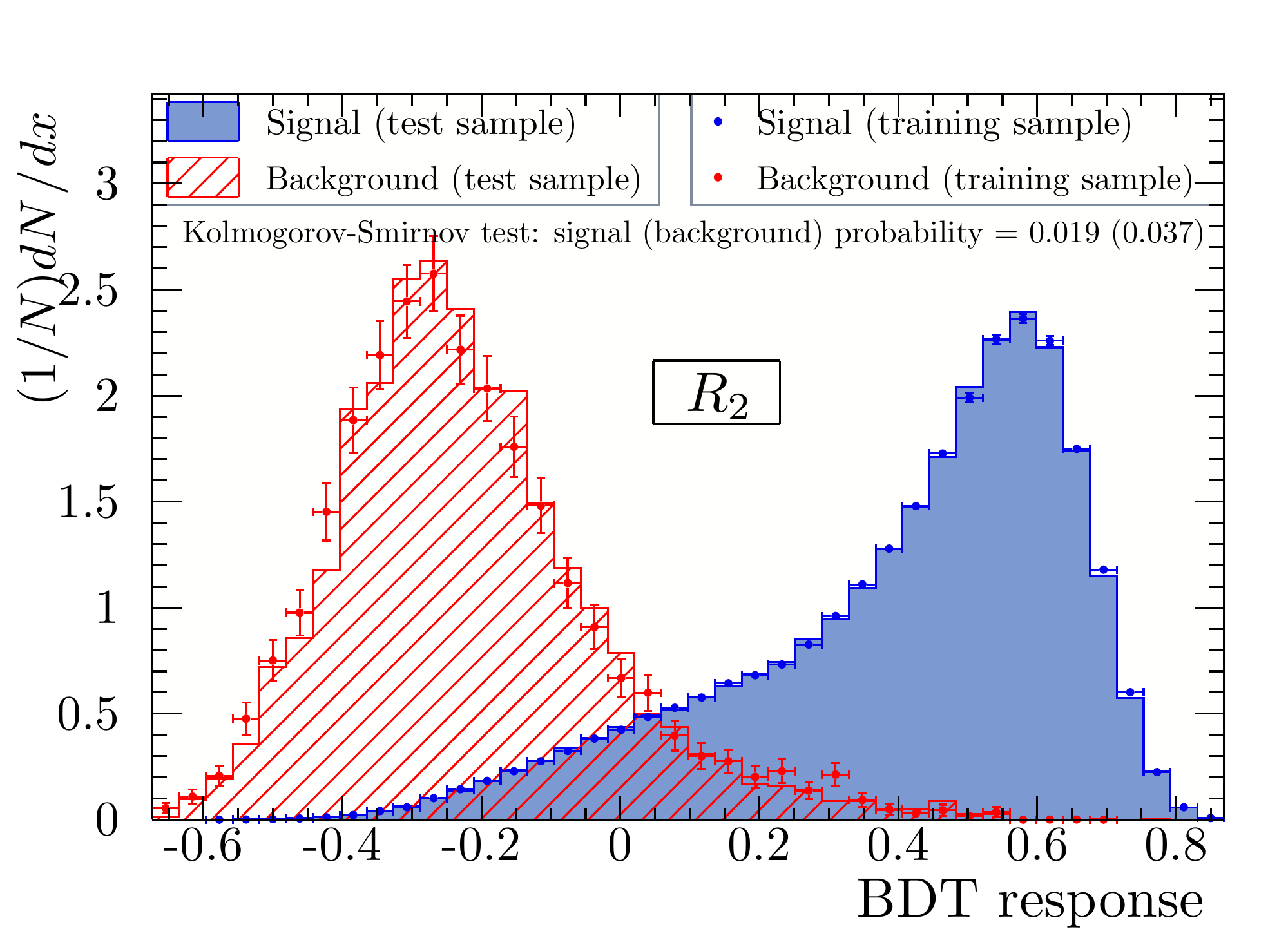}} 
  \subfloat {\label{significance_R2}\includegraphics[width=0.495\textwidth]{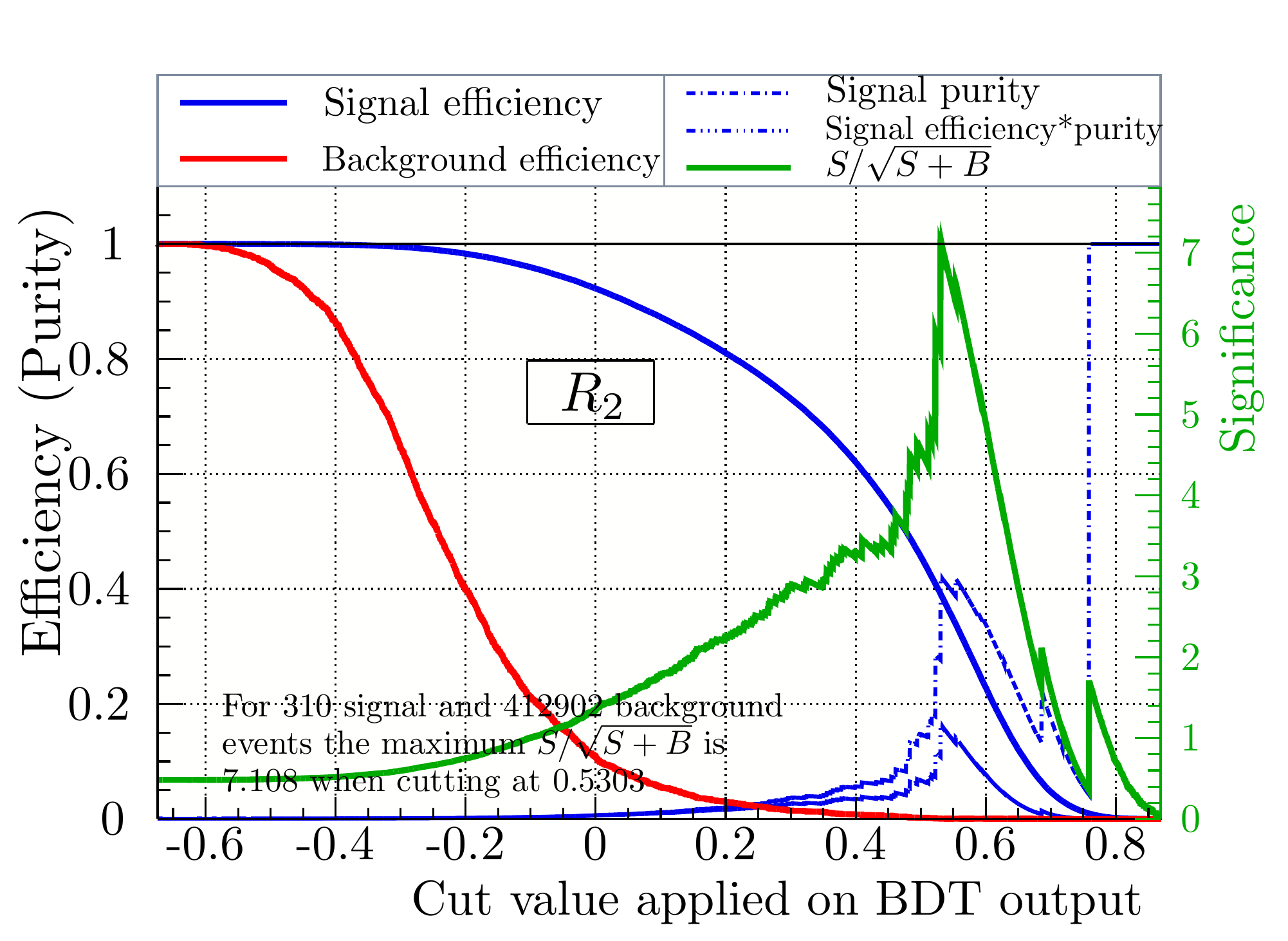}}  
  \caption{The top-left plot depicts the distribution (normalized) of the BDT output for the training and testing samples for both the signal $S_3$ (blue) and background (red) classes. The right plot depicts signal $S_3$ (blue) and background (red) efficiencies, as well as statistical significance ($\frac{N_S}{\sqrt{N_S+N_B}}$) as a function of the cut applied to BDT output. The same for the $R_2$ model is shown in the bottom left and bottom right plots.}
\label{plot:MVA-Output-Signi}
\end{figure}

\begin{table}[H]
\centering
\resizebox{\columnwidth}{!}{%
 \begin{tabular}[b]{|c|c|c|c|c|c|c|c|c|c|c|c|}
\hline
Variable     & $\slashed{E}_T$ & $\text{M}_{\text{eff}}$ & $\Delta R (J_0, J_1)$ & $\Delta \phi(J_1,\slashed{E}_T)$ & $M(J_0)$  & $\tau_{32}(J_1)$ &  $\tau_{32}(J_0)$ & $\Delta \phi(J_0,\slashed{E}_T)$ & $M(J_1)$ & $\tau_{31}(J_1)$ & $\tau_{31}(J_0)$ \\ %
\hline\hline
$S_3$ & 59.98 & 49.43 & 23.44 & 21.42 & 5.99 & 4.15 & 3.99 & 3.83 & 2.33 & 0.99 & 0.95 \\
\hline
\hline\hline
$R_2$ & 59.33 & 50.63 & 21.97 & 20.87 & 6.42 & 5.85 & 4.82 & 4.36 & 2.49 & 1.49 & 1.11 \\
\hline
 \end{tabular} 
}
\caption{Before employing at MVA, the method unspecific relative importance (separation power) of the individual variables.}
\label{relative_imp}
\end{table}
The separation power of different kinematic variables for the two models used in MVA, is presented in Tab.~\ref{relative_imp}. This table shows that the order of the variables in for distinguishing the leptoquark signal from the overwhelming background are the MET, $M_{\text{eff}}$, relative separation between the fatjets in $\eta-\phi$ plane, and azimuthal separation between the subleading fatjet and MET. Due to improper selection of various (BDT-specific) parameters during training, the BDT method may result in overtraining. Overtraining can be prevented if the Kolmogorov-Smirnov probability is checked throughout training. We train the algorithm separately for the $S_3$ and $R_2$ models and ensure that there is no overtraining in our analysis. The top left panel of Fig.~\ref{plot:MVA-Output-Signi} shows the normalised distribution of the BDT output for the signal $S_3$ (blue) and its background (red) for both training and testing samples, whereas the bottom left plot shows the same for the $R2$ model. We observe that for both models, signal and background are well separated. In the same figure, top right plot illustrates the signal $S_3$ (blue) and background (red) efficiencies, as well as statistical significance (green) as a function of the cut applied to BDT output, while the bottom right plot depicts the same for the $R_2$ model. 
\par
\begin{table}[htbp!]
\begin{center}
\resizebox{0.8\columnwidth}{!}{%
 \begin{tabular}{|c|c|c|c|c|c|c|}
\hline
 & $N_S^{bc}$ (fb) & BD$T_{opt}$ & $N_S$ (fb) & $N_B$ (fb) & $\frac{N_S}{\sqrt{N_S+N_B}}$ ($\frac{N_S}{\sqrt{N_B}})$, 3 $\text{ab}^{-1}$ & $\frac{N_S}{N_B}$  \\ 
\hline\hline
$S_3$ & 0.1047 & 0.4080 & 0.03403 & 0.04850 & 6.5 (8.5) & 0.702 \\
\hline
$R_2$ & 0.1033 & 0.5303 & 0.04047 & 0.05677 & 7.1 (9.3) & 0.713 \\
\hline
$N_{SM}$ & 137.634 & \multicolumn{5}{| c |}{} \\
\hline
 \end{tabular}
} 
\caption{The table shows the effectiveness of the present search in terms of statistical significance for $S_3$ and $R_2$ models. Before applying any cuts to the BDT output, the total number of events for different models and the combined background are $N_S^{bc}$ and $N_{SM}$, respectively (as shown in Tab.~\ref{tab:cut-flow}). For the 14 TeV LHC, after employing an optimum cut (BD$T_{opt}$) on the BDT response, the surviving number of signal and background events are provided by $N_S$ and $N_B$ (in fb), respectively. For quick access, the statistical significances corresponding to 3 $\text{ab}^{-1}$ luminosity are also shown.}
\label{tab:MVA-result}
\end{center}
\end{table}
The statistical significance of the two models at 3 $\text{ab}^{-1}$ integrated luminosity at the 14 TeV LHC and the signal-to-background ratio are shown in Tab.~\ref{tab:MVA-result}. There, $N_S^{bc}$ \footnote{Although we use full NLO events, if one uses LO events but normalizes with the total NLO cross-section, $N_S^{bc}$ number for both models decreases by around $2\%$.} and $N_{SM}$ represent the total number of events for the signal and background before applying any cut to the BDT output, while $N_S$ and $N_B$ represent the same after applying an optimal cut BDT$_{\rm opt}$ to the BDT response. We observe that both models at the HL-LHC have discovery potential for the 1.3 TeV scalar leptoquark. The $5\sigma$-discovery and $2\sigma$-exclusion limits of these two models at the HL-LHC are presented in Tab.~\ref{Tab:Discovery-exclusion}. There is a slight difference in the discovery potential of these two models because of their polarization.   Note that explicitly the polarization variables have a negligible role compared to other variables such as MET, $\Delta R(J_0,J_1)$, etc., as given in Tab.~\ref{relative_imp} for discovering the leptoquark signal at the LHC. However, we see significant differences in the $J_0$ and $J_1$ mass distribution and slight in N-subjettiness distributions because of the different chirality of the top quarks of the two models. So, once the LQ signal is discovered at the LHC, we can use the polarization variables to distinguish these two models, which are described in detail in the next section. In our analysis, we find that a scalar LQ of mass 1270 GeV or smaller can be rejected with $2\sigma$ with an integrated luminosity of 140 $\text{fb}^{-1}$, which is compatible with the existing ATLAS search and analysis. We also find that a luminosity around 1600 $\text{fb}^{-1}$ is required for the $5\sigma$ discovery of 1.3 TeV scalar LQ. 
%
\begin{table}[h]
\begin{center}
 \begin{tabular}{|c|c|c|}
\hline
$\mathcal{L}= 3 ab^{-1}$ & $S_3^{\frac{2}{3}}$ & $R_2^{\frac{2}{3}}$ \\
\hline
5$\sigma$ discovery & 1380 GeV & 1370 GeV \\
\hline
2$\sigma$ exclusion & 1520 GeV & 1520 GeV \\
\hline
\end{tabular}
\caption{Discovery and exclusion reach at 14 TeV LHC for 3 $ab^{-1}$ luminosity.}
\label{Tab:Discovery-exclusion}
\end{center}
\end{table}

\section{Distinguishing two models}
\label{sec:distinguigh_models}
If a leptoquark signature is observed at the collider in some particular final state, the next goal will be to distinguish different models in order to probe its genesis. In the above section, we have seen that pair production in both $S_3^{\frac{2}{3}}$ and $R_2^{\frac{2}{3}}$ models can finally give two fatjets plus large missing energy signature. The leptoquarks in these two models decay to top quarks of different helicities.
Top quark's polarization can be probed by studying the distribution of some particular kinematic variables of its decay products, which can in turn allow us to probe the type of the leptoquark. In the following subsection, we discuss some such polarization variables that can address the leptoquark identity.

\subsection{Polarization Variables}
\label{sec:pol_var}
There are different variables which can exhibit dependence on top quark polarization. In the following, we discuss a few of them.
\subsubsection{Angular variable in the rest frame of (anti-)top}
\label{kin_dist}

In the rest frame of top quark, if $\theta_i$ be the angle between the decay particle $i$ and the direction of boost of the top quark
, the differential distribution of the decay width $\Gamma$ with respect to the angular variable $\rm \cos{\theta_i}$ is given by,
\begin{equation}
\dfrac{1}{\Gamma}\dfrac{d \Gamma} {{d \cos{\theta_ i}}} = \dfrac{1}{2} (1 + {\rm{P_t}\ k_i\ \cos{\theta_i}}),\label{eq:pol_dis} 
\end{equation}
where $\rm P_t$ is the top quark polarization, which is $+1$ for the right-handed top and $-1$ for the left-handed top. $k_i$ is the spin analyzing power of $i$-th decay particle. In Tab.~\ref{tab:spin_analyzing}, we show the spin analyzing power of different decay particles. In Appendix~\ref{app:dist_daught_top}, the spin analyzing power of bottom quark is derived. Similar distribution in the anti-top rest frame can be written as, 
\begin{equation}
\dfrac{1}{\bar{\Gamma}}\dfrac{d \bar{\Gamma}}{{d \cos{\bar{\theta}_{\bar i}}}} =  \dfrac{1}{2} (1 + {\rm{\bar{P}_{\bar{t}}}\ \bar{k}_{\bar i}\ \cos{\bar{\theta}_{\bar i}}}),
\end{equation}
where the entities with bar are the corresponding quantities for the anti-top quark.
 Here as well, $\bar{P}_{\bar{t}}$ is +1 for right-handed anti-top and -1 for left-handed antitop. $\bar{k}_{\bar i}$ is given by $\bar{k}_{\bar i} = -k_i$. So it is evident that the distribution of the $i$-th decay particle coming for the right-handed top will be same as the distribution of ${\bar i}$-th decay product of the left-handed antitop. As we are producing leptoquark pair which will decay to top and antitop with opposite helicities, this feature will ensure the distributions of b and $\bar{b}$ for a model are the same.


\begin{table}[H]
\begin{center}
 \begin{tabular}{|c|c|c|}
\hline
Daughters & b & $W^{+}$ \\
\hline
$k_i$ & -0.41 & +0.41 \\
\hline
\end{tabular}
\caption{Spin analyzing power of bottom quark and $W^+$ coming from top decay.}
 \label{tab:spin_analyzing}
\end{center}
\end{table}

\begin{figure}[H]
\centering
\includegraphics[angle=0,scale=0.8]{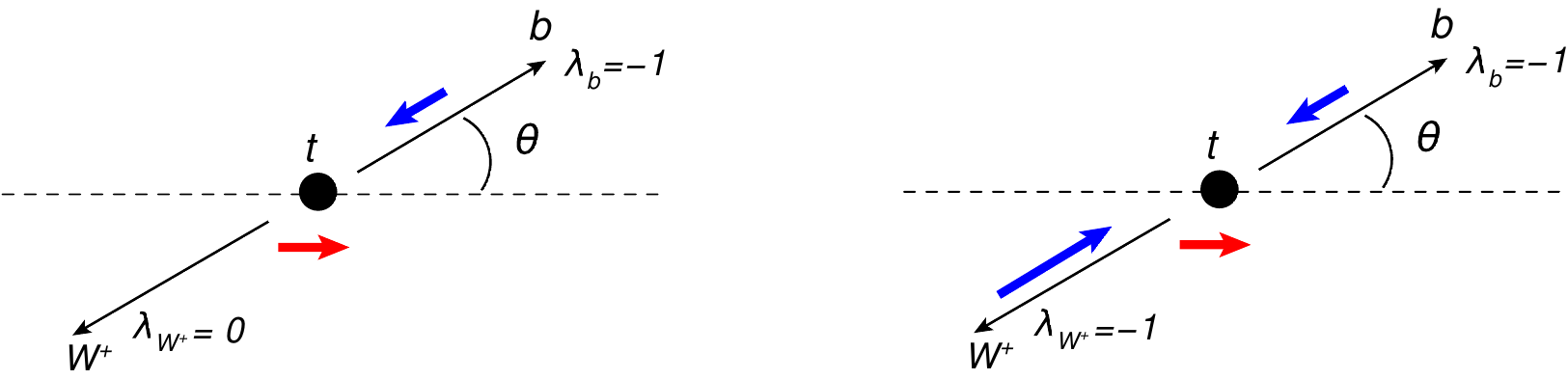}
\caption{Decay diagram of right-handed top quark in its rest frame. Black dot represents top quark.
Thick colored arrows denote spin of the particles. For b-quark, essentially $\lambda_b=-1$ component gets produced and the other component $\lambda_b=+1$ is heavily suppressed, because of its small mass. 
 The top decays to $\lambda_{W^{+}}=0$ and $\lambda_{W^{+}}=-1$ helicity components of $W^+$ 70\% and 30\% times, respectively.  As the other transverse component W boson, i.e. $\lambda_{W^{+}}=1$,  requires right-handed b-quark to conserve spin, it is also suppressed. So in effect only the two diagrams shown here contribute to the right-handed top quark decay.
}  
 \label{fig:top-decay-pol}
\end{figure}

 The decay of top quark gives rise to mostly left-handed ($\lambda_{b}=-1$) b-quark and the other component, {\em i.e.} the right-handed one, is heavily suppressed because of small mass of b-quark\footnote{This happens as the decay is governed by weak interaction, which couples to only left-handed fermions in the massless limit.}. It is known that the top quark decays 70\% of the time to longitudinal ($\lambda_{W^{+}}=0$) and 30\% of the time to one of the transverse ($\lambda_{W^+}=-1$) component of the W boson\ds{~\cite{Czarnecki:2010gb,ATLAS:2012nhi}}\footnote{The top quark decay to other transverse component ($\lambda_{W^+} = +1 $) is almost negligible, as this requires  right-handed b-quark (which is heavily suppressed) to conserve spin angular momentum.}. So for top quark, essentially only two decay configurations exist. In Fig.~\ref{fig:top-decay-pol}, to illustrate, we show these two configurations for decay of a right-handed top quark in its frame. To conserve the total spin in the decay process, the total spin of the b-quark and W boson system must be equal to $\dfrac{1}{2}$. Moreover, we can write the spin state of the b-quark and W boson system in the basis of $|+\rangle\strut_{\hat{z}}$ and $|-\rangle\strut_{\hat{z}}$ states\footnote{$|+\rangle\strut_{\hat{n}} = \cos{\frac{\Theta}{2}} |+\rangle\strut_{\hat{z}} + \sin{\frac{\Theta}{2}} e^{i\Phi}  |-\rangle\strut_{\hat{z}}$, where $\hat{n}$ is a unit vector along ($\Theta$,$\Phi$) direction. 
} 
(with positive z-axis along the top boost direction). So to conserve third component of spin, only $|+\rangle\strut_{\hat{z}}$ component can contribute, as the top quark spin is along the the boost direction. For the left diagram, the total spin of b-quark and W boson system makes an angle  $(180 - \theta)$ with the boost direction, whereas for the right diagram it makes angle $\theta$. So the left diagram follows a $\sin^2{\frac{\theta}{2}}$ distribution, whereas the right diagram follows a $\cos^2{\frac{\theta}{2}}$ distribution\footnote{${(\cos{\frac{\Theta}{2}})}^2|_{\Theta=180-\theta} = \sin^2{\frac{\theta}{2}}  $}. Obviously, the weighted sum of these two distributions should lead to Eq.~\ref{eq:pol_dis} \footnote{0.7 $\sin^2{\frac{\theta}{2}}$ + 0.3 $\cos^2{\frac{\theta}{2}}$ = $0.5- 0.2\ \cos{\theta}$ }.

\subsubsection{Energy variables in the Lab  frame}

In the literature\ds{~\cite{PrasathV:2014omf,Papaefstathiou:2011kd,Allahverdi:2015mha,Shelton:2008nq}}, two most discussed energy variables for the polarisation study are $z=\frac{E_b}{E_t}$ and $u=\frac{E_l}{(E_l+E_b)}$. However, the variable $z=\frac{E_b}{E_t}$, which is the fraction of energy of the top quark carried by the b quark in the lab frame, is only the relevant one here as the W boson originating from top quark decays hadronically in our study.
The variable ``z" and $\rm{cos\; \theta_b}$ are fully correlated and they are related by the following relation~\cite{Shelton:2008nq} (see Appendix~\ref{app:rel_Cos_and_z}),
\begin{equation}
\rm{\cos{\theta_b}} = \dfrac{1}{\beta_t}\Big( \dfrac{2 m_t^2}{m_t^2 - m_W^2} z -1 \Big),
\label{eq:cos-z}
\end{equation}
 where $\beta_t$ represents the boost of the top quark in the lab frame.  The distribution of decay width with respect to z (using Eq.~\ref{eq:pol_dis} and Eq.~\ref{eq:cos-z}) can be given as~\cite{Papaefstathiou:2011kd},
\begin{equation}
\dfrac{1}{\Gamma}\dfrac{d\Gamma} {dz} = \dfrac{1}{\beta_t} \dfrac{m_t^2}{m_t^2 - m_W^2}\ \Big(1- P_t\ k_f  \dfrac{1}{\beta_t} + P_t\ k_f \dfrac{1}{\beta_t} \dfrac{2 m_t^2}{m_t^2 - m_W^2} z \Big) \,.
\end{equation}

The similar expression will hold for antitop particle with every element replaced by their corresponding barred element.

\subsubsection{Distributions of polarization variables}

\begin{figure}[H]
\centering
\includegraphics[angle=0,width=0.45\linewidth,height=0.37\linewidth]{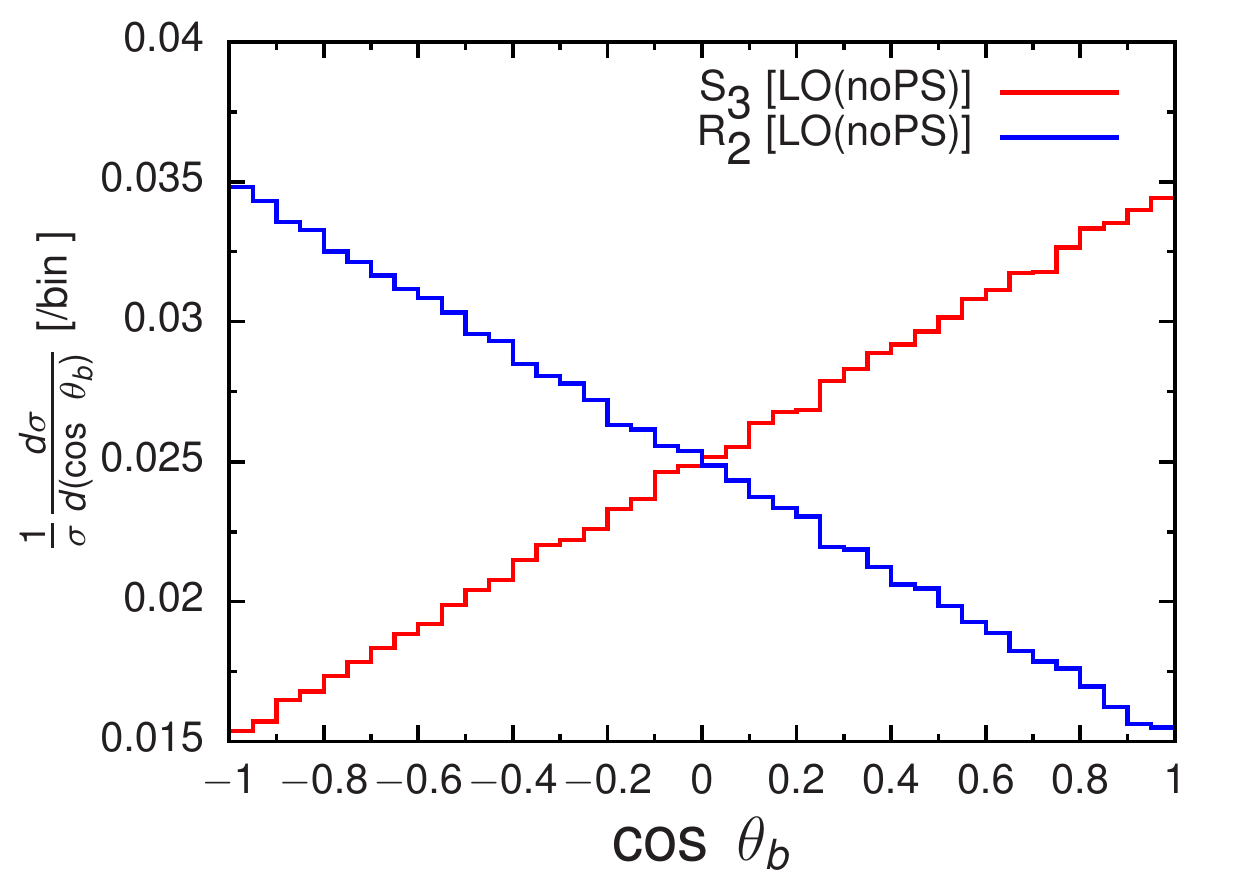}
\includegraphics[angle=0,width=0.45\linewidth,height=0.37\linewidth]{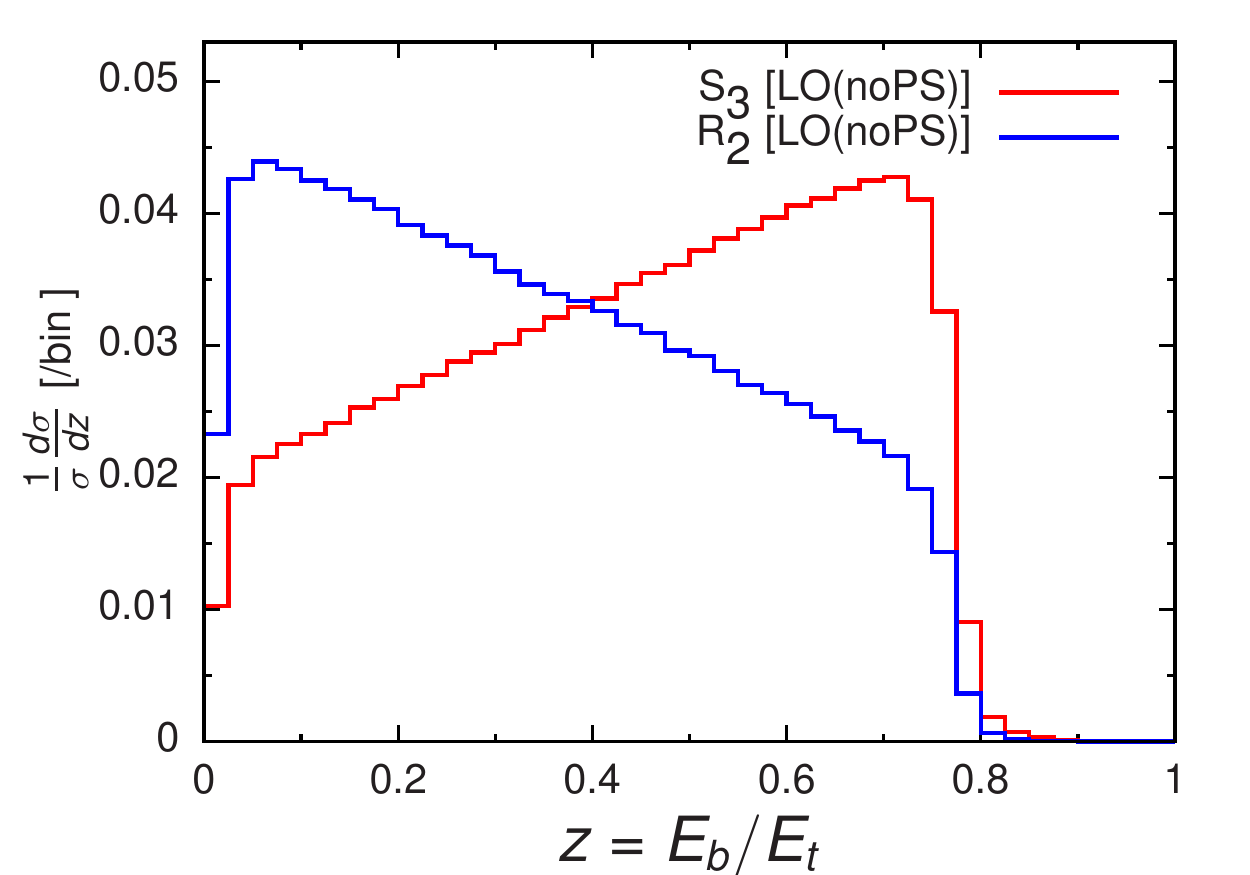}
\caption{The distributions of $\rm \cos\ {\theta_b}$ and $\dfrac{E_b}{E_t}$ at LO without parton shower. The mass of the leptoquark has been taken to be $\rm 1300\ GeV$.}  
 \label{fig:LO-pol-var-truth-level}
\end{figure}

\begin{figure}[H]
\centering
\includegraphics[angle=0,width=0.45\linewidth,height=0.34\linewidth]{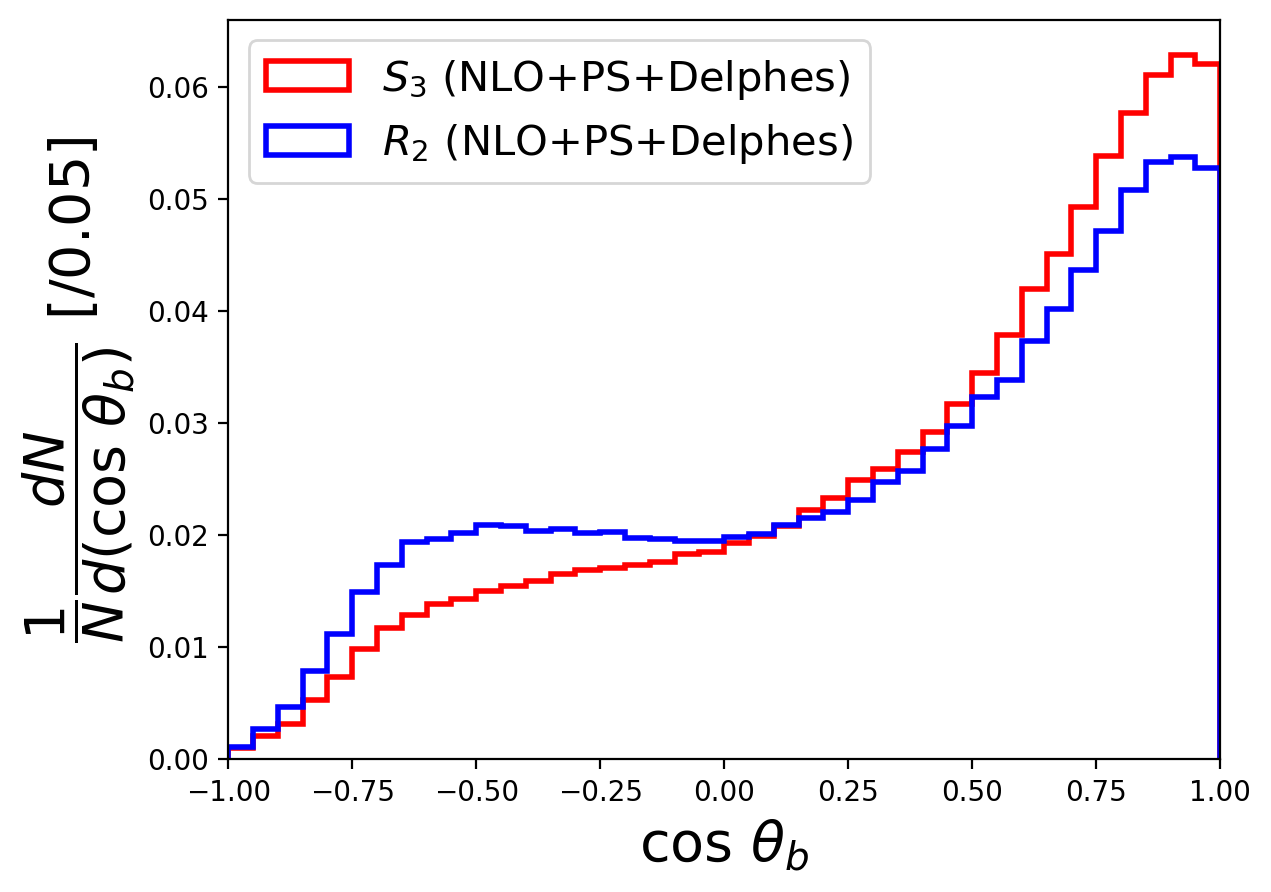}
\includegraphics[angle=0,width=0.45\linewidth,height=0.34\linewidth]{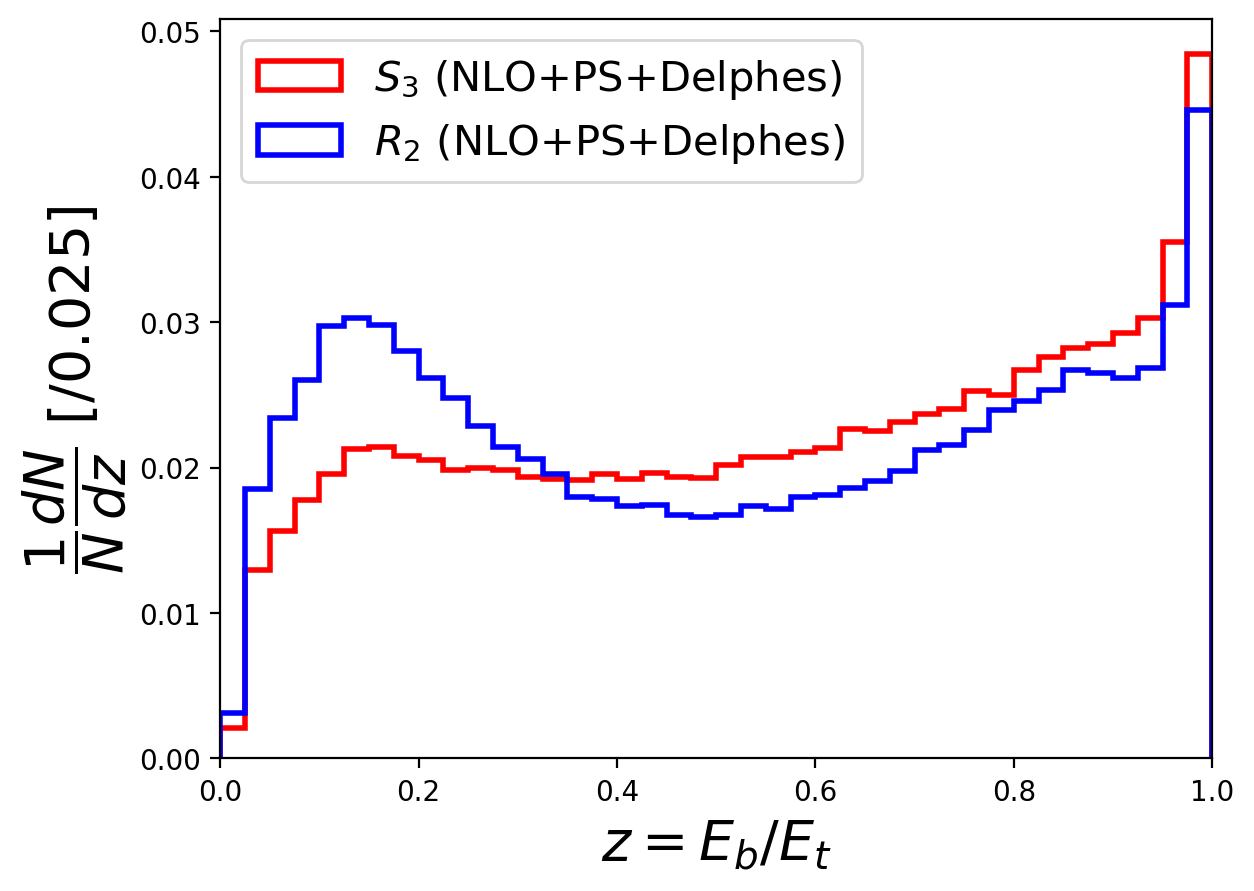}
\caption{The distributions of $\rm \cos\ {\theta_b}$ and $\dfrac{E_b}{E_t}$ after Delphes simulation and applying cuts up to C4 mentioned in Subsection.~\ref{subsec:Event Selection}. The effect of radiation causes significant changes in the distribution compared to truth level results. For $\rm \cos\ {\theta_b} \sim 1$ and for z around 0.8 and more, the distributions are strikingly different from the truth level results because of the contamination in the b-jet from W decay products, owing to very large boost of top quark.}  
 \label{fig:NLO-pol-var-Delphes-level}
\end{figure}

In Fig.~\ref{fig:LO-pol-var-truth-level}, we show truth level normalized distributions of $\rm \cos{\theta_b}$  and $\dfrac{E_b}{E_t}$ at LO at the left and right subfigures, respectively\footnote{We discussed in the Sec.~\ref{kin_dist}, b and $\bar{b}$ jets have the same distributions for a model. For an event, now onwards by b we will mean either b or $\bar{b}$ jet and t will mean corresponding top or antitop fatjet.}. The distribution with respect to $\rm \cos{\theta_b}$ can be understood from Eq.~\ref{eq:pol_dis}. Therefore in $S_3$ model, for most of the events in the rest frame of the top quark, the b-quark moves in the same direction as the boost of the top quark. Obviously, the opposite happens for the $R_2$ model.  For the $z=\frac{E_b}{E_t}$ variable, we see for the $S_3$ and $R_2$ models, the distribution peak near the right and left end of the plots, respectively. This can also be understood from the $\rm \cos{\theta_b}$ distribution. As for the $R_2$ model, in the rest frame, for majority of events, the b-quarks move in the direction opposite to the boost and their energy $E_b$ will be less. Therefore the distribution in this case peaks towards the left. The reverse happens for the $S_3$ model. Another interesting thing to observe in the right figure is that the cross-section is zero after $z=0.8$. This happens because all the top quark energy cannot be carried by the b-quark only, as the W boson needs at least its rest mass energy, $M_W$. In Fig.~\ref{fig:NLO-pol-var-Delphes-level}, we show these distributions after including NLO calculation, showering effect, and applying various cuts up to C4 (discussed in Subsection~\ref{subsec:Event Selection}) in Delphes simulation. Here the distribution of b-jet is found to be different from that of b-quark because of showering effects and formation of jets. Near the boost direction, i.e. near $\cos\ \theta_b \sim 1$, the difference between the b-quark and b-jet distributions is striking as there b-jet gets contaminated with the particles originating from W boson because of very large boost of top quark.


\subsection{Log-likelihood ratio test}
In this subsection, we study the prospect of distinguishing two models, if in the future, a scalar leptoquark of mass $1300$ GeV is observed. It will take around 1600 $\rm fb^{-1}$ of data for a 5$\sigma$ discovery. At this mass, for $\mathcal{L}=3000 \rm fb^{-1}$, with the optimized cuts chosen by BDT, the number of signal and background events are found to be (102,145) for the $S_3$ model and (121,170) for the $R_2$ model \footnote{multiplying luminosity with the cross-sections given in Tab.~\ref{tab:MVA-result} gives these event numbers.}. For these number of events we find the distribution of events with respect to $\frac{E_b}{E_t}$.
We use log-likelihood ratio (LLR) hypothesis test for distinguishing two models\footnote{We have also checked with $\chi^2$ hypothesis test and got similar kind of results.}. The likelihood function is given by the product of Poisson distribution functions at all bins. That is, for $O_i$ being the observed data and $E_i$ being the expected data, the likelihood function $\cal{L}$ is given as, 
\begin{align}
\cal{L(\rm{E|O})} &=\prod_{i=1}^{n} e^{-{E_i}} {E_i}^{O_i}/\Gamma(O_i+1) \,.
\end{align}
The exclusion significance of a model M1, when another model M2 is observed, is given as
\begin{align}
Z_{M1|M2} &= \sqrt{ -2ln\frac{\cal{L(\rm{M1|M2})}}{\cal{L(\rm{M2|M2})}}} \,.
\end{align}

\begin{figure}[H]
\centering
\includegraphics[angle=0,width=0.45\linewidth,height=0.35\linewidth]{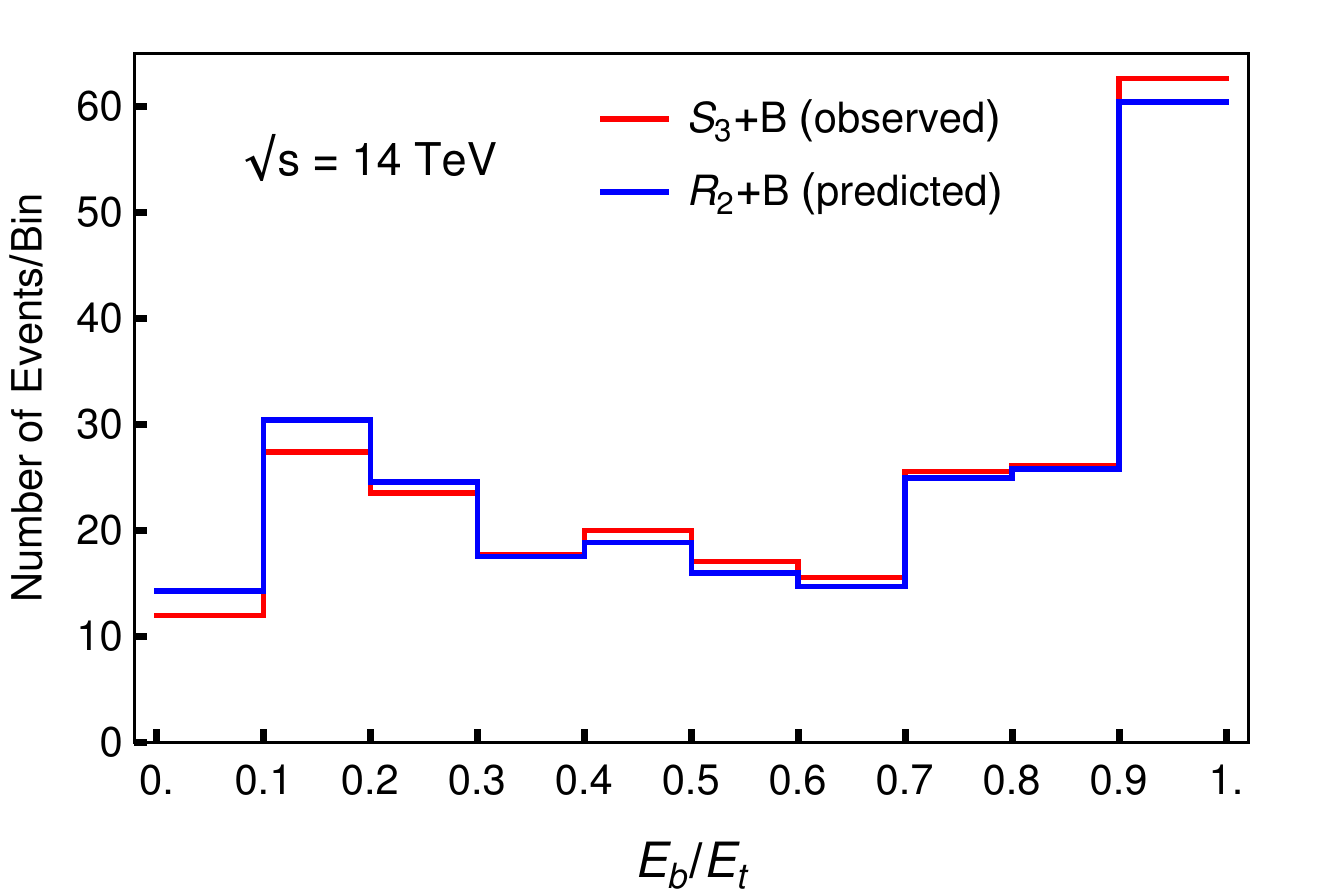}
\includegraphics[angle=0,width=0.45\linewidth,height=0.35\linewidth]{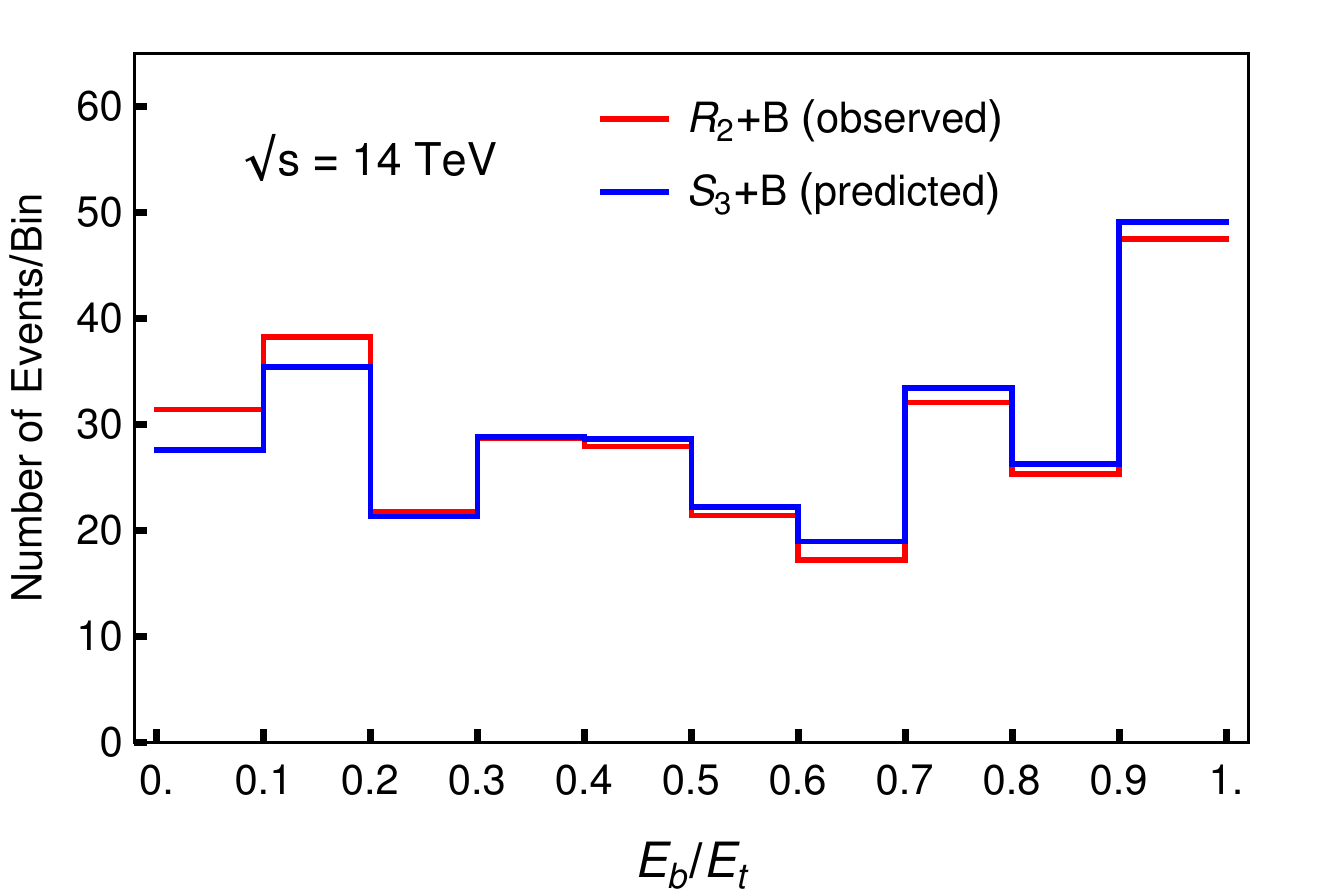}
\caption{The signal+background event distributions in $\frac{E_b}{E_t}$ for observed and predicted models data after applying an optimal BDT cut (given in Tab.~\ref{tab:MVA-result}) with 3000 $\rm fb^{-1}$. To find the events for the predicted model, the signal events of it are passed through the same BDT model used for finding the event numbers of the observed model.}  
 \label{fig:Delphes_Eb_by_ET}
\end{figure}

\begin{figure}[H]
\centering
\includegraphics[angle=0,width=0.45\linewidth,height=0.35\linewidth]{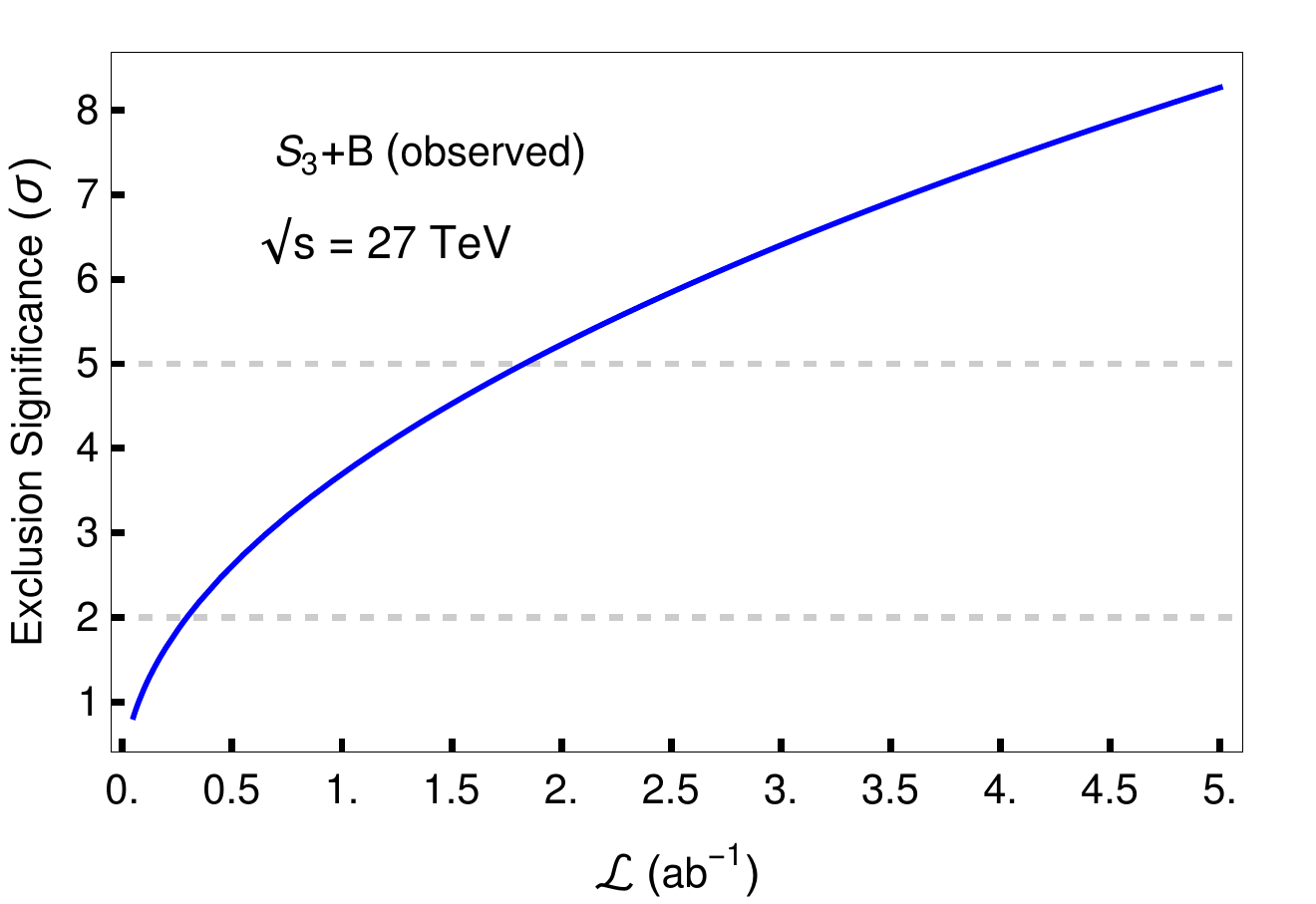}
\includegraphics[angle=0,width=0.45\linewidth,height=0.35\linewidth]{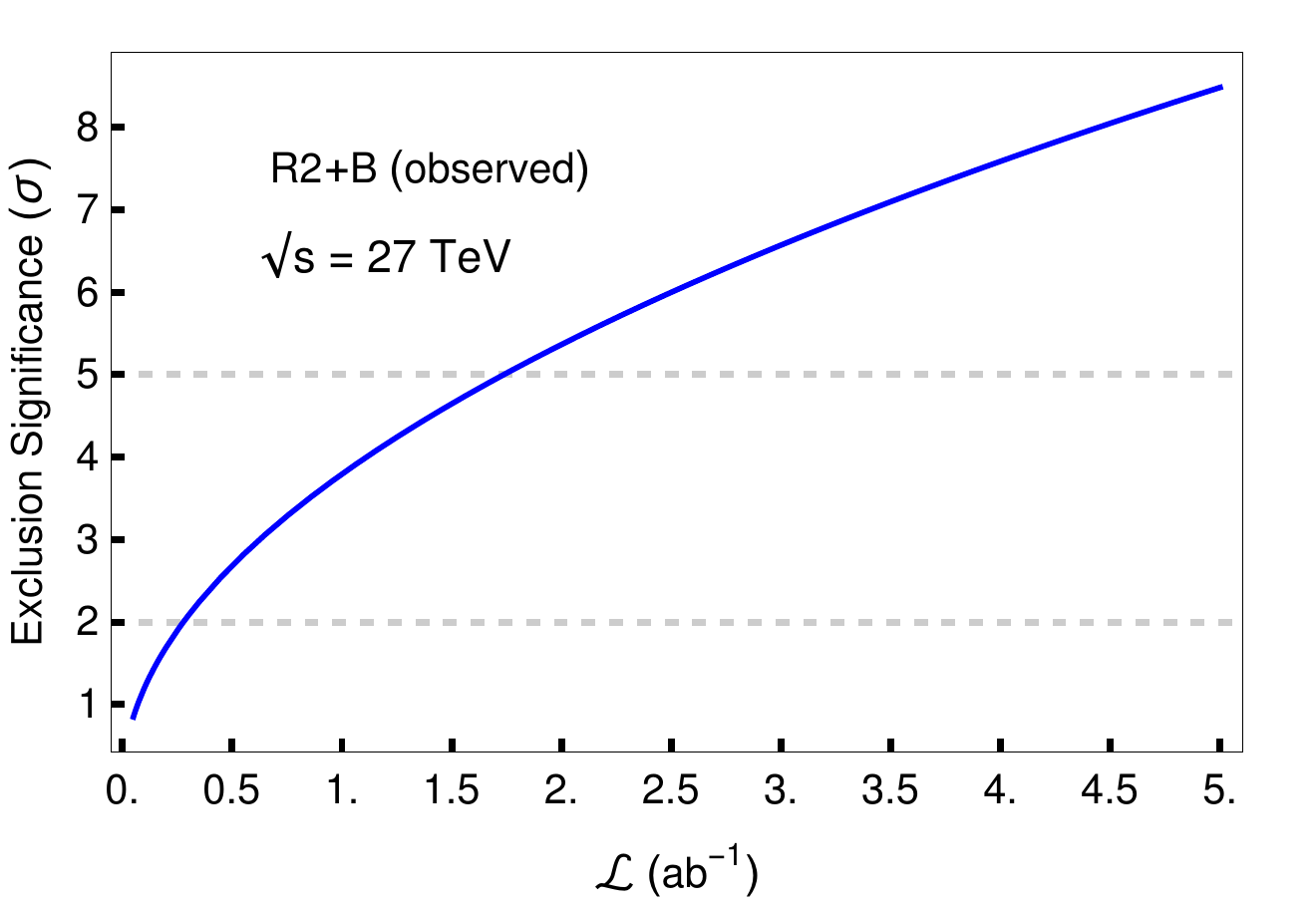}
\caption{ The exclusion significance vs. required luminosity at 27 TeV collider by projecting the distributions at 14 TeV collider to 27 TeV collider.  The mass of the leptoquark has been taken to be $M_{LQ}= 1300\ \rm{GeV}$.}  
 \label{fig:Req_Luminosity_27TeV}
\end{figure}

We have considered both the scenarios when either of the models is observed and the other one is predicted for which we want to find the exclusion significance. To find distribution for events numbers for the predicted model, the signal events of it are scanned through  the same BDT-model used for the observed model. In Fig.~\ref{fig:Delphes_Eb_by_ET}, we show $\frac{E_b}{E_t}$ distribution for event numbers for observed and predicted models at 14 TeV LHC with 3 $\rm ab^{-1}$ of data.
 For the analysis, we have taken first 8 bins, starting from the left, of the $\frac{E_b}{E_t}$ distribution\footnote{For the bins around z=0.8 and above, the b-jet energy is not very well measured. In this region, because of very high boost of top quark, b-jet gets contaminated with the other two light
jets, originated from the hadronic decay of top quark.}, given in Fig.~\ref{fig:Delphes_Eb_by_ET}. We obtain an exclusion significance (Z) of $0.98\ \sigma$, when $S3+B$ is taken as observed at the LHC and $R2+B$ is considered as the predicted one. For the reverse case, we obtain Z value as $1.01\ \sigma$, see Tab.~\ref{tab:exclusion_significance}. As the exclusion significance is quite low, it shows that two models can not be distinguished well at the LHC. However, it is prompting to see whether these two models can be distinguished at 27 TeV (HE-LHC) collider for the same mass of the leptoquark. To do this study, we assume that the shape of the signal and individual background distributions will remain same at the 27 TeV LHC as that of the 14 TeV collider. We then scale the distributions by overall factors after calculating their total cross-sections at these two different center of mass energy colliders. In Fig.~\ref{fig:Req_Luminosity_27TeV}, we show the plot for exclusion significance vs. required luminosity at the HE-LHC. We find that with moderate amount of luminosity (around 1800 $\rm fb^{-1}$) at this collider, either of the models can be excluded at 5$\sigma$ significance when the other one appears as observed. In the last column of Tab.~\ref{tab:exclusion_significance}, we show the exclusion significances for 3 $\rm ab^{-1}$ data at this collider.

%
%
%
%
%

\begin{table}[H]
\begin{center}
 \begin{tabular}{|c|c|c|c|c|}
\hline
 \multirow{ 2}{*}{ $\mathcal{L}$}  &  \multirow{ 1}{*}{ predicted} &  \multirow{ 1}{*}{ observed } & Rejection Prob. (Z)  & Rejection Prob. (Z) \\
  &   &   & (14 TeV)  & (27 TeV) \\
\hline
\multirow{ 2}{*}{$3 ab^{-1}$} & $R_2+B$ & $S_3+B$ & 0.98\ $\sigma$  & 6.45\ $\sigma$ \\
\cline{2-5}
& $S_3+B$ & $R_2+B$ & 1.01\ $\sigma$  & 6.59\ $\sigma$  \\
\hline
\end{tabular}
\caption{Probability of excluding one model when other model is the observed model at 14 TeV LHC and 27 TeV HE-LHC with $\mathcal{L}=3\ ab^{-1}$.}
 \label{tab:exclusion_significance} 
\end{center}
\end{table}

\section{Conclusions}
\label{sec:conclsn}
TeV-scale leptoquarks that can emerge from various models are well-motivated and phenomenologically interesting to be searched at high-energy collider experiments. Present work investigates the pair production of third-generation $\frac{2}{3}e$-charged scalar leptoquark at the LHC using NLO QCD accuracy, matched to parton shower for precise probing. Among different potential scalar leptoquark models, two primary interests - $S_3$ and $R_2$ can be probed by looking at their decay into a top with a tau neutrino, thus producing a compelling signature of a pair of top-like fatjets along with substantial missing transverse energy. Here tops, created from heavy leptoquarks, are naturally boosted and therefore considering them as boosted jets is quite meaningful.

With a precise understanding of jet physics, it is now possible to study the intrinsic substructure and properties of such jets, thereby pointing out the origin of these jets with a high degree of accuracy. Therefore the considered channel has excellent potential for separating the tiny signal from the overwhelming SM background. Parton shower effects are included in our study and its usefulness in the low transverse momentum region is seen in Fig.~\ref{fig:PSeffectoverNLOFO}. We also demonstrate that the factorization and renormalization scale uncertainties for the  NLO+PS events are much lower than that of LO+PS events (see Fig.~\ref{fig:scale_variation} and Tab.~\ref{tab:cross-section}).

For accurate prediction, we include all the relevant background processes with two to four extra QCD radiations and normalize them using the available higher-order QCD-corrected production cross-section. Different high-level variables, such as MET, $M_{\text{eff}}$, $\Delta R(J_0, J_1)$, $\Delta \phi(J_i,\slashed{E}_T)$, jet substructure based pruned jet mass and N-subjettiness are proved to be efficacious
to pinpoint the signal. Multivariate analysis is carried out individually for these two models and we show that at the 14 TeV LHC with an integrated luminosity of 3000 $fb^{-1}$, the leptoquarks of mass 1380 GeV can be discovered ($5\sigma$), and up to 1520 GeV can be excluded ($2\sigma$).

Among the two scalar leptoquark models considered here, it is interesting to note that the top quarks resulting from the decay of leptoquarks possess different helicities. Most of the high-level variables utilized for multivariate analysis are not sensitive to this polarization. Only the jet mass variables acquire some minor effect due to the modified distribution pattern in the decay process. However, these are insignificant enough, thereby providing almost equivalent mass constraints for both models.

We further construct different polarization sensitive variables to distinguish these scalar leptoquark models of the same charge. We exhibit the effectiveness of such variables in terms of  ({\em i}) an angular variable in the top quark's rest frame, ({\em ii}) the ratio of the energy variables $\frac{E_b}{E_t}$. Such effects are demonstrated at the truth level and after including parton shower and (fat)jet formation (see Figs.~\ref{fig:LO-pol-var-truth-level}, \ref{fig:NLO-pol-var-Delphes-level}).
Significant distortion is noticeable following detector simulation and (fat)jet formation. This is primarily attributed to the contamination and poor measurement efficiency of the b-jet momenta within a highly collimated top-like fatjet.
The log-likelihood-ratio (LLR) hypothesis test is used to distinguish the models in the presence of combined background events. We find that the statistical exclusion significance remains low at around $1 \sigma$ confidence level at the LHC.
However, it is shown that the 27 TeV collider can play a promising role and it is estimated that the required luminosity would be around 300 $fb^{-1}$ (1800 $fb^{-1}$) to distinguish these two models with $2 \sigma$ ($5 \sigma$) significance.


\section*{Acknowledgements}
We thank Rinku Maji and Saurabh K. Shukla for fruitful discussions.
 Computational works are performed using the HPC resources (Vikram-100 HPC) and the TDP project at PRL.

\appendix
\section{Appendix}
\label{appen}

\subsection{Distribution of daughter of top quark}
\label{app:dist_daught_top}
Here we will find the differential distribution of decay width of right handed top in its rest frame. The distribution for left handed particle can be obtained similarly.

\begin{figure}[H]
\begin{center}
\includegraphics[angle=0,scale=0.8]{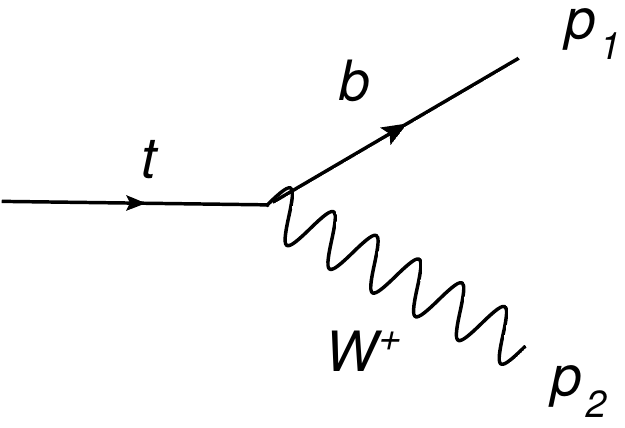}\\
\caption{The Feynman diagram for top decay.}  
 \label{fig:feynman_dia_top_decay}
\end{center}
\end{figure}

The matrix element can be written as 
\begin{align}
M &= \bar{u}_b(p_1) \frac{i g}{\sqrt{2}} \gamma_\mu P_L u_t(m_t) \epsilon_{W^{+}}^{\mu*}(p_2)
\end{align}

So, $|M|^{2}$ can be written as 
\begin{align}
|M|^{2} &= \frac{g^2}{2}\bar{u}_b(p_1)  \gamma_\mu P_L u_t(m_t) \bar{u}_t(m_t) P_R \gamma_\nu  u_b(p_1)\epsilon_{W^{+}}^{\mu*}(p_2)\epsilon_{W^{+}}^{\nu}(p_2) \notag \\
\sum_{final\ spins} |M|^{2} &= -\frac{g^2}{2}Tr[ \gamma_\mu P_L u_t(m_t) \bar{u}_t(m_t) P_R \gamma_\nu  (\slashed{p_1}+m_b)](g^{\mu\nu} - \frac{p^{\mu}p^{\nu}}{M_W^2}) \label{eq:top_decay_spin_sum} 
\end{align}

In the Weyl basis, $\gamma^{\mu} = \begin{pmatrix}
0 & \sigma^{\mu} \\
\bar{\sigma}^{\mu} & 0 
\end{pmatrix}$ and  $\gamma^{5} = \begin{pmatrix}
-I & 0 \\
0 & I 
\end{pmatrix}$, where $\sigma^{\mu}= (I, \boldsymbol{\sigma})$ and $\bar{\sigma}^{\mu}= (I, -\boldsymbol{\sigma})$. The spinor in the rest frame is given by  $$u^{s}_t({m_t}) =\sqrt{m_t}\begin{pmatrix}
\xi^{s} \\
\xi^{s} 
\end{pmatrix}$$

Using above expressions, Eq.~\ref{eq:top_decay_spin_sum} can be written as
\begin{align}
\sum_{final\ spins} |M|^{2} &= -\frac{g^2 m_t}{2}Tr[ 
\begin{pmatrix}
0 & 0 \\
\bar{\sigma}_{\mu} & 0 
\end{pmatrix}
\begin{pmatrix}
\xi^{s}\xi^{s\dagger} & \xi^{s}\xi^{s\dagger} \\
\xi^{s}\xi^{s\dagger} & \xi^{s}\xi^{s\dagger} 
\end{pmatrix} 
\begin{pmatrix}
0 & 0 \\
\bar{\sigma}_{\nu} & 0 
\end{pmatrix}  \slashed{p_1}](g^{\mu\nu} - \frac{p^{\mu}p^{\nu}}{M_W^2}) \notag\\
&= -\frac{g^2 m_t}{2}Tr[ 
\bar{\sigma}^{\mu} 
\xi^{s}\xi^{s\dagger} 
\bar{\sigma}_{\mu} 
 {\sigma} \cdot p_1 - \frac{\bar{\sigma}\cdot p_2 
\xi^{s}\xi^{s\dagger} 
\bar{\sigma}\cdot p_2 
 {\sigma} \cdot p_1}{M_W^2}] \label{eq:top_decay_spin_sum_2}
\end{align}

For the spin up top $\xi^{s}\xi^{s\dagger}$ can be written as $\frac{I+\sigma^3}{2}$. So Eq.~\ref{eq:top_decay_spin_sum_2} can be written as 
\begin{align}
\sum_{final\ spins} |M|^{2} &= -\frac{g^2 m_t}{4}Tr[ 
\bar{\sigma}^{\mu} 
(I+\sigma^3) 
\bar{\sigma}_{\mu} 
 {\sigma} \cdot p_1 - \frac{\bar{\sigma}\cdot p_2 
(I+\sigma^3) 
\bar{\sigma}\cdot p_2 
 {\sigma} \cdot p_1}{M_W^2}] \label{eq:top_decay_spin_sum_3}
\end{align}

Using $\sigma^i\sigma^j + \sigma^j\sigma^i = 2 \delta^{ij} I$, the following can be proven:

\begin{align}
\bar{\sigma}^{\mu}\bar{\sigma}_{\mu} &= -2I  \label{eq:App-rel1} \\
\bar{\sigma}^{\mu} \sigma^3 \bar{\sigma}_{\mu} &= 2\sigma^3 \label{eq:App-rel2} \\
\bar{\sigma}\cdot p_2 \bar{\sigma}\cdot p_2  &= (p_2^0)^2 I + \vec{p_2}^2 I + 2\sigma^i p_2^i  p_2^0 \label{eq:App-rel3}  \\
\bar{\sigma}\cdot p_2  \sigma^3 \bar{\sigma}\cdot p_2  &= (p_2^0)^2  \sigma^3 + 2 p_2^3 p_2^0 I - \sigma^3 \vec{p_2}^2  + 2\sigma^i p_2^3  p_2^i \label{eq:App-rel4} 
\end{align}

Again the following relations can be proven easily
\begin{align}
Tr(\sigma^i\sigma^j) &= 2 \delta^{ij} \label{eq:App-rel5} \\
Tr(\sigma^i) &= 0 \label{eq:App-rel6} \\
Tr(\sigma \cdot a) &= 2 a^{0}\label{eq:App-rel7} 
\end{align}

The different parts of Eq.~\ref{eq:top_decay_spin_sum_3} can be obtained using Eq.~\ref{eq:App-rel1}-Eq.~\ref{eq:App-rel7}. After using them, we have

\begin{align}
Tr[\bar{\sigma}^{\mu}\bar{\sigma}_{\mu}\sigma \cdot p_1] &= -4p_1^0   \\
Tr[\bar{\sigma}^{\mu} \sigma^3 \bar{\sigma}_{\mu}\sigma \cdot p_1] &= -4p_1^3  \\
Tr[\bar{\sigma}\cdot p_2 \bar{\sigma}\cdot p_2\sigma \cdot p_1]  &= 2p_1^0(p_2^0)^2  + 2p_1^0\vec{p_2}^2  -4 \vec{p_1}\cdot \vec{p_2}  p_2^0 \label{eq:tr-part3} \\
Tr[\bar{\sigma}\cdot p_2  \sigma^3 \bar{\sigma}\cdot p_2\sigma \cdot p_1]  &= - 2(p_2^0)^2 p_1^3  + 4 p_2^3 p_2^0 p_1^0 +  2p_1^3 \vec{p_2}^2  - 4 p_2^3  \vec{p_1}\cdot \vec{p_2}  \label{eq:tr-part4} 
\end{align}

Using energy conservation in the top rest frame,
\begin{align}
m_t &= |\vec{p}_1| + \sqrt{{\vec{p}_1}^2+m_W^2}  \notag\\ 
 |\vec{p}_1| &= \frac{ m_t^2  - m_W^2}{2 m_t } 
\label{eq:p1mod} 
\end{align}

Eq.~\ref{eq:tr-part3} can be written as 
 
\begin{align}
& 2p_1^0(p_2^0)^2  + 2p_1^0\vec{p_2}^2  -4 \vec{p_1}\cdot \vec{p_2}  p_2^0 \notag \\
=& 2p_1^0(m_W^2 + 2\vec{p_1}^2)  +4 \vec{p_1}\cdot \vec{p_1} (m_t- p_1^0) \notag \\
=& 2|\vec{p}_1| (m_W^2  +2 |\vec{p}_1| m_t) = 2|\vec{p}_1| m_t^2
\end{align}

Eq.~\ref{eq:tr-part4} can be written as 
 
\begin{align}
 & - 2(p_2^0)^2 p_1^3  + 4 p_2^3 p_2^0 p_1^0 +  2p_1^3 \vec{p_2}^2  - 4 p_2^3  \vec{p_1}\cdot \vec{p_2} \notag\\
=& - 2 p_1^3( (p_2^0)^2 -\vec{p_2}^2 )   + 4 p_2^3 ( p_2^0 p_1^0  - \vec{p_1}\cdot \vec{p_2} )\notag\\
=& - 2 p_1^3( m_W^2 )   - 2 p_1^3 (m_t^2 -m_W^2) = - 2 p_1^3 m_t^2
\end{align}

Using the above formulae in Eq.~\ref{eq:top_decay_spin_sum_3}, we have
\begin{align}
 \sum_{final\ spins} |M|^{2}  &= -\frac{g^2 m_t}{4}[ -4p_1^0 -4p_1^3 - \frac{ 2|\vec{p}_1| m_t^2 - 2 p_1^3 m_t^2}{M_W^2}] \notag\\
  &= {g^2 m_t}[ (1+ \frac{m_t^2}{2M_W^2}) + (1- \frac{m_t^2}{2M_W^2}) \cos{\theta_b} ] \notag\\
    &= {g^2 m_t}(1+ \frac{m_t^2}{2M_W^2})[ 1  + \frac{2M_W^2- m_t^2}{2M_W^2+ m_t^2} \cos{\theta_b} ] \notag\\
&= {g^2 m_t}(1+ \frac{m_t^2}{2M_W^2})[ 1  + k_b\ \cos{\theta_b} ], \notag
\end{align}

where $k_b=\frac{2M_W^2- m_t^2}{2M_W^2+ m_t^2}= -0.4$, spin analyzing power of b-quark. 

The differential distribution of decay width for right handed top quark is given by
\begin{align}
\frac{d\Gamma}{d \cos{\theta_b}} &= \frac{1}{2 m_t^2} \frac{|\vec{p}_1|}{ 8 \pi}\sum_{final\ spins} |M|^{2} \notag\\ 
 &= \frac{1}{2 m_t^2} \frac{1}{ 8 \pi} \frac{ m_t^2  - m_W^2}{2 m_t }{g^2 m_t}(1+ \frac{m_t^2}{2M_W^2})[ 1  + k_b\ \cos{\theta_b} ]\notag\\ 
 &= \frac{g^2}{32\pi }\frac{(m_t^2-m_W^2)(2M_W^2+ m_t^2)}{m_t^2m_W^2}[ 1  + k_b\ \cos{\theta_b} ]\notag
\end{align}

For the spin down top, $\xi^{s}\xi^{s\dagger}$ can be written as $\frac{I-\sigma^3}{2}$. So it is easy follow that in the above expression there will be a minus sign in front of $k_b$ for this case.
\subsection{Relation between $\rm \cos\ {\theta'_b}$ and z}
\label{app:rel_Cos_and_z}
In the following, quantities in the lab frame will be denoted by unprimed symbols, whereas in the top rest frame they will be denoted by primed symbols\footnote{Note in the main text, we did not use any prime for the angle in the rest frame. So the $\rm \cos\ {\theta'_b}$ here is same as $\rm \cos\ {\theta_b}$ in the main text.}. So in the rest frame of the top quark the angle of bottom quark's direction of motion with the boost direction of top quark is given by 
\begin{align}
{\rm \cos\ {\theta'_b}} = \frac{p_b^{z'}}{| \vec{p}_b\,\!'|},
\label{eq:costhetaprime}
\end{align}
where z and $z'$ axes are along the direction of motion of the top quark in the lab frame.

Using Lorentz transformation between two frames with $\beta=\frac{|\vec{p}_t|}{E_t}$ and  $\gamma=\frac{1}{\sqrt{1-\beta^2}}=\frac{E_t}{m_t}$
\begin{align}
p_b^{z'} = -\gamma \beta E_b + \gamma p_b^z
\label{eq:rel_pbzprime_and_Ebandpbz} 
\end{align}

Using energy conservation in the lab frame,
\begin{align}
E_t  &=  E_b + \sqrt{({\vec{p}_t - \vec{p}_b})^2+m_W^2} \notag\\
{p_b^z} &= -\frac{m_t^2 + m_b^2 - m_W^2 - 2 E_t E_b}{2 |\vec{p}_t| }  \notag\\
  &= \frac{ E_b}{\beta} - \frac{m_t^2 + m_b^2 - m_W^2}{2 \beta \gamma m_t  }
\label{eq:rel_pbz_and_Eb} 
\end{align}

Using Eq.~\ref{eq:rel_pbz_and_Eb} in Eq.~\ref{eq:rel_pbzprime_and_Ebandpbz}, we have

\begin{align}
p_b^{z'} &= -\gamma \beta E_b + \gamma  \frac{ E_b}{\beta} - \frac{m_t^2 + m_b^2 - m_W^2}{2 \beta  m_t} \notag\\
 &= \frac{z E_t}{\gamma\beta} - \frac{m_t^2 + m_b^2 - m_W^2}{2 \beta  m_t} \notag\\
 &= \frac{z m_t}{\beta} - \frac{m_t^2 + m_b^2 - m_W^2}{2 \beta  m_t} \notag
\end{align}

Assuming $m_b=0$, 
\begin{align}
 p_b^{z'} &= \frac{1}{\beta}(z m_t - \frac{m_t^2-m_W^2}{2 m_t})
\label{eq:rel_pbzprime_and_Ebandpbz_new_mb0} 
\end{align}

Using energy conservation in the top rest frame,
\begin{align}
m_t &= E_b' + \sqrt{{\vec{p}_b\,\!'}^2+m_W^2}  \notag\\ 
 | \vec{p}_b\,\!'| &= \frac{ m_t^2  - m_W^2}{2 m_t } 
\label{eq:rel_pbmodprime} 
\end{align}

Using Eq.~\ref{eq:rel_pbzprime_and_Ebandpbz_new_mb0} and Eq.~\ref{eq:rel_pbmodprime}, in Eq.~\ref{eq:costhetaprime}, we have 

\begin{align}
{\rm \cos{\theta'_b}} &= \frac{1}{\beta}(\frac{2 m_t^2}{m_t^2-m_W^2} z - 1)
\label{eq:costhetaprime_new}
\end{align}

\bibliographystyle{JHEP}
\bibliography{ref.bib}
\end{document}